\documentclass{article}
\usepackage{bm} 
\usepackage{amsmath} 
\usepackage{amssymb} 
\usepackage{amsthm} 
\usepackage[boxed]{algorithm}
\usepackage[noend]{algpseudocode}
\usepackage{pifont} 
\usepackage{braket}

\newcommand{\labs}{\left|}
\newcommand{\rabs}{\right|}
\newcommand{\lnorm}{\left\|}
\newcommand{\rnorm}{\right\|}
\newcommand{\lcurly}{\left\{}
\newcommand{\rcurly}{\right\}}
\newcommand{\lbrac}{\left[}
\newcommand{\rbrac}{\right]}
\newcommand{\lpar}{\left(}
\newcommand{\rpar}{\right)}

\algnewcommand\algorithmicinput{\textbf{Input:}}
\algnewcommand\Input{\item[\algorithmicinput]}
\algnewcommand\algorithmicoutput{\textbf{Output:}}
\algnewcommand\Output{\item[\algorithmicoutput]}
\algnewcommand\algorithmicalgorithm{\textbf{Algorithm:}}
\algnewcommand\Algorithm{\item[\algorithmicalgorithm]}

\newtheorem{theorem}{Theorem}[section]
\newtheorem{lemma}{Lemma}[section]
\newtheorem{proposition}{Proposition}[section]
\newtheorem{corollary}{Corollary}[section]

\newtheorem{remark}{Remark}[section]
\newtheorem{definition}{Definition}[section]
\newtheorem{problem}{Problem}[section]

\DeclareMathOperator{\tr}{tr}

\DeclareMathOperator{\diag}{diag}

\DeclareMathOperator{\poly}{poly}
\DeclareMathOperator{\polylog}{polylog}

\DeclareMathOperator{\gap}{\Delta}

\newcommand\vecb{\boldsymbol{\mathrm{b}}}

\newcommand\vece{\boldsymbol{\mathrm{e}}}
\newcommand\vecf{\boldsymbol{\mathrm{f}}}

\newcommand\vecl{\boldsymbol{\mathrm{l}}}

\newcommand\vecq{\boldsymbol{\mathrm{q}}}

\newcommand\vecu{\boldsymbol{\mathrm{u}}}
\newcommand\vecv{\boldsymbol{\mathrm{v}}}
\newcommand\vecw{\boldsymbol{\mathrm{w}}}

\newcommand\vecz{\boldsymbol{\mathrm{z}}}

\newcommand\vecftilde{\widetilde{\boldsymbol{\mathrm{f}}}}

\newcommand\vecltilde{\widetilde{\boldsymbol{\mathrm{l}}}}

\newcommand\vecztilde{\widetilde{\boldsymbol{\mathrm{z}}}}

\newcommand\vecuhat{\widehat{\boldsymbol{\mathrm{u}}}}

\newcommand\veczhat{\widehat{\boldsymbol{\mathrm{z}}}}

\newcommand\matA{\boldsymbol{\mathrm{A}}}
\newcommand\matB{\boldsymbol{\mathrm{B}}}
\newcommand\matC{\boldsymbol{\mathrm{C}}}
\newcommand\matD{\boldsymbol{\mathrm{D}}}
\newcommand\matE{\boldsymbol{\mathrm{E}}}

\newcommand\matG{\boldsymbol{\mathrm{G}}}
\newcommand\matH{\boldsymbol{\mathrm{H}}}
\newcommand\matI{\boldsymbol{\mathrm{I}}}

\newcommand\matK{\boldsymbol{\mathrm{K}}}

\newcommand\matP{\boldsymbol{\mathrm{P}}}
\newcommand\matQ{\boldsymbol{\mathrm{Q}}}
\newcommand\matR{\boldsymbol{\mathrm{R}}}
\newcommand\matS{\boldsymbol{\mathrm{S}}}
\newcommand\matT{\boldsymbol{\mathrm{T}}}
\newcommand\matU{\boldsymbol{\mathrm{U}}}
\newcommand\matV{\boldsymbol{\mathrm{V}}}
\newcommand\matW{\boldsymbol{\mathrm{W}}}
\newcommand\matX{\boldsymbol{\mathrm{X}}}

\newcommand\matZ{\boldsymbol{\mathrm{Z}}}

\newcommand\matDelta{\boldsymbol{\mathrm{\Delta}}}

\newcommand\matLambda{\boldsymbol{\mathrm{\Lambda}}}

\newcommand\matPi{\boldsymbol{\mathrm{\Pi}}}

\newcommand\matAtilde{\widetilde{\boldsymbol{\mathrm{A}}}}

\newcommand\matDtilde{\widetilde{\boldsymbol{\mathrm{D}}}}

\newcommand\matHtilde{\widetilde{\boldsymbol{\mathrm{H}}}}

\newcommand\matQtilde{\widetilde{\boldsymbol{\mathrm{Q}}}}

\newcommand\matTtilde{\widetilde{\boldsymbol{\mathrm{T}}}}
\newcommand\matUtilde{\widetilde{\boldsymbol{\mathrm{U}}}}
\newcommand\matVtilde{\widetilde{\boldsymbol{\mathrm{V}}}}

\newcommand\matLambdatilde{\widetilde{\boldsymbol{\mathrm{\Lambda}}}}

\newcommand\matPitilde{\widetilde{\boldsymbol{\mathrm{\Pi}}}}
\newcommand\matSigmatilde{\widetilde{\boldsymbol{\mathrm{\Sigma}}}}

\newcommand\matBbar{\bar{\boldsymbol{\mathrm{B}}}}

\newcommand\matUbar{\bar{\boldsymbol{\mathrm{U}}}}
\newcommand\matVbar{\bar{\boldsymbol{\mathrm{V}}}}

\newcommand\matHhat{\widehat{\boldsymbol{\mathrm{H}}}}

\newcommand\matQhat{\widehat{\boldsymbol{\mathrm{Q}}}}

\newcommand\matUhat{\widehat{\boldsymbol{\mathrm{U}}}}

\newcommand\matLambdahat{\widehat{\boldsymbol{\mathrm{\Lambda}}}}

\newcommand{\fl}{\boldsymbol{\mathsf{fl}}}
\newcommand{\umach}{\textbf{\textup{u}}}

\newcommand{\QR}{\mathsf{QR}}
\newcommand{\MM}{\mathsf{MM}}
\newcommand{\BTMM}{\mathsf{MM_{BT}}}
\newcommand{\TRID}{\mathsf{TRID}}
\newcommand{\DIAGONALIZE}{\mathsf{DIAGONALIZE}}
\newcommand{\GAP}{\mathsf{GAP}}
\newcommand{\HALVE}{\mathsf{HALVE}}
\newcommand{\FLINF}{\mathsf{INF}}
\newcommand{\cmm}{\beta}
\newcommand{\cfmm}{\xi}
\newcommand{\err}{\mathsf{err}}
\newcommand{\matmulexponent}{\omega}
\newcommand{\flopcost}{\mathcal{F}}
\newcommand{\projectormatrix}{\matPi}
\newcommand{\projectormatrixtilde}{\matPitilde}
\newcommand{\geneigmatrix}{\Lambda}
\newcommand{\fmmalgo}{FMM}
\usepackage[utf8]{inputenc}
\usepackage{listings}
\usepackage{fullpage}
\usepackage{mathdots} 
\usepackage{enumerate}

\title{Deterministic complexity analysis of Hermitian eigenproblems}

\author{Aleksandros Sobczyk 
\\IBM Research and ETH Zurich
\\Zurich, Switzerland}
\date{}

\usepackage{graphicx}
\usepackage{xcolor}
\usepackage{fullpage} 
\usepackage{hyperref} 

\RequirePackage[T1]{fontenc}  
\RequirePackage[tt=false, type1=true]{libertine} 
\RequirePackage[varqu]{zi4} 
\RequirePackage[libertine]{newtxmath} 

\usepackage{setspace}
\usepackage{todonotes}
\hypersetup{
    colorlinks,
    linkcolor={blue},
    citecolor={blue},
    urlcolor={blue}
}

\begin{document}

\maketitle
\begin{abstract}
In this work we revisit the arithmetic and bit complexity of Hermitian eigenproblems. Recently, [BGVKS, FOCS 2020] proved that a (non-Hermitian) matrix $\matA$ can be diagonalized with a randomized algorithm in $O(n^{\omega}\log^2(\tfrac{n}{\epsilon}))$ arithmetic operations, where $\omega\lesssim 2.371$ is the square matrix multiplication exponent, and [Shah, SODA 2025] significantly improved the bit complexity for the Hermitian case. Our main goal is to obtain similar deterministic complexity bounds for various Hermitian eigenproblems. 
In the Real RAM model, we show that a Hermitian matrix can be diagonalized deterministically in $O(n^{\omega}\log(n)+n^2\mathrm{polylog}(\tfrac{n}{\epsilon}))$ arithmetic operations, improving the classic deterministic $\widetilde O(n^3)$ algorithms, and derandomizing the aforementioned state-of-the-art. 
The main technical step is a complete, detailed analysis of a well-known divide-and-conquer tridiagonal eigensolver of Gu and Eisenstat [GE95], when accelerated with the Fast Multipole Method, asserting that it can accurately diagonalize a symmetric tridiagonal matrix in nearly-$O(n^2)$ operations.
In finite precision, we show that an algorithm by Sch\"onhage [Sch72] to reduce a Hermitian matrix to tridiagonal form is stable in the floating point model, using $O(\log(\tfrac{n}{\epsilon}))$ bits of precision. This leads to a deterministic algorithm to compute all the eigenvalues of a Hermitian matrix in $O\left( n^{\omega}\flopcost\left(\log(\tfrac{n}{\epsilon})\right) + n^2\polylog(\tfrac{n}{\epsilon})\right)$ bit operations, where $\flopcost(b)\in\widetilde O(b)$ is the bit complexity of a single floating point operation on $b$ bits. This improves the best known $\widetilde{O}(n^3)$ deterministic and $O\left( n^{\omega}\log^2(\tfrac{n}{\epsilon})\mathcal{F}\left(\log(\tfrac{n}{\epsilon})\right)\right)$ randomized complexities. We conclude with some other useful subroutines such as computing spectral gaps, condition numbers, and spectral projectors, and with some open problems.
\end{abstract}
\newpage
\tableofcontents
\newpage
\section{Introduction}
    \label{section:introduction}
    Eigenproblems arise naturally in many applications. Given a matrix $\matA$, the goal is to compute (a subset of) the eigenvalues $\lambda$ and/or the eigenvectors $\vecv$, which satisfy
    \begin{align*}
        \matA\vecv = \lambda\vecv.
    \end{align*}
    The properties of the given matrix, as well as the quantities that need to be computed can vary depending on the application, giving rise to different variants of the eigenproblem.    These include $(i)$ \textit{eigenvalue problems}, such as the approximation of eigenvalues, singular values, spectral gaps, and condition numbers, $(ii)$ \textit{eigenspace problems}, which refer to the approximation of eigenvectors, spectral projectors, and invariant subspaces, and $(iii)$ \textit{diagonalization problems}, which involve the (approximate) computation of all the eigenvalues and eigenvectors of a matrix (or pencil), i.e., a full spectral factorization and/or the Singular Value Decomposition (SVD).

    In this work we focus on algorithms for \textit{Hermitian eigenproblems}, i.e., the special case where the input matrix is Hermitian. Our motivation to dedicate the analysis to this special class is twofold. 
    First, they arise in many fundamental applications in Machine Learning, Spectral Graph Theory, and Scientific Computing. 
    For example, the SVD, which is ubiquitous for many applications such as low rank approximations \cite{papadimitriou1998latent,frieze2004fast,drineas2008relative,clarkson2017low}, directly reduces to a Hermitian eigenproblem. 
    Second, the spectral theorem states that a Hermitian matrix $\matA$ can always be written in a factored form $\matA=\matQ\matLambda\matQ^*$, where $\matQ$ is unitary and $\matLambda$ is diagonal with real entries. 
    This alleviates several difficulties of the non-Hermitian case, which can lead to efficient dedicated algorithms. 

    Algorithms for eigenproblems have been studied for decades, some of the earliest being attributed to Jacobi \cite{jacobi1846leichtes,golub2000eigenvalue}. We refer to standard textbooks for an overview of the rich literature  \cite{demmel1997applied,golub2013matrix,parlett1998symmetric,bhatia2007perturbation,saad2011numerical}. 
    Some landmark works include the power method \cite{mises1929praktische}, the Lanczos algorithm \cite{lanczos1950iteration}, and the paramount QR algorithm \cite{francis1961qr,francis1962qr,kublanovskaya1962some}, which  has been recognized as one of the ``top ten algorithms of the twentieth century'' \cite{dongarra2000guest}, signifying the importance of the eigenproblem in science and engineering.   
    Given a Hermitian tridiagonal matrix $\matT$ with size $n\times n$, the algorithm can compute a set of approximate eigenvalues in $ O(n^2\log(\tfrac{n}{\epsilon}))$ arithmetic operations, based on the seminal analyses of Wilkinson \cite{wilkinson1968global} and Dekker and Traub \cite{dekker1971shifted}. A set of approximate eigenvectors can be also returned in $ O(n^3\log(\frac{n}{\epsilon}))$ operations. In conjunction with the classic unitary similarity transformation algorithm of Householder \cite{householder1958unitary}, the shifted-QR algorithm has heavylifted the computational burden of solving eigenvalue problems for decades, both in theory and in practice. 
    A detailed bit complexity analysis was provided recently in \cite{banks2021global1,banks2022global2,banks2022global3}.
    
    Despite the daunting literature, several details regarding the computational complexity of many algorithms remain unclear. 
    It is well-known, for example, that the cubic arithmetic complexity to compute the eigenvalues of a dense matrix is not optimal: a classic work by Pan and Chen \cite{pan1999complexity} showed that the eigenvalues can be approximated in $O(n^\omega)$ arithmetic operations, albeit without detailing how to also compute eigenvectors, or to fully diagonalize a matrix. 
    Here $\omega\geq 2$ is the \textit{matrix multiplication exponent}, i.e. the smallest number such that two $n\times n$ matrices can be multiplied in $O(n^{\omega+\eta})$ arithmetic operations, for any $\eta>0$. The current  best known upper bound is $\omega< 2.371339$  \cite{alman2024more}, and we will write $n^{\omega}$ instead of $n^{\omega+\eta}$ hereafter for simplicity. 
    Recently, Banks, Garza-Vargas, Kulkarni, and Srivastava \cite{banks2022pseudospectral} described a numerically stable randomized algorithm to compute a provably accurate diagonalization, in $\widetilde O(n^{\omega})$ operations, improving the previous  best-known bounds, specifically, $\widetilde O(n^3)$ (Hermitian) and $O(n^{10})$ (non-Hermitian) \cite{armentano2018stable}. \cite{shah2025hermitian} further improved the analysis for the Hermitian case, and several works have studied extensions and related applications \cite{demmel2024inverse,demmel2024generalized,schneider2024pseudospectral,sobczyk2024invariant}. 
    To date, we are not aware of any \emph{deterministic} algorithm  that achieves the same arithmetic (or bit) complexity with provable approximation guarantees, even for the Hermitian case.
    In this work, we proceed step-by-step and analyze several variants of the aforementioned eigenvalue, eigenspace, and diagonalization problems, in different \textit{models of computation}, and report complexity upper bounds with provable approximation guarantees.
    
    \subsection{Models of computation and complexity}
    From the Abel-Ruffini theorem it is known that the eigenvalues and/or the eigenvectors of matrices with size larger than $n=5$ can not be computed exactly. Even in exact arithmetic, they can only be approximated. Before analyzing algorithms, we first need to clarify what are the quantities of interest, to define how accuracy is measured, and what is the underlying model of computation. The main two models that we use to analyze algorithms are the Real RAM and the Floating Point model, described below. 

    \begin{description}
    \item[Exact real arithmetic (Real RAM):]
        For the Real RAM model we follow the  definition of \cite{erickson2022smoothing}. The machine has a memory (RAM) and registers that store real numbers in infinite precision. Moving real numbers between the memory and the registers takes constant time. A processor can execute arithmetic operations $\{+,-,\times,/,\sqrt\cdot,>\}$ on the real numbers stored in the registers exactly, without any errors, in constant time. Other functions such as $\log(x),\exp(x),$ and trigonometric functions, are not explicitly available, but they can be  efficiently approximated up to some additive error, e.g. with a truncated polynomial expansion, using a polylogarithmic number of basic arithmetic operations.
        Bit-wise operations on real numbers are forbidden, since this can give the machine unreasonable computing power \cite{hartmanis1974power,schonhage1979power,erickson2022smoothing}. 

    \item[Floating point:]
    In this model, a real number $\alpha\in\mathbb{R}$ is rounded to a floating point number $
        \fl(\alpha) = s\times 2^{e-t} \times m,$ where $s=\pm 1$, $e$ is the exponent, $t$ is the bias, and $m$ is the significand.
    Floating point arithmetic operations also introduce rounding errors, i.e., for two floating point numbers $\alpha$ and $\beta$, each  operation $\odot \in\{+,-,\times,/\}$ satisfies:
    \begin{align}
        \fl(\alpha\odot\beta) = (1+\theta)(\alpha\odot\beta),
        \qquad
        \text{and also }
        \qquad
        \fl(\sqrt{a})=(1+\theta)\sqrt{a},
        \label{eq:elementwise_flop_errors}
    \end{align} 
    where $|\theta|\leq\umach$, and $\umach$ is the machine precision. Assuming a total of $b$ bits for each number, every floating point operation costs $\flopcost(b)$ bit operations, where typically $\flopcost(b)\in O(b^2)$, or even $\widetilde O(b)$, with more advanced algorithms \cite{schonhage1971fast,furer2007faster,harvey2021integer}.
    More details can be found in Appendix \ref{appendix:floating_point_arithmetic}.
    \end{description}

    In Section \ref{section:hermitian_eigenvalues} we will also use as a subroutine an algorithm from \cite{bini1991parallel,bini1998computing}, which was originally analyzed in Boolean RAM model. We describe in detail how to use it in the corresponding section.

    \paragraph*{Arithmetic and boolean complexity} Given a model of computation, the \textit{arithmetic complexity} of an algorithm is quantified in terms of the arithmetic operations executed. The \textit{boolean} (or \textit{bit}) \textit{complexity}, accounts for the total number of boolean operations (i.e. boolean gates with maximum fan-in two).

\subsection{Notation}
Matrices and vectors are denoted with bold capital and small letters, respectively. 
$\lVert \matA\rVert$ is the spectral norm of $\matA$, $\lVert \matA\rVert_F$ its Frobenius norm, $\kappa(\matA)=\|\matA\|\|\matA^{\dagger}\|$ is the condition number, and $\Lambda(\matA)$ its spectrum. For the complexity analysis we denote the geometric series $S_{x}(m)=\sum_{l=1}^m (2^{x-2})^l$. As already mentioned, for simplicity, the complexity of multiplying two $n\times n$ matrices will be denoted as $O(n^\omega)$, instead of $O(n^{\omega+\eta})$, for arbitrarily small $\eta>0$. 
We also use the standard notation $\omega(a,b,c)$ for the complexity of multiplying two rectangular matrices with sizes $n^{a}\times n^b$ and $n^b\times n^c$ in time $O(n^{\omega(a,b,c)})$, and therefore $\omega:=\omega(1,1,1)$. For example, for $a=c=1$ and $b=2$, the best known bound for $\omega(1,2,1)\approx 3.250035$ \cite{alman2024more}, which is slightly better than naively performing $n$ square multiplications in  $O(n\cdot n^{\omega})\lesssim O(n^{3.371339})$. $\flopcost(b)$ denotes the bit complexity of a single floating point operation on $b$ bits. 

\subsection{Methods and Contributions}
    
    \subsubsection{Real RAM}
    \label{paragraph:intro_real_ram} We start with the analysis in exact arithmetic, which is the simplest model to analyze numerical algorithms. The goal is to obtain end-to-end upper bounds for the arithmetic complexity of approximately diagonalizing symmetric tridiagonal matrices and, ultimately, dense Hermitian matrices, as well as to approximate the SVD. To measure the accuracy, we follow the notion of \textit{backward-stability} (or \textit{backward-approximation}) for Hermitian diagonalization from 
    \cite{nakatsukasa2013stable}.
    Formally, the following problems are considered.
    \begin{problem} Backward-approximate diagonalization problems in exact arithmetic.
        \label{problem:problems_in_exact_arithmetic}
        \begin{enumerate}[(i)]
            \item \textbf{Symmetric arrowhead/tridiagonal diagonalization:} Given a symmetric arrowhead or tridiagonal matrix $\matA$ with size $n\times n$, $\|\matA\|\leq 1$, and accuracy $\epsilon\in(0,1)$, compute a diagonal matrix $\matLambdatilde$ and a matrix $\matUtilde$, such that $\matUtilde=\matU+\matE_{\matU}$, where $\matU$ is orthogonal and $\|\matE_{\matU}\|\leq \epsilon$, and $\lnorm \matA-\matU\matLambdatilde\matU^\top \rnorm \leq \epsilon$.
            \item \textbf{Hermitian diagonalization:} Given a Hermitian matrix $\matA$ with size $n\times n$, $\|\matA\|\leq 1$, and accuracy $\epsilon\in(0,1)$, compute a diagonal matrix $\matLambdatilde$ and a matrix $\matUtilde$, such that $\matUtilde=\matU+\matE_{\matU}$, where $\matU$ is unitary and $\|\matE_{\matU}\|\leq \epsilon$, and $\lnorm \matA-\matU\matLambdatilde\matU^* \rnorm \leq \epsilon$.
            \item \textbf{SVD:} Given a matrix $\matA\in\mathbb{C}^{m\times n}$, $m=n^k$, $k\geq 1$, $\|\matA\|\leq 1$, and accuracy $\epsilon\in(0,1)$ compute a diagonal matrix $\matSigmatilde$, an $m\times n$ matrix $\matUtilde=\matU+\matE_{\matU}$, and an $n\times n$ matrix $\matVtilde=\matV+\matE_{\matV},$ such that $\matU^*\matU=\matV^*\matV=\matI$, $\lnorm \matE_{\{\matU,\matV\}}\rnorm \leq \epsilon$, and $\lnorm \matA-\matU\matSigmatilde\matV^* \rnorm \leq \epsilon$.
        \end{enumerate}
    \end{problem}
 
    To obtain rigorous complexity guarantees for these problems, with respect to all the involved parameters, we start from Problem \ref{problem:problems_in_exact_arithmetic}-(i), namely, diagonalization of symmetric tridiagonal matrices. To that end, we revisit the so-called fast tridiagonal eigensolvers, which aim to reduce the complexity from cubic to $\widetilde O(n^2)$ operations. 
    Many such algorithms have been studied in the literature \cite{dongarra1987fully,dhillon1997new,bini1992practical,bini1991parallel,gill1990n,vogel2016superfast,ou2022superdc,stor2015accurate,barlow1993error}, most of which are based on the divide-and-conquer (DC) strategy of Cuppen \cite{cuppen1980divide}. 
    The algorithms in all of the aforementioned works have been rigorously analyzed, however, explicit complexity bounds in terms of strictly solving Problem \ref{problem:problems_in_exact_arithmetic}-(i) are not detailed. 
    We resolve this by providing an end-to-end complexity analysis of the algorithm of Gu and Eisenstat \cite{gu1995divide}. In their original work, the authors outlined how to accelerate several parts of the algorithm with the Fast Multipole Method (FMM) \cite{rokhlin1985rapid}, which could eventually lead to a final complexity of $\widetilde O(n^2)$. However, the actual analysis of this approach and the FMM details were not provided. 
    \cite{musco2018stability} further extended the analysis in floating point, but it also relies on a numerically stable FMM implementation, which is not detailed. In this work, we use the elegant FMM analysis of \cite{gu1993stable,livne2002n,cai2020stable}, which is particularly suited for the problems considered.
    It is detailed in the following Proposition \ref{proposition:fmm}. \begin{proposition}[FMM]
    \label{proposition:fmm}
    There exists an algorithm, which we refer to as $(\epsilon,n)$-approximate FMM (or $(\epsilon,n)$-\fmmalgo, for short), which takes as input:
    \begin{itemize}
        \item a kernel function $k(x)\in\lcurly \log(|x|), \frac{1}{x}, \frac{1}{x^2} \rcurly$,
        \item $2n+m$ real numbers: $\{x_1,\ldots,x_m\}\cup \{c_1,\ldots,c_n\}\cup\{y_1,\ldots,y_n\}$, and a constant $C$, such that $m\leq n$ and for all $i\in[m],j\in[n]$ it holds that
        \begin{align*}
            |x_i|,|c_j|,|y_j|<C
            \qquad
            \text{ and }
            \qquad
            |x_i-y_j|\geq \Omega(\poly(\tfrac{\epsilon}{n})).
        \end{align*}
    \end{itemize}
    It returns $m$ values $\widetilde f(x_1),\ldots,\widetilde f(x_m)$ such that
    $
        \labs
            \widetilde f(x_i)-f(x_i)
        \rabs
        \leq \epsilon,
    $
    for all $i\in[m]$, where $f(x) = \sum_{j=1}^n c_j k(x_i-y_j)$,
    in a total of $O\lpar 
        n\log^{\cfmm}(\tfrac{n}{\epsilon})
    \rpar$ arithmetic operations, where $\cfmm\geq 1$ is a small constant that is independent of $\epsilon,n$.
    \begin{proof}
        The result follows from the analysis of \cite{gu1993stable,livne2002n,cai2020stable}. A detailed description can be found in the restatement, Proposition \ref{proposition:fmm_appendix}, in Appendix \ref{appendix:fast_multipole_method}.
    \end{proof}
\end{proposition}
By taking advantage of the FMM, the analysis from Section \ref{section:tridiagonal_diagonalization} and the supporting Appendices \ref{section:arrowhead_preliminaries} and \ref{appendix:fast_multipole_method} leads to the following Theorem \ref{theorem:alg_recursive_diagonalization}, whose proof can be found in Section \ref{section:proof_of_tridiagonal_diagonalization}.    
\begin{theorem}
    \label{theorem:alg_recursive_diagonalization}
    Given an unreduced symmetric tridiagonal matrix $\matT$ with size $n\times n$, $\|\matT\|\leq 1,$ an accuracy $\epsilon\in(0,1/2)$, and an $(\epsilon,n)$-\fmmalgo\   implementation as in Proposition \ref{proposition:fmm},  the recursive algorithm of \cite{gu1995divide} (Algorithm \ref{algorithm:recursive_diagonalization}), returns an approximately orthogonal matrix $\matUtilde$ and a diagonal matrix $\matLambdatilde$ such that
    \begin{align*}
        \lnorm \matT - \matUtilde\matLambdatilde\matUtilde^\top \rnorm \leq \epsilon,
        \quad
        \lnorm \matUtilde^\top \matUtilde - \matI \rnorm \leq \epsilon/n^2,
    \end{align*}
    or, stated alternatively,
    \begin{align*}
        \matUtilde=\matU+\matE_{\matU},
        \quad
        \matU^\top\matU=\matI,
        \quad
        \lnorm \matE_{\matU} \rnorm \leq \epsilon/n^2,
        \quad
        \lnorm \matT - \matU\matLambdatilde\matU^\top \rnorm \leq \epsilon,
        \quad
    \end{align*}
    using a total of $O\lpar n^2\log^{\cfmm+1}(\tfrac{n}{\epsilon})\rpar$ arithmetic operations and comparisons, where $\xi\geq 1$ is a small constant that depends on the specific FMM implementation and it is independent of $\epsilon,n$.
   
\end{theorem}
\begin{table}[htb]
    \centering
    \caption{Comparison of algorithms for the Problems \ref{problem:problems_in_exact_arithmetic} (i)-(iii) in the Real RAM model. Here $\xi\geq 1$ is a small constant which depends on the FMM implementation (see Prop. \ref{proposition:fmm}). The algorithms marked with (\textbf{R}) are randomized and succeed with high probability (at least $1-1/\poly(n)$).}
    \label{table:main_results_exact_arithmetic_diagonalization}
    {
    \begin{tabular}{lll}
    \hline
      & Arithmetic Complexity  & Comment  \\\hline\hline
    
        Prob. \ref{problem:problems_in_exact_arithmetic}-(i)
        &  
        &
    \\
         
        Arrowhead/Trid.
        &  
        &
    \\   
        diagonalization
        &  
        &
    \\   Refs. \cite{cuppen1980divide,o1990computing,gu1995divide}& 
        $O(n^3)+\widetilde O(n^2)$ 
        &
        Conjectured $ \widetilde O(n^2)$ 
    \\    
        Theorem \ref{theorem:alg_recursive_diagonalization}
        & 
        $O\lpar n^2\log^{\cfmm+1}(\tfrac{n}{\epsilon})\rpar$ 
        &
        Req. FMM (Prop. \ref{proposition:fmm})
    \\\hline
        Prob. \ref{problem:problems_in_exact_arithmetic}-(ii)
        &  
        &
    \\      
        Hermitian
        &  
        &
    \\      
        diagonalization
        &  
        &
    \\ 
        Refs. \cite{banks2022pseudospectral,shah2025hermitian} (\textbf{R}) 
        & 
        $O\lpar n^{\omega}\log^2(\tfrac{n}{\epsilon})\rpar$ 
        &
        -
    \\
        Ref. \cite{benor2018quasi} (\textbf{R}) & 
        $\widetilde O\lpar n^{\omega+1}\rpar$ 
        &
        Conjectured  $\widetilde O(n^{\omega})$
    \\
        Refs. \cite{dekker1971shifted,wilkinson1968global,hoffmann1978new} & 
        $O\lpar n^{3}\log(\tfrac{n}{\epsilon})\rpar$ 
        &
        Shifted-QR
    \\
        Ref. \cite{nakatsukasa2013stable} & 
        $\widetilde O\lpar n^{3}\rpar$ 
        &
        Req. separated spectrum
    \\
        Ref. \cite{pan1999complexity} 
        & 
        $O\lpar n^{\omega} + n\polylog(\tfrac{n}{\epsilon})\rpar$ 
        &
        Only eigenvalues
    \\    
        Corollary \ref{corollary:hermitian_diagonalization}
        & 
        $O\lpar n^\omega\log(n) + n^2\log^{\cfmm+1}(\tfrac{n}{\epsilon})\rpar$ 
        &
        Req. FMM (Prop. \ref{proposition:fmm})
    \\\hline
        Prob. \ref{problem:problems_in_exact_arithmetic}-(iii), SVD
        &  
        &
    \\  Shifted-QR on $\matA^*\matA$ & 
        $O\lpar n^{\omega(1,k,1)} + n^3\log(\tfrac{n}{\epsilon}) \rpar$ 
        &
        Partial error analysis 
    \\
        Refs. \cite{banks2022pseudospectral,shah2025hermitian,kacham2024faster} (\textbf{R}) 
        & 
        $O\lpar 
            n^{\omega(1,k,1)} 
            +
            n^{\omega}\log^2(\tfrac{n}{\epsilon})
        \rpar$ 
        &
        Partial error analysis
    \\    
        Theorem \ref{theorem:svd}
        & 
        $O\lpar n^{\omega(1,k,1)} + n^\omega\log(n) 
        +
        n^2\polylog(\tfrac{n\kappa(\matA)}{\epsilon})\rpar$ 
        &
        Req. FMM (Prop. \ref{proposition:fmm})
        \\\hline
    \end{tabular}
    }
\end{table}    

This result has a direct application to the dense Hermitian diagonalization problem. 
The best-known complexities of deterministic algorithms, with end-to-end analysis, are at least cubic. 
For example, the aforementioned shifted-QR algorithm requires $O(n^3\log(n/\epsilon))$ arithmetic operations, even for tridiagonal matrices. 
The cubic barrier originates from the accumulation of the elementary rotation matrices to form the final eigenvector matrix. 
The so-called \emph{spectral divide-and-conquer} methods (see e.g. \cite{demmel1997applied,nakatsukasa2013stable}) have also at least cubic complexities in the deterministic case. 
The two main difficulties in their analysis are the basis computation of spectral projectors, for which the best-known deterministic complexity bounds are $\Omega(n^3)$, e.g. via Strong Rank-Revealing QR (RRQR) factorization \cite{gu1996efficient}, and the choice of suitable splitting points, which relies on the existence of spectral gaps. 

Randomness can help to overcome certain difficulties in the analysis. \cite{demmel2007fastla} analyzed a numerically stable Randomized URV decomposition, which can be used to replace RRQR for basis computations in spectral DC algorithms. A highly parallelizable Hermitian diagonalization algorithm with end-to-end analysis was proposed in \cite{benor2018quasi}. While the reported sequential arithmetic complexity is $O(n^{\omega+1})$, the authors conjectured that it can be further reduced to $\widetilde O(n^{\omega})$.  The first end-to-end $\widetilde O(n^\omega)$ complexity upper bound was achieved in
\cite{banks2022pseudospectral}. One of the main techniques was to use random perturbations to ensure that the pseudospectrum is well-separated, which helps to find splitting points in spectral DC. \cite{shah2025hermitian} further improved the analysis for Hermitian matrices.

Theorem \ref{theorem:alg_recursive_diagonalization} can be directly used to obtain a simple and deterministic solution for the  Hermitian diagonalization problem. 
Specifically, Corollary \ref{corollary:hermitian_diagonalization} states that the problem can be solved in $O(n^{\omega}\log(n) + n^2\polylog(n/\epsilon))$ arithmetic operations, which is both faster and fully deterministic. 
This is achieved by combining Theorem \ref{theorem:alg_recursive_diagonalization} with the (rather overlooked) algorithm of Schönhage \cite{schonhage1972unitare}, who proved that a Hermitian matrix can be reduced to tridiagonal form with unitary similarity transformations in $O(n^\omega\log(n)c(\omega))$ arithmetic operations, where $c(\omega)=O(1)$ if $\omega$ is treated as a fixed constant larger than $2$, while $c(\omega)=O(\log(n))$ if it turns out that $\omega=2$. 

As a consequence, Theorem \ref{theorem:svd} reports similar deterministic complexity results for the SVD. The SVD is a fundamental computational kernel in many applications, such as Principal Component Analysis \cite{jolliffe2002principal}, and it is also widely used as a subroutine for many other advanced algorithms, e.g. \cite{papadimitriou1998latent,frieze2004fast,drineas2008relative,boutsidis2014optimal,boutsidis2014randomized,cohen2015dimensionality,boutsidis2016optimal,clarkson2017low}, to name a few. A straightforward algorithm to compute it is to first form the Gramian matrix $\matA^*\matA$, and then diagonalize it. Other classic SVD algorithms, such as the widely adopted Golub-Kahan bidiagonalization and its variants \cite{golub1965calculating}, or polar decomposition-based methods \cite{nakatsukasa2013stable,nakatsukasa2010optimizing}, avoid the computation of $\matA^*\matA$ for numerical stability reasons, but they also rely on a diagonalization algorithm as a subroutine.  
\cite{kacham2024faster} elaborated on a matrix multiplication time SVD algorithm, by using \cite{banks2022pseudospectral} to diagonalize $\matA^\top \matA$, albeit without fully completing the backward stability analysis. 
The following Theorem \ref{theorem:svd}, which is proved in Appendix \ref{appendix:svd_analysis}, states our main result. 

\begin{theorem}[SVD]
    \label{theorem:svd}
    Let $\matA\in\mathbb{C}^{m\times n}$, $m=n^k$, $k\geq 1$. Assume that $1/n^c\leq \|\matA\|\leq 1$, for some constant $c$. Given accuracy $\epsilon\in(0,1/2)$ and an $(\epsilon,n)$-\fmmalgo\   (see Prop. \ref{proposition:fmm}), we can compute three matrices $\matUtilde\in\mathbb{C}^{m\times n},\matSigmatilde\in\mathbb{R}^{n\times n},\matVtilde\in\mathbb{C}^{n\times n}$ such that $\matSigmatilde$ is diagonal with positive diagonal elements and
    \begin{align*}
        \matA=\matUtilde\matSigmatilde\matVtilde^*,
        \quad
        \lnorm \matUtilde^* \matUtilde - \matI \rnorm \leq \epsilon,
        \quad
        \lnorm \matVtilde^* \matVtilde - \matI \rnorm \leq \epsilon/(\kappa(\matA)n^2)\ll \epsilon,
    \end{align*}
    or, stated alternatively,
    \begin{align*}
        \lnorm \matA-\matU\matSigmatilde\matV^*\rnorm \leq \epsilon,
        \quad
        \matUtilde&=\matU+\matE_{\matU},
        \quad
        \lnorm \matE_{\matU }\rnorm \leq \epsilon,
        \quad
        \matU^*\matU=\matI,
        \\
        \matVtilde&=\matV+\matE_{\matV},
        \quad
        \lnorm \matE_{\matV}\rnorm \leq \epsilon/n^2,
        \quad
        \matV^*\matV=\matI.
    \end{align*}
    The algorithm requires a total of at most $O\lpar n^{\omega(1,k,1)} + n^\omega\log(n)+ n^2\polylog(\tfrac{n\kappa(\matA)}{\epsilon})\rpar$ arithmetic operations, where $\kappa(\matA)=\|\matA\|\|\matA^\dagger\|$.
\end{theorem}

To summarize this section, in Table \ref{table:problems_in_finite_precision} we list all the deterministic arithmetic complexities achieved in the Real RAM model, for all the aforementioned problems, and compare with some important existing algorithms.

\subsubsection{Finite precision}
\label{paragraph:intro_finite_precision} Similar deterministic complexity upper bounds are obtained for several problems in finite precision. In particular, we study the following problems, for which we seek to bound the boolean complexity, i.e., the total number of bit operations.
\begin{problem}
    \label{problem:problems_in_finite_precision}
    Main problems in finite precision.
    \begin{enumerate}[(i)]
        \item \textbf{Tridiagonal reduction:} Given a Hermitian matrix $\matA$ with floating point elements, reduce $\matA$ to tridiagonal form using (approximate) unitary similarity transformations. In particular, return a tridiagonal matrix $\matTtilde$, and (optionally)  an approximately unitary matrix $\matQtilde$, such that \begin{align*}
            \lnorm \matQtilde\matQtilde^*-\matI\rnorm \leq \epsilon,
            \quad
            \text{and}
            \quad
            \lnorm \matA - \matQtilde\matTtilde\matQtilde^* \rnorm \leq \epsilon \lnorm \matA \rnorm,
            \end{align*} 
        \item \textbf{Hermitian eigenvalues:} Given a Hermitian matrix $\matA$, $\|\matA\|\leq 1$, and accuracy $\epsilon\in(0,1)$, compute a set of approximate eigenvalues $\widetilde\lambda_i$ such that $\labs \widetilde\lambda_i-\lambda_i(\matA)\rabs \leq \epsilon$.
    \end{enumerate}
\end{problem}
\begin{table}[htb]
    \caption{Boolean complexity for Problems \ref{problem:problems_in_finite_precision}, for matrix size $n\times n$ and accuracy $\epsilon\in(0,1)$. }
    \label{table:problems_in_finite_precision}
    \centering
    \noindent
        \begin{tabular}{lll}
        \hline
          & Boolean Complexity 
          & Comment
        \\\hline\hline
        Prob. \ref{problem:problems_in_finite_precision}-(i)        
        &
        &
    \\ 
    Tridiag. Reduction
        &
        &
    \\
        Refs. \cite{householder1958unitary,higham2002accuracy}
        & 
        $O\lpar 
            n^3  
            \flopcost\lpar 
                \log\left(\tfrac{n}{\epsilon}\right) 
            \rpar\rpar$ 
        &
        Standard Householder reduction
    \\    
        Theorem  \ref{theorem:stable_tridiagonal_reduction} 
        & 
        $O\lpar 
            n^{\omega}\log(n)
            \flopcost 
            \lpar 
                \log(\tfrac{n}{\epsilon}) 
            \rpar 
        \rpar$
        &
        \cite{schonhage1972unitare} with stable fast QR \cite{demmel2007fastla}
    \\\hline
            Prob. \ref{problem:problems_in_finite_precision}-(ii) 
            &
            &
        \\
            Herm. Eigenvalues
            &
            &
        \\
            Ref. \cite{shah2025hermitian}& 
            $O\lpar n^{\omega}\log^2(\tfrac{n}{\epsilon})
            \flopcost\Big( 
                \log(\tfrac{n}{\epsilon})
            \Big)
            \rpar$
            &
            Randomized, $\Pr[\text{success}]\geq 1-\frac{1}{n}$
        \\    
            Theorem  \ref{theorem:hermitian_eigenvalues}
            & 
            $O\lpar
                n^{\omega}\flopcost\Big( \log(\tfrac{n}{\epsilon}) \Big)
                +
                n^2\polylog(\tfrac{n}{\epsilon})
            \rpar$ 
            &
            Deterministic, Thm. \ref{theorem:stable_tridiagonal_reduction} + \cite{bini1998computing}
        \\
        \hline
        \end{tabular}
    \label{table:fp_results}
\end{table}
Regarding deterministic algorithms, with end-to-end-analysis, the standard approach is to first reduce the Hermitian matrix to tridiagonal form with Householder transformations \cite{householder1958unitary}, which can be done stably in $O(n^3)$ arithmetic operations using $O(\log(n/\epsilon))$ bits of precision; see e.g. \cite{higham2002accuracy}. 
Thereafter, there is a plethora of algorithms (e.g. the ones mentioned in the previous section) for the eigenvalues of the tridiagonal matrix, with varying complexities and stability properties. However, the total boolean complexity cannot be lower than $\Omega(n^3\flopcost(\log(\frac{n}{\epsilon})))$ due to the Householder reduction step.
Other well-known deterministic and numerically stable algorithms in the literature also require at least $n^3$ arithmetic operations to compute all the eigenvalues \cite{paige1980accuracy,demmel1992jacobi,nakatsukasa2013stable}, and at least $\polylog(n,1/\epsilon)$ bits of precision in a floating point machine. The arithmetic complexity of the algorithm of \cite{pan1999complexity} scales as $O(n^\omega)$ with respect to the matrix size $n$, but the boolean complexity can increase up to $O(n^{\omega+1})$ in rational arithmetic. \cite{louis2016accelerated} described a randomized algorithm to compute only the largest eigenvalue in nearly $O(n^\omega)$ bit complexity. The fastest algorithm to compute all the eigenvalues of a Hermitian matrix is \cite{shah2025hermitian}, which requires $O(n^\omega\polylog(n/\epsilon))$ boolean operations and succeeds with high probability.

Randomized eigenvalue algorithms have also been studied in the sketching/streaming setting \cite{andoni2013eigenvalues,needell2022testing,swartworth2023optimal}.
The (optimal) algorithm of \cite{swartworth2023optimal} has not been analyzed in finite-precision, but, due to its simplicity, it should be straightforward to achieve. 
The  algorithm approximates all the eigenvalues of a Hermitian matrix $\matA$ up to additive error $\epsilon\|\matA\|_F$. However, to reduce the error to a spectral-norm bound $\epsilon\|\matA\|$, the algorithm internally needs to diagonalize a matrix with size $\Omega(n)$, and therefore it does not provide any improvement against any other Hermitian eigenvalue solver.
Nevertheless, our main results can be directly applied as the main eigenvalue subroutine of this algorithm, and to help analyze its bit complexity.

To improve the aforementioned Hermitian eigenvalue algorithms,
we first prove in Theorem \ref{theorem:stable_tridiagonal_reduction} it is proved that Schönhage's algorithm is numerically stable in the floating point model of computation. Thereafter, we carefully combine it with the algorithms of \cite{bini1991parallel,bini1992practical,bini1998computing} which provably and deterministically approximate all the eigenvalues of a symmetric tridiagonal matrix in $\widetilde O(n^2)$ boolean operations. The latter algorithm is analyzed in the Boolean RAM model, therefore, in order to use it we need to convert the floating point elements of the tridiagonal matrix that is returned by Theorem \ref{theorem:stable_tridiagonal_reduction} to integers, which can be done efficiently under reasonable assumptions on the initial number of bits used to represent the floating point numbers. 
This is described in detail in the proof of our main Theorem \ref{theorem:hermitian_eigenvalues}. Our result derandomizes and slightly improves the final bit complexity of the algorithm of \cite{shah2025hermitian}, which has the currently best known bit complexity for this problem. 
Table 
\ref{table:fp_results} summarizes this discussion.

As a direct consequence of Theorem \ref{theorem:hermitian_eigenvalues}, we also provide the analysis for several other useful subroutines related to eigenvalue/eigenvector computations, including:
\begin{enumerate}[(i)]
    \item \textbf{Singular values and condition number}: In Proposition \ref{proposition:alg_sigmak} we describe how to obtain relative error approximations of singular values. In Corollary \ref{corollary:alg_cond} we show how to compute the condition number.
    \item\textbf{Definite pencil eigenvalues}: In Corollary \ref{corollary:hermitian_definite_pencil_eigenvalues} we demonstrate how to extend Theorem \ref{theorem:hermitian_eigenvalues} to compute the eigenvalues of Hermitian-definite pencils. 
    \item\textbf{Spectral gaps}: In Corollary \ref{corollary:alg_deterministic_spectral_gap} we show how to compute the spectral gap and the midpoint between any two eigenvalues of a Hermitian-definite pencil. Our algorithm is deterministic and it requires significantly less bits of precision than the algorithm of \cite{sobczyk2024invariant}, who described a randomized algorithm for this problem that is slightly faster than applying \cite{banks2022pseudospectral} as a black-box, but it only computes a single spectral gap.
    \item\textbf{Spectral projector}: Corollary \ref{corollary:spectral_projector} details how to compute spectral projectors on invariant subspaces of Hermitian-definite pencils, which are important for many applications.
\end{enumerate}
\subsection{Outline}
The paper is organized as follows. In Section \ref{section:tridiagonal_diagonalization} we analyze the algorithm of \cite{gu1995divide} when implemented with the FMM and its applications (see also  Appendices \ref{section:arrowhead_preliminaries}, \ref{appendix:fast_multipole_method}, and \ref{appendix:tridiagonal_diagonalization}). In Section \ref{section:tridiagonal_reduction_stability} it is proved that Sch\"onhage's algorithm is numerically stable in floating point, and it is used as a preprocessing step to compute the eigenvalues of Hermitian matrices. For the proof we use the technical lemmas that are proved in the supporting Appendix \ref{appendix:tridiagonal_reduction}. In Section \ref{section:applications_tridiagonal_reduction} we mention some direct applications to compute singular values, pencil eigenvalues, spectral gaps, and spectral projectors. We finally conclude and state some open problems in Section \ref{section:conclusion}.

\subsection*{Acknowledgements}{
I am grateful to Efstratios Gallopoulos, Daniel Kressner, and David Woodruff for helpful discussions.}

\section{Diagonalization of symmetric tridiagonal and arrowhead matrices in Real RAM}
\label{section:tridiagonal_diagonalization}
Our main target is to compute the eigenvalues and the eigenvectors of tridiagonal symmetric matrices in nearly linear time. To derive the desired result, we provide an analysis the divide-and-conquer algorithm of Gu and Eisenstat \cite{gu1995divide}, when implemented with the FMM from Proposition \ref{proposition:fmm}.

The algorithm of \cite{gu1995divide} first ``divides'' the problem by partitioning the (unreduced) tridiagonal matrix $\matT$ as follows:
\begin{align*}
    \matT = \begin{pmatrix}
        \matT_1 & \beta_{k+1}\vece_k & \\
        \beta_{k+1}\vece_k^\top & \alpha_{k+1} & \beta_{k+2}\vece_1^\top \\
         & \beta_{k+2}\vece_1 & \matT_2
    \end{pmatrix}.
\end{align*}
If one has access to the spectral decomposition of $\matT_1$ and $\matT_2$, i.e. $\matT_1=\matQ_1\matD_1\matQ_1^\top$ and $\matT_2=\matQ_2\matD_2\matQ_2^\top$, then $\matT$ can be factorized as
\begin{align}
    \label{eq:tridiagonal_to_arrowhead}
    \begin{pmatrix}
        & \matQ_1 & \\
        1 & & \\
        & & \matQ_2
    \end{pmatrix}    
    \begin{pmatrix}
        \alpha_{k+1}& \beta_{k+1}\vecl_1^\top   & \beta_{k+2}\vecf_2^\top\\
        \beta_{k+1}\vecl_1 & \matD_1 & \\
         \beta_{k+2}\vecf_2& & \matD_2
    \end{pmatrix}
    \begin{pmatrix}
        & 1 & \\
        \matQ_1^\top & & \\
        & & \matQ_2^\top
    \end{pmatrix}
    =
    \matQ\matH\matQ^\top,
\end{align}
where $\vecl_1^\top$ is the last row of $\matQ_1$ and $\vecf_2^\top$ is the first row of $\matQ_2$, and $\matH$ has the so-called \textit{arrowhead} structure. Thus, given this form, to compute the spectral decomposition of $\matT$, it suffices to diagonalize $\matH$. One can then apply recursively the algorithm to compute the spectral decompositions of $\matT_1$ and $\matT_2$, and, finally, at the ``conquering stage,'' combine the solutions with the eigendecomposition of $\matH$.

For the individual steps to be computed efficiently, we will need to use the FMM. Specifically, we will need to use it to evaluate the functions that are listed in Equations \eqref{eq:fmm_1}-\eqref{eq:fmm_5}, where the evaluation points will also satisfy certain criteria that are detailed in Lemma \ref{lemma:arrowhead_preliminaries}. 
We ensure that these criteria are met by using a deflation pre-processing step (see also Appendix \ref{appendix:arrowhead_deflation}), which allows us to take advantage of the FMM, specifically, Proposition \ref{proposition:fmm}. 

\subsection{Symmetric arrowhead diagonalization}
The first step is to provide an end-to-end complexity analysis for the arrowhead diagonalization algorithm of \cite{gu1995divide}, when implemented with the FMM. We start with an $n\times n$ arrowhead matrix of the form
\begin{align}
    \label{eq:arrowhead}
    \matH=\begin{pmatrix}
        \alpha & \vecz^\top \\
        \vecz & \matD
    \end{pmatrix},
\end{align}
where $\matD$ is a diagonal matrix, $\vecz$ is a vector of size $n-1$ and $\alpha $ is a scalar. Without loss of generality we assume that $\|\matH\|\leq 1$. The main result is stated in Theorem \ref{theorem:arrowhead_diagonalization}. 

In order to prove Theorem \ref{theorem:arrowhead_diagonalization}, we expand in detail the following methodology which is outlined in \cite{gu1995divide}, by proving several technical lemmas  in Appendices \ref{section:arrowhead_preliminaries} and \ref{appendix:fast_multipole_method} that leverage the FMM.
\begin{enumerate}
    \item Deflation: The matrix $\matH$ is preprocessed to ensure that it satisfies the following:
    $
        d_{i+1}-d_i \geq \tau, \text{ and } |\vecz_i|\geq \tau,
    $
    where $\tau\in(0,1)$. This assumption implies several useful properties, described in Lemma \ref{lemma:arrowhead_preliminaries}, and it allows us to efficiently utilize the FMM in the subsequent steps. The deflation procedure is described in Proposition \ref{proposition:arrowhead_deflation}.
    \item Eigenvalues: The eigenvalues of the deflated matrix can be conveniently approximated as the roots of the corresponding secular equation (Eq. \eqref{eq:arrowhead_secular_equation}):
    $
        f(\lambda)=\lambda-\alpha+\sum_{j=2}^n \tfrac{\vecz_j^2}{d_j-\lambda}.
    $
    An FMM-based root finder is detailed in Lemma \ref{lemma:fmm_approximate_eigenvalues}.
    \item From Lemma \ref{lemma:arrowhead_reconstruction_from_shaft_and_eigenvalues}, the approximate eigenvalues returned by Lemma \ref{lemma:fmm_approximate_eigenvalues} are the exact eigenvalues of another arrowhead matrix $\matHhat=\begin{pmatrix}
        \widehat\alpha & \widehat\vecz^\top \\
        \widehat\vecz & \matD
    \end{pmatrix}$. There is an analytical expression for the elements of $\matHhat$ (see also  \cite{boley1977inverse,gu1995divide}). Lemma \ref{lemma:fmm_approximate_shaft} describes how to compute those elements with the FMM.
    \item Given the exact eigenvalues and the approximate elements of the matrix $\matHhat$, we can focus on its eigenvectors. In particular, in Lemma \ref{lemma:fmm_approximate_inner_products} it is shown how to use the FMM to approximate the inner products between the eigenvectors of $\matHhat$ and some arbitrary unit vector $\vecb.$ 
    Computing such inner products with all the columns of the identity, we obtain the final approximate eigenvector matrix of $\matH$ and, ultimately, an approximate diagonalization of $\matH$.
\end{enumerate}
\begin{remark}
    We note that, if we are only interested in a full diagonalization, Lemma \ref{lemma:fmm_approximate_shaft} is redundant, i.e., we can naively compute the elements exactly without the FMM with the same complexity. However, it is useful if we need to compute only a few matrix-vector products with the eigenvector matrix. Lemma \ref{lemma:fmm_approximate_inner_products}, details how to approximate such matrix-vector products efficiently with the FMM. 
\end{remark}

\begin{theorem}
    \label{theorem:arrowhead_diagonalization}
    Given a symmetric arrowhead matrix $\matH\in\mathbb{R}^{n\times n}$ as in Eq. \eqref{eq:arrowhead}, with $\|\matH\|\leq 1$, an accuracy parameter $\epsilon\in(0,1)$, a matrix $\matB$ with $r$ columns $\matB_i,i\in[r]$, where $\|\matB_i\|\leq 1$, and an $(\epsilon,n)$-\fmmalgo\   implementation (see Prop. \ref{proposition:fmm}), we can compute a diagonal matrix $\matLambdatilde$, and a matrix
    $\matQtilde_{\matB}$, such that 
    \begin{align*}
        \lnorm \matH-\matQ\matLambdatilde\matQ^\top \rnorm &\leq \epsilon,
        \quad
        \labs \lpar \matQ^\top\matB - \matQtilde_{\matB} \rpar_{i,j}\rabs \leq  \epsilon/n^2,
    \end{align*}
    where 
    $
        \matQ\in\mathbb{R}^{n\times n}
    $ is (exactly) orthogonal, in 
    $
        O\lpar nr\log^{\cfmm+1}(\tfrac{n}{\epsilon})\rpar
    $
    arithmetic operations and comparisons, where $\xi\geq 1$ is a small constant that depends on the specific FMM implementation and it is independent of $\epsilon,n$.

    Alternatively, if only want to compute a set of approximate values $\widetilde\lambda_1,\ldots,\widetilde\lambda_n$, such that $|\lambda_i(\matH)-\widetilde\lambda_i|\leq \epsilon$, the complexity reduces to $O\lpar n\log(\frac{1}{\epsilon})\log^{\cfmm}(\frac{n}{\epsilon})\rpar$ arithmetic operations.
    \begin{proof}
        The full proof is provided in the restatement in Theorem \ref{theorem:arrowhead_diagonalization_appendix} in the Appendix. Briefly, we start with a deflation step in order to ensure that the properties of Lemma \ref{lemma:arrowhead_preliminaries} are satisfied. Next, we compute the eigenvalues using the bisection algorithm of Lemma \ref{lemma:fmm_approximate_eigenvalues}. Given the eigenvalues, we can use Lemmas \ref{lemma:fmm_approximate_shaft} to compute the elements of the ``reconstructed'' matrix $\matHhat$, and finally use Lemma \ref{lemma:fmm_approximate_inner_products} to approximate inner products with the eigenvectors of $\matHhat$, which give the final approximate eigenvectors of $\matH$.
    \end{proof}
\end{theorem}

\subsection{Tridiagonal diagonalization}
\label{section:proof_of_tridiagonal_diagonalization}

Given the analysis for arrowhead diagonalization, we can now proceed to tridiagonal matrices.
The next lemma bounds the error of the reduction to arrowhead form when the spectral factorizations of the matrices $\matT_1$ and $\matT_2$ in Equation \eqref{eq:tridiagonal_to_arrowhead} are approximate rather than exact. This will be used as an inductive step for the final algorithm.
\begin{lemma}
\label{lemma:tridiagonal_assembly}
Let $\epsilon\in(0,1/2)$ be a given accuracy parameter and $\matT = \begin{pmatrix}
    \matT_1 & \beta_{k+1}\vece_k & \\
    \beta_{k+1}\vece_k^\top & \alpha_{k+1} & \beta_{k+2}\vece_1^\top \\
     & \beta_{k+2}\vece_1 & \matT_2
\end{pmatrix}$ be a tridiagonal matrix with size  $n\geq 3$ and $\|\matT\|\leq 1$, where $\matT_1=\matU_1\matD_1\matU_1^\top$ and $\matT_2=\matU_2\matD_2\matU_2^\top$ be the exact spectral factorizations of $\matT_1$ and $\matT_2$. Let $\matUtilde_1,\matDtilde_1,\matUtilde_2,\matDtilde_2$ be approximate spectral factorizations of $\matT_1,\matT_2$. If these factors satisfy
    \begin{align*}
        \lnorm
            \matT_{\{1,2\}} - \matUtilde_{\{1,2\}}\matDtilde_{\{1,2\}}\matUtilde_{\{1,2\}}^\top
        \rnorm 
        \leq \epsilon_1,
        \quad
        \lnorm \matUtilde_{\{1,2\}}\matUtilde_{\{1,2\}}^\top -\matI \rnorm &\leq \epsilon_1/n,
    \end{align*}
    for some $\epsilon_1\in(0,1/2)$, where $\matDtilde_{\{1,2\}}$ are both diagonal, then, assuming an $(\epsilon,n)$-\fmmalgo\   implementation as in Prop. \ref{proposition:fmm}, we can compute a diagonal matrix $\matDtilde$ and an approximately orthogonal matrix $\matUtilde$ such that
    \begin{align*}
        \lnorm\matUtilde^\top\matUtilde-\matI\rnorm\leq 3(\epsilon_1+\epsilon)/n,
        \quad \text{and} \quad
        \lnorm 
            \matT-\matUtilde\matDtilde\matUtilde^\top
        \rnorm \leq 2\epsilon_1+7\epsilon,
    \end{align*}
    in a total of $O\lpar n^2\log^{\cfmm+1}(\tfrac{n}{\epsilon})\rpar$ arithmetic operations and comparisons, where $\xi\geq 1$ is a small constant that depends on the specific FMM implementation and it is independent of $\epsilon,n$.
    \begin{proof}
        The proof is deferred in the Appendix (see the restatement in Lemma \ref{lemma:tridiagonal_assembly_appendix}).
    \end{proof}
\end{lemma}

Lemma \ref{lemma:tridiagonal_assembly} gives rise to the following recursive algorithm. We can finally proceed with the proof of Theorem \ref{theorem:alg_recursive_diagonalization}, which gives the complexity of Algorithm \ref{algorithm:recursive_diagonalization}.
\begin{algorithm}[htb]
    \caption{Recursive algorithm based on \cite{gu1995divide} to diagonalize a symmetric tridiagonal matrix.}
    \label{algorithm:recursive_diagonalization}
    \small
    \begin{algorithmic}[1]
        \Statex \textbf{Algorithm}: $[\matUtilde,\matLambdatilde]\leftarrow \DIAGONALIZE(\matT,\epsilon)$
        \If{$n\leq 2$}
            \State Compute $\matUtilde,\matLambdatilde$ to be the  exact diagonalization of $\matT$.
        \Else {\bf :}
            \State Partition $\matT = \begin{pmatrix}
                \matT_1 & \beta_{k+1}\vece_k & \\
                \beta_{k+1}\vece_k^\top & \alpha_{k+1} & \beta_{k+2}\vece_1^\top \\
                 & \beta_{k+2}\vece_1 & \matT_2
            \end{pmatrix}$.
            \State $[\matUtilde_1,\matDtilde_1]\leftarrow \DIAGONALIZE(\matT_1,\epsilon)$.
            \State $[\matUtilde_2,\matDtilde_2]\leftarrow \DIAGONALIZE(\matT_2,\epsilon)$.
            \State Assemble $\matUtilde,\matLambdatilde$ from $\matT,\matUtilde_1,\matDtilde_1,\matUtilde_2,\matDtilde_2$ using Lemma \ref{lemma:tridiagonal_assembly} with parameter $\epsilon$.
        \EndIf
        \State \Return $\matUtilde,\matLambdatilde$.
    \end{algorithmic}
\end{algorithm}

\begin{proof}[Proof of Theorem \ref{theorem:alg_recursive_diagonalization}]
        Let $\matT^{(p)}$ be a matrix at recursion depth $p$. We can always divide $\matT^{(p)}$ such that the sizes of $\matT_{\{1,2\}}^{(p)}$ differ by at most $1$. If we keep dividing the matrix $\matT^{(p)}$ in this manner, then the recursion tree has depth at most $d=\lceil\log(n)\rceil$.
        
        The correctness follows by induction. Consider the base case $d$ where   $\matT^{(d)}$ has size at most $5\times 5$, which means that $\matT^{(d)}_{1,2}$ have size $1\times 1$ or $2\times 2$. We can compute the exact diagonalization of $\matT^{(d)}_{1,2}$, and therefore they satisfy the requirements of Lemma \ref{lemma:tridiagonal_assembly} for $\epsilon_1=0$. Thus, if we apply Lemma \ref{lemma:tridiagonal_assembly} with parameter $\epsilon'$ to compute the matrices $\matUtilde^{(d)}$ and $\matDtilde^{(d)}$,  they will satisfy the following bounds:
        \begin{align*}
        \mathsf{err}_{\matUtilde}(d)&:=
        \lnorm\matUtilde^{(d)\top}\matUtilde^{(d)}-\matI\rnorm\leq 3(\epsilon_1+\epsilon')/n=3\epsilon'/n,
        \\
        \mathsf{err}_{\matT}(d)&:=
        \lnorm 
            \matT-\matUtilde^{(d)}\matDtilde^{(d)}\matUtilde^{(d)\top}
        \rnorm \leq 7\epsilon'.
    \end{align*}
    We need to bound the error at the root (depth $p=0$). As long as  the error at depth $p-1$ is smaller than $1/2$, the the error at depth $p$ is bounded by
    \begin{align*}
        \mathsf{err}_{\matUtilde}(p) &\leq 3\mathsf{err}_{\matUtilde}(p+1)+3\epsilon'/n
        <
        4\mathsf{err}_{\matUtilde}(p+1)+4\epsilon'/n, 
        \\
        \mathsf{err}_{\matT}(p) &\leq 2\mathsf{err}_{\matT}(p+1)+7\epsilon'
        < 8\mathsf{err}_{\matT}(p+1)+8\epsilon'.
    \end{align*}
    Solving the recursion we have 
    \begin{align*}
        \mathsf{err}_{\matUtilde}(0) &< \sum_{p=1}^{\lceil\log(n)\rceil}4^p\epsilon'/n = O(n^2)\frac{\epsilon'}{n}=O(n)\epsilon',
        \quad
        \mathsf{err}_{\matT}(0) < \sum_{p=1}^{\lceil\log(n)\rceil}8^p\epsilon' = O(n^3)\epsilon'.
    \end{align*} 
    It thus suffices to run the algorithm with initial error $\epsilon'=c\epsilon/n^3$ for some small constant $c$.

    The complexity analysis for $\epsilon'$ follows. For size $n$, the complexity is given by
    \begin{align*}
        T(n) \leq 2T(\tfrac{n}{2}) 
        + 
        C n^2\log^{\cfmm+1}(\tfrac{n}{\epsilon'})
        .
    \end{align*}
    Solving the recursion yields $T(n)=
    O\lpar
        n^2\log^{\cfmm+1}(\tfrac{n}{\epsilon'})
    \rpar
    =
    O\lpar
        n^2\log^{\cfmm+1}(\tfrac{n}{\epsilon})
    \rpar
    $.

    The final matrices $\matUtilde=\matUtilde^{(0)}$ and $\matLambdatilde=\matDtilde^{(0)}$ satisfy 
    $\lnorm\matUtilde^{(d)\top}\matUtilde^{(d)}-\matI\rnorm\leq \epsilon/n^2$
    and
    $\lnorm \matT-\matUtilde\matLambdatilde\matUtilde^\top \rnorm
        \leq
        \epsilon$, which means that:
    \begin{align*}
        \lnorm \matLambdatilde \rnorm
        \leq
        \lnorm 
            \matUtilde^{-1} 
        \rnorm^2 
        \lpar 
            \lnorm \matT\rnorm + 
            \lnorm \matT-\matUtilde\matLambdatilde\matUtilde^\top \rnorm
        \rpar
        \leq
        \frac{1}{1-\epsilon/n^2} (1 + \epsilon).
    \end{align*}
    We can then write $\matUtilde=\matU+\matE_{\matU}$, which gives
    \begin{align*}
        \lnorm
        \matT-\matU\matLambdatilde\matU^\top
        \rnorm
        &\leq
        \lnorm
        \matT-\matUtilde\matLambdatilde\matUtilde^\top
        \rnorm
        +
        2\lnorm\matLambdatilde\rnorm\lnorm\matE_{\matU}\rnorm
        +
        \lnorm\matLambdatilde\rnorm\lnorm\matE_{\matU}\rnorm^2
        \\
        &\leq
        \epsilon + \frac{(1+\epsilon)}{1-\epsilon/n^2}\frac{\epsilon}{n^2} + \frac{1+\epsilon}{1-\epsilon/n^2}\lpar \frac{\epsilon}{n^2}\rpar^2
        =
        \epsilon \lpar
            1
            +
            \frac{(1+\epsilon)(1+\epsilon/n^2)}{n^2-\epsilon}
        \rpar.
    \end{align*}
    Rescaling $\epsilon$ by a small constant slightly larger than one gives the alternative statement.
    \end{proof}

    \subsection{Hermitian diagonalization}
    Given an algorithm to diagonalize tridiagonal matrices, the following corollary is immediate.
    \begin{corollary}
    \label{corollary:hermitian_diagonalization}
    Let $\matA$ be a Hermitian matrix of size $n$ with $\|\matA\|\leq 1$. Given accuracy $\epsilon\in(0,1/2)$, and an $(\epsilon,n)$-\fmmalgo\   implementation of Prop. \ref{proposition:fmm}, we can compute a matrix $\matQtilde$ and a diagonal matrix $\matLambdatilde$ such that
    \begin{align*}
        \lnorm \matA - \matQtilde\matLambdatilde\matQtilde^* \rnorm \leq \epsilon,
        \quad
        \lnorm \matQtilde^* \matQtilde - \matI \rnorm \leq \epsilon/n^2.
    \end{align*}
    The algorithm requires a total of $O\lpar n^\omega\log(n) + n^2\log^{\cfmm+1}(\tfrac{n}{\epsilon})\rpar$ arithmetic operations and comparisons, where $\xi\geq 1$ is a small constant that depends on the specific FMM implementation and it is independent of $\epsilon,n$.
    \begin{proof}
        A full proof, as well as an alternative statement of the result, can be found in the restatement, Corollary \ref{corollary:hermitian_diagonalization_appendix}, in Appendix \ref{appendix:hermitian_diagonalization_analysis}.
    \end{proof}
    \end{corollary}

\section{Stability of tridiagonal reduction}
\label{section:tridiagonal_reduction_stability}
In this section we analyze the numerical stability and the boolean complexity of  Schönhage's algorithm the floating point model. For this we will use the following stable matrix multiplication and backward-stable QR factorization algorithms as subroutines from \cite{demmel2007fastla,demmel2007fastmm}. 
The corresponding definitions and imported results for these subroutines are deferred to Appendix \ref{appendix:tridiagonal_reduction_subroutines}.

\subsection{Matrix nomenclature}
Schönhage \cite{schonhage1972unitare} used a block variant of Rutishauser's algorithm \cite{rutishauser1963jacobi} to reduce a matrix to tridiagonal form, where elementary rotations are replaced with block factorizations; 
see also \cite{bischof2000framework,ballard2012communication,ballard2015avoiding} for similar methodologies. 
We start with a $n\times n$ block-pentadiagonal matrix $\matA^{(k,s,t)}$, where $k\in \big\{0,\ldots,\log(n)-2\big\}$ (we assume without loss of generality that $n$ is a power of two). 
The matrix $\matA^{(k,s,t)}$ is partitioned in $b_k\times b_k$ blocks of size $n_k\times n_k$ each, where $b_k=\tfrac{n}{n_k}$ and $n_k=2^{k}$. 
The integer $s\in 2,\ldots,b_k$ t denotes that all the blocks $\matA_{i,i-2}$ and $\matA_{i-2,i}$, for all $i=2,\ldots,s$ are equal to zero. $s=2$ is a special case to denote a full block pentadiagonal matrix. 
The integer $t\in \big\{s+2,\ldots, b_k\big\}$  denotes that the matrix has two additional nonzero blocks in the third off-diagonals, specifically at positions $\matA_{t,t-3}$ and $\matA_{t-3,t}$. 
These blocks are often called the ``bulge'' in the literature.
When $t=0$, there is no bulge. As a consequence, the matrix $\matA^{(k,2,0)}$ is full block-pentadiagonal, while the matrix $\matA^{(k,b_k,0)}$ is block-tridiagonal. An illustration of these definitions is shown in Equation \eqref{eq:block_pentadiagonal}. A box is placed around the bulge on the second matrix.
\begingroup
\small
\setlength\arraycolsep{0.8pt}
\begin{align}
    \label{eq:block_pentadiagonal}
    \overbrace{
        \begin{pmatrix}
            \matA_{1,1}  & \matA_{1,2}  &\matA_{1,3}   & 0             & 0           & 0           & 0           & 0           \\
            \matA_{2,1}  & \matA_{2,2}  &\matA_{2,3}   & \matA_{2,4}   & 0           & 0           & 0           & 0           \\
            \matA_{3,1}  & \matA_{3,2}  &\matA_{3,3}   & \matA_{3,4}   & \matA_{3,5} & 0           & 0           & 0           \\
            0            & \matA_{4,2}  &\matA_{4,3}   & \matA_{4,4}   & \matA_{4,5} & \matA_{4,6} & 0           & 0           \\
            0            & 0            &\matA_{5,3}   & \matA_{5,4}   & \matA_{5,5} & \matA_{5,6} & \matA_{5,7} & 0           \\
            0            & 0            & 0            & \matA_{6,4}   & \matA_{6,5} & \matA_{6,6} & \matA_{6,7} & \matA_{6,8} \\
            0            & 0            & 0            & 0             & \matA_{7,5} & \matA_{7,6} & \matA_{7,7} & \matA_{7,8} \\
            0            & 0            & 0            & 0             & 0           & \matA_{8,6} & \matA_{8,7} & \matA_{8,8} \\
        \end{pmatrix}
    }^{
        \matA^{(k,2,0)}
    },
    \ 
    \overbrace{
    \begin{pmatrix}
        \matA_{1,1}  & \matA_{1,2}  & 0            & 0             & 0           & 0           & 0           & 0           \\
        \matA_{2,1}  & \matA_{2,2}  &\matA_{2,3}   & 0             & 0           & 0           & 0           & 0           \\
        0            & \matA_{3,2}  &\matA_{3,3}   & \matA_{3,4}   & \matA_{3,5} & \boxed{\matA_{3,6}} & 0           & 0           \\
        0            & 0            &\matA_{4,3}   & \matA_{4,4}   & \matA_{4,5} & \matA_{4,6} & 0           & 0           \\
        0            & 0            &\matA_{5,3}   & \matA_{5,4}   & \matA_{5,5} & \matA_{5,6} & \matA_{5,7} & 0           \\
        0            & 0            &\boxed{\matA_{6,3}}   & \matA_{6,4}   & \matA_{6,5} & \matA_{6,6} & \matA_{6,7} & \matA_{6,8} \\
        0            & 0            & 0            & 0             & \matA_{7,5} & \matA_{7,6} & \matA_{7,7} & \matA_{7,8} \\
        0            & 0            & 0            & 0             & 0           & \matA_{8,6} & \matA_{8,7} & \matA_{8,8} \\
    \end{pmatrix}}
    ^{
        \matA^{(k,4,6)}
    }
    .
\end{align}
\normalsize
\endgroup

\subsection{Rotations}
The algorithm defines two types of block rotations $R_i$ and $R_i'$, which are unitary similarity transformations, with the following properties.
\begin{definition}[Rotations]
    The algorithm of \cite{schonhage1972unitare} uses the following two types of rotations:
    \begin{enumerate}
    \item $R_i(\matA^{(k,i,0)})$, for $i=2,\ldots,b_k-1$, operates on a block-pentadiagonal matrix without a bulge. It transforms the matrix $\matA^{(k,i,0)}$ to a matrix $\matA^{(k,i+1,i+3)}$. In paricular, the block $\matA_{i,i-1}$ becomes upper triangular, the block $\matA_{i+1,i-1}$ becomes zero, and a new bulge block arises at $\matA_{i,i+3}$. Due to symmetry, $\matA_{i-1,i}$ becomes lower triangular, $\matA_{i-1,i+1}$ is eliminated, and $\matA_{i+3,i}$ becomes non-zero. 
    \item $R_j'(\matA^{(k,s,j+1)})$, for some $j=s+1,\ldots,b_k-1$, operates on a block-pentadiagonal matrix with a bulge at positions $\matA_{j+1,j-2}$, $\matA_{j-2,j+1}$. It transforms the matrix $\matA^{(k,s,j+1)}$ to a matrix $\matA^{(k,s,j+3)}$, such that the bulge is moved two positions ``down-and-right'', i.e. the blocks $\matA_{j-2,j+1}$ and  $\matA_{j+1,j-2}$ become zero and the blocks $\matA_{j,i+3}$ and $\matA_{j,j+3}$ become the new bulge. In addition, the matrices $\matA_{j,j-2}$ and $\matA_{j+1,j-1}$ become upper triangular, and, by symmetry, the matrices $\matA_{j-2,j}$ and $\matA_{j-1,j+1}$ become lower triangular.
\end{enumerate}
\end{definition}

An example of the aforementioned rotations is illustrated in Equations \eqref{eq:rotation_r_i} and \eqref{eq:rotation_r_i_prime} in the Appendix, and in Lemmas \ref{lemma:rotation_r_i_floating_point} and \ref{lemma:rotation_r_i_prime_floating_point} it is proved that both types can be stably implemented in floating point using fast QR factorizations.

\subsection{Recursive bandwidth halving}
Using Lemmas \ref{lemma:rotation_r_i_floating_point} and \ref{lemma:rotation_r_i_prime_floating_point}, we can analyze the following Algorithm \ref{algorithm:halve}, which halves the bandwidth of a matrix. Its complexity and stability properties are stated in Lemma \ref{lemma:bandwidth_halving_floating_point_appendix} in the Appendix.
Applying this algorithm recursively gives the main Theorem \ref{theorem:stable_tridiagonal_reduction}.
\begin{algorithm}[htb]
    \caption{Halves the bandwidth of a Hermitian matrix with unitary rotations.}
    \label{algorithm:halve}
    \small
    \begin{algorithmic}[1]
        \Statex \textbf{Algorithm}: $[\matQhat^{(k)},\matA^{(k-1,2,0)}]\leftarrow \HALVE(\matA^{(k,2,0)},k,n)$
        \State Set $n_k=2^k$, $b_k=n/n_k$.
        \For{$i=2,\ldots, b_k$}
            \State Compute $\matQ_{i,i},\matA^{(k,i+1,i+3)} \leftarrow R_i(\matA^{(k,i,0)})$.
            \For{$j=i+2,\ldots,b_k$ with step $2$}
                \State Compute $\matQ_{i,j},\matA^{(k,i+1,j+3)} \leftarrow R'_j(\matA^{(k,i+1,j+1)})$.
            \EndFor
            \State Stack together all the matrices $\matQ_{i,j}$ to form $\matQ_i$.
        \EndFor
        \State Assemble the matrix $\matQhat^{(k)}$ by multiplying the matrices $\matQ_{i}$.
        \State \Return $\matQhat^{(k)},\matA^{(k-1,2,0)}$.
    \end{algorithmic}
\end{algorithm}

\begin{theorem}
    \label{theorem:stable_tridiagonal_reduction}
    There exists a floating point implementation of the tridiagonal reduction algorithm of \cite{schonhage1972unitare}, which takes as input a Hermitian matrix $\matA$, and returns a tridiagonal matrix $\matTtilde$, and (optionally) an approximately unitary matrix $\matQtilde$. If the machine precision $\umach$ satisfies
    $
        \umach \leq \epsilon\frac{1}{cn^{\beta+4}},
    $
    where $\epsilon\in(0,1)$, $c$ is a constant, and $\cmm$ is the same as in Corollary \ref{corollary:alg_qr}, which translates to $O(\log(n)+\log(1/\epsilon))$ bits of precision, then the following hold:
    \begin{align*}
        \lnorm \matQtilde\matQtilde^*-\matI\rnorm \leq \epsilon,
        \quad
        \text{and}
        \quad
        \lnorm \matA - \matQtilde\matTtilde\matQtilde^* \rnorm \leq \epsilon \lnorm \matA \rnorm.
    \end{align*}
    The algorithm executes at most $O\lpar
                    n^2 S_{\omega}(\log(n))
    \rpar$ floating point operations to return only $\matTtilde$, where $S_x(m)=\sum_{l=1}^m (2^{x-2})^l$.
    If $\matA$ is banded with $1\leq d\leq n$ bands, the floating point operations reduce to $O(n^2 S_{\omega}(\log(d))$.
    If $\matQtilde$ is also returned, the complexity increases to $O(n^2C_{\omega}(\log(n)))$, 
    where $C_{x}(n) := 
    \sum_{k=2}^{\log(n)-2}
    \lpar
        S_{x}(\log(n)-1) - S_{x}(k)
    \rpar$. If $\omega$ is treated as a constant $\omega\approx 2.371$ the corresponding complexities are $O(n^{\omega}), O(n^2d^{\omega-2}),$ and $O(n^{\omega}\log(n))$, respectively.
    \begin{proof}
        The proof is deferred to the Appendix (see the restatement in Theorem \ref{theorem:stable_tridiagonal_reduction_appendix}).
    \end{proof}
\end{theorem}

\subsection{Eigenvalues of Hermitian matrices}
\label{section:hermitian_eigenvalues}
We now have all the prerequisites to compute  the eigenvalues Hermitian matrices in nearly matrix multiplication time in finite precision.
For this we can use the eigenvalue solver of \cite{bini1991parallel}, which has $\widetilde O(n^2)$ boolean complexity, albeit in the Boolean RAM model. Specifically, the algorithm accepts as input symmetric tridiagonal matrices with bounded integer entries. 

\begin{theorem}[Imported from \cite{bini1991parallel,bini1998computing}]
Let $\matT$ be a symmetric tridiagonal matrix with integer elements bounded in magnitude by $2^m$ for some $m$. Let $\epsilon=2^{-u}\in(0,1)$ be a desired accuracy. Algorithm 4.1 of \cite{bini1991parallel} computes a set of approximate eigenvalues $\widetilde\lambda_i\in\mathbb{R}$ (which are returned as rationals) such that
$
    \labs \widetilde\lambda_i-\lambda_i(\matT) \rabs < \epsilon.
$
The algorithm requires
$
    O\lpar
        n^2b\log^2(n)\log(nb)(\log^2(b)+\log(n))\log(\log(nb))
    \rpar
$
boolean operations, where $b=m+u$.
\label{theorem:bini_pan_tridiagonal_eigenvalues}
\end{theorem}

\begin{theorem}
    \label{theorem:hermitian_eigenvalues}
    Let $\matA$ be a (banded) Hermitian matrix, with $\|\matA\|\leq 1$, $1\leq d\leq n-1$ off-diagonals, and let $\epsilon\in(0,1)$ be an  accuracy parameter. Assume that the elements of $\matA$ are floating point numbers on a machine with precision $\umach$, $t=\log(1/\umach)$ bits for the significand, and $p=O(\log(\log(n)))$ bits for the exponent. There exists an algorithm that returns a set of $n$ approximate eigenvalues $\widetilde\lambda_1,\ldots,\widetilde\lambda_n$ such that
    \begin{align*}
        \labs \widetilde\lambda_i - \lambda_i(\matA) \rabs
        \leq
        \epsilon
    \end{align*}
    using at most 
    \begin{align*}
        O\lpar
            n^2S_{\matmulexponent}(\log(d))\cdot \flopcost(\log(\tfrac{n}{\epsilon}))
            +
            n^2\polylog(\tfrac{n}{\epsilon})
        \rpar
    \end{align*}
    boolean operations, where $\flopcost(b)$ is the bit complexity of a floating point operation on $b$ bits, and  $n^2S_{\matmulexponent}(\log(d))=O(n^2d^{\matmulexponent-2})$ if $\omega$ is treated as a constant greater than two.
    \begin{proof}
        First, recall that each element of a floating point matrix $\matA$ is represented by a floating point number $\matA_{i,j}=\pm 1\times 2^{e_{i,j}}\times m_{i,j}$, where $e_{i,j}\in[-2^{p},2^p]$ is its exponent, $p$ is the number of exponent bits, and $m_{i,j}$ is the significand, which is an integer with $t=\log(1/\umach)$ bits. To represent all the elements of $\matA$ with normalized floating point numbers it is sufficient to use $p=O\lpar \log(\log(A))\rpar$, where $A=\max\lcurly\|\matA\|_{\max}, \frac{1}{\|\matA\|_{\min}}\rcurly$. By assumption, $\|\matA\|_{\max}\leq \|\matA\|\leq 1$. It is common to assume that $\frac{1}{\|\matA\|_{\min}}\in\poly(n)$, which means that $O(\log(\log(n)))$ bits for the exponent are sufficient.

        To compute the eigenvalues of $\matA$ we work as follows.
        We first run the algorithm of Theorem \ref{theorem:stable_tridiagonal_reduction} to reduce $\matA$ to tridiagonal form $\matTtilde$, with parameter $\epsilon$ and
        $
            O\lpar
                \log(\tfrac{n}{\epsilon})
            \rpar
        $
        bits of precision, and a total of $O(n^2S_{\matmulexponent}(\log(d)))$ floating point operations, which is equal to $O(n^2d^{\matmulexponent-2})$ if $\omega$ is treated as a constant.
        It holds that
        \begin{align*}
            \labs 
                \lambda_i(\matTtilde) - \lambda_i(\matA)
            \rabs
            \leq \epsilon'\|\matA\|
            \leq
            \epsilon \frac{\|\matA\|}{n2^{2^{p+1}}}
            \leq
            \epsilon,
        \end{align*}
        where in the last inequality we used that $\|\matA\|\leq n\|\matA_{\max}\|\leq n 2^{2^{p+1}}$.
        
        Therefore, it suffices to approximate the eigenvalues of $\matTtilde$. For this we first transform $\matTtilde$ to a tridiagonal matrix with integer values and use the algorithm of \cite{bini1991parallel} to compute its eigenvalues.
        A floating point matrix can be transformed to one with integer entries as follows. 
        Assuming that we have $p$ bits for the floating point exponent, and $t=\log(1/\umach)$ bits for the significand, the floating point numbers with $O(p+t)$ bits can be transformed to integers of $O(2^{2p}+t)$ bits by multiplying all the elements with $2^{2^p+t}$. 
        This is achieved by first allocating a tridiagonal matrix with $3n-2$ integers with $O(2^{2p}+t)=O(\log^2(n)+\log(n/\epsilon))$ bits each, that are initially set to zero. 
        We then visit each floating point element $\matTtilde_{i,j}$ of the original matrix, which is represented as a number $\pm 1\times 2^{e_{i,j}}\times m_{i,j}$, where $e_{i,j}\in[-2^{p},2^p]$ is its exponent and $m_{i,j}$ is the significand, which is an integer with $t$ bits. 
        We copy the $t$ bits of $m_{i,j}$ at the positions $e_{i,j}+2^p,e_{i,j}+2^p+1,\ldots, e_{i,j}+2^p+t$ of the corresponding element of the new matrix. 
        This takes $O(n(2^{2p}+t))=O(n(\log^2(n)+\log(n/\epsilon)))$ boolean operations in total to allocate the new matrix and copy the elements. 
        
        Now that we have an integer valued symmetric tridiagonal matrix $\matT',$ with $O(2^{2p}+t)=O(\log^2(n)+\log(n/\epsilon))$ bits per element, we can compute its eigenvalues up to additive accuracy $\epsilon$ with Theorem \ref{theorem:bini_pan_tridiagonal_eigenvalues}. 
        Specifically, we set $\epsilon''=\epsilon\cdot 2^{2^p+t}$, and run the algorithm with the matrix $\matT'$ as input and $u=\log(\tfrac{1}{\epsilon''})$. 
        Let \begin{align*}
            b = u+m = \log\lpar\frac{1}{\epsilon''}\rpar + 2^{p+1} + t = \log\lpar
                    \frac{1}{\epsilon 2^{2^p+t}}
                \rpar + \log \lpar
                    2^{2^{p+1}+t}
                \rpar
            = 
            O(\log(\tfrac{n}{\epsilon})),
        \end{align*}
        where the last inequality follows from the assumption that $p=O(\log(\log(n)))$.
        The returned eigenvalues $\lambda_i'$ satisfy 
        \begin{align*}
        \labs
            \lambda_i'
            -
            \lambda_i(\matT')
        \rabs            
        \leq
        \epsilon''
        \Rightarrow
        \labs
            \frac{\lambda_i'}{2^{2^p+t}}
            -
            \lambda_i(\matTtilde)
        \rabs
        \leq
        \epsilon.
        \end{align*}
        Therefore, if we set $\widetilde\lambda_i=\frac{\lambda_i'}{2^{2^p+t}}$, we finally obtain that $
            \labs
                \widetilde\lambda_i
                -
                \lambda_i(\matA)
            \rabs
            \leq
            2\epsilon.$
        Rescaling $\epsilon$ gives the final bound.
        The algorithm runs in
        \begingroup
        \allowdisplaybreaks
        \begin{align*}
            &B= O\bigg(
                n^2b\log^2(n)\log(nb)(\log^2(b)+\log(n))\log(\log(nb))
            \bigg)
            \\
            &=
            O\bigg(
                n^2
                \log(\tfrac{n}{\epsilon})
                \log^2(n)
                \log(n\log(\tfrac{n}{\epsilon}))                
                \lbrac
                    \log^2(\log(\tfrac{n}{\epsilon}))
                    +
                    \log(n)
                \rbrac
                \log(\log(n\log(\tfrac{n}{\epsilon})))
            \bigg)
        \end{align*}
        \endgroup
        boolean operations.
    \end{proof}
\end{theorem}

\section{Further applications of stable tridiagonal reduction and Hermitian eigenvalue solver}
\label{section:applications_tridiagonal_reduction}
In this section we state some applications of the tridiagonal reduction algorithm to some eigenproblems.

\subsection{Singular values and condition number}
\begin{proposition}
    \label{proposition:alg_sigmak}
        Given a matrix $\matA\in\mathbb{C}^{n\times n},$ with $\|\matA\|\leq 1$, an integer $k\in[n]$, and accuracy $\epsilon\in(0,1)$, we can compute a value $\widetilde\sigma_{k}\in (1\pm\epsilon)\sigma_k(\matA)$ in
        \begin{align*}
            O\lpar
                \lbrac
                    n^{\omega}\flopcost(\log(\tfrac{n}{\epsilon\sigma_k}))
                    +
                    n^2\polylog(\tfrac{n}{\epsilon\sigma_k})
                \rbrac
                \log(\log(\epsilon\sigma_k))
            \rpar
        \end{align*}
        boolean operations, deterministically.\\
        \textbf{Note:} If $\|\matA\|>1$, we can scale with $1/(n\|\matA\|_{\max})$ or $1/\|\matA\|_F$. The complexity is unaffected. 
    \begin{proof}
    The problem is solved by iteratively calling Theorem \ref{theorem:hermitian_eigenvalues} on $\matA\matA^*$, squaring the accuracy in each iteration until the desired singular value is approximated. 
    First, we divide $\matA$ by two, to ensure that $\|\matA\|\leq 1/2$. Then we construct $\matAtilde=  \MM(\matA,\matA^*)=\matA\matA^*/4+\matE^{ \MM}$. 
    We start with $\epsilon_0=1/2$. 
    From Weyl's inequality, $|\lambda_k(\matAtilde)-\lambda_k(\matA\matA^*/4)| \leq \epsilon_0/2$.

    Next, we call Theorem \ref{theorem:hermitian_eigenvalues} to compute all the eigenvalues of $\matAtilde$ up to accuracy $\epsilon_0/2$. The algorithm returns approximate eigenvalues $\widetilde\lambda_1^{(0)},\ldots,\widetilde\lambda_n^{(0)}$ such that
    \begin{align*}
        \labs \widetilde\lambda_i^{(0)} - \lambda_i(\matAtilde) \rabs
        \leq
        \epsilon_0/2
        \quad
        \Rightarrow
        \quad
        \labs \widetilde\lambda_i^{(0)} - \lambda_i(\matA\matA^*/4)
        \rabs \leq \epsilon_0,
    \end{align*}
    using at most 
    $
        O\lpar
            n^\omega \flopcost(\log(\tfrac{n}{\epsilon_0}))
            +
            n^2\polylog(\tfrac{n}{\epsilon_0})
        \rpar
    $
    boolean operations. 
    
    We keep calling the above steps iteratively, by squaring $\epsilon_t$ in every iteration $t$, i.e., $\epsilon_t=\epsilon_0^{2^t}$. Let $\lambda_i=\lambda_i(\matA\matA^*/4)$.
    We run the algorithm until $\epsilon_t \leq \frac{\epsilon}{2}|\widetilde\lambda_k^{(t)}|.$ The latter condition ensures that
    \begin{align*}
        \epsilon_t 
        \leq
        \frac{\epsilon}{2} |\lambda_k^{(t)}|
        \leq
        \frac{\epsilon}{2}\lpar
            \lambda_k
            +
            \epsilon_t
        \rpar
        \Rightarrow
        \epsilon_t\leq \epsilon\lambda_k,
    \end{align*}
    where in the last we used the assumption that $\epsilon\in(0,1/2)$.
    Since at each iteration $\lambda_k-\epsilon_t\leq \widetilde\lambda_k^{(t)}$, then the following condition is sufficient for termination:
    \begin{align*}
        \epsilon_t \leq \frac{\epsilon}{2}
        (\lambda_k-\epsilon_t) 
        \Rightarrow 
        \epsilon_t \leq \frac{2}{5}\epsilon\lambda_k.
    \end{align*}
    This is reached in at most $t=O(\log(\log(\epsilon\lambda_k)))$ iterations. From the cost of the matrix multiplications and from Theorem \ref{theorem:hermitian_eigenvalues}, each iteration costs
    \begin{align*}
        O\lpar mn^{\omega-1}\flopcost(\log(\tfrac{mn}{\epsilon_t}))
        +
        n^2S_{\omega}
            (\log(n))
            \cdot
            \flopcost(\log(\tfrac{n}{\epsilon_t}))
            +
            n^2\polylog(n/\epsilon_t)
        \rpar,
    \end{align*}
    boolean operations,
    which is maximized in the last iteration. This gives a total cost of at most
    \begin{align*}O
        \lpar 
            t
            \lbrac
            mn^{\omega-1}\flopcost(\log(\tfrac{mn}{\epsilon\sigma_k}))
            +
            n^2
                S_{\omega}(\log(n))
                \flopcost(\log(\tfrac{n}{\epsilon\sigma_k}))
                +
                \polylog(\tfrac{n}{\epsilon\sigma_k})
            \rbrac
        \rpar
    \end{align*}
    boolean operations, since $\lambda_k=\sigma_k^2/4$, which is equal to
    \begin{align*}
        O\lpar
            \lbrac
                mn^{\omega-1}\flopcost(\log(\tfrac{mn}{\epsilon\sigma_k}))
                +
                n^2\polylog(\tfrac{n}{\epsilon\sigma_k})
            \rbrac
            \log(\log(\epsilon\sigma_k))
        \rpar
    \end{align*}
    for constant $\omega\approx 2.371$.
    We finally return $\widetilde\sigma_k = 2\sqrt{\widetilde\lambda_{k}^{(t)}}$, which satisfies the advertised bound.
    \end{proof}
\end{proposition}
\begin{corollary}
    \label{corollary:alg_cond}
    Let $\matA\in\mathbb{C}^{n\times n}$, $\kappa=\kappa(\matA)$, and $\delta\in(0,1/2)$. We can compute $\widetilde\kappa$ which $\kappa\leq \widetilde\kappa\leq  3n\kappa$, in 
    \begin{align*}
        O\lpar
            \lbrac
                n^{\omega}\flopcost(\log(n\kappa))
                +
                n^2\polylog(n\kappa)
            \rbrac
            \log(\log(n\kappa))
        \rpar
    \end{align*}
    boolean operations deterministically. 
    \begin{proof}
        We first set  $\widetilde\Sigma=n\|\matA\|_{\max}$. Assuming $O(\log(n))$ bits for the elements of $\matA$, this takes $O(n^2\flopcost(\log(n))$ boolean operations (for floating point comparisons). We can then scale $\matA'\leftarrow \matA/M$, where $M$ is again the smallest power of $2$ that is larger than $2\widetilde\Sigma$.
        This gives $\frac{1}{2n}\leq \|\matA'\|\leq \frac{1}{2}$, and also $\sigma_{\min}(\matA')
        \in
        \lbrac \tfrac{\sigma_{\min}(\matA)}{2n\|\matA\|}, \tfrac{\sigma_{\min}(\matA)}{2\|\matA\|} \rbrac = 
        \lbrac 
        \tfrac{1}{2n\kappa(\matA)}, \tfrac{1}{2\kappa(\matA)} \rbrac
        $.
        As in the previous case, we  call Proposition \ref{proposition:alg_sigmak} on $\matA'$ with $k=1$ and error $\epsilon=1/2$, which returns $\widetilde\sigma_{\min}'\in(1\pm \tfrac{1}{2}) 
        \sigma_{\min}(\matA')$. It requires
        \begin{align*}
            O\lpar
                \lbrac
                    n^{\omega}\flopcost(\log(n\kappa(\matA)))
                    +
                    n^2\polylog(n\kappa(\matA))
                \rbrac
                \log(\log(n\kappa(\matA)))
            \rpar
        \end{align*}
        boolean operations, and we finally  set $\widetilde\kappa= \frac{1}{\widetilde\sigma_{\min}'}\in \lbrac
            \kappa(\matA), 3n\kappa(\matA)
        \rbrac$.
    \end{proof}
\end{corollary}

\subsection{Definite pencil eigenvalues}
Using the proposed algorithms we can compute the eigenvalues of a Hermitian definite pencil.
\begin{corollary}
    \label{corollary:hermitian_definite_pencil_eigenvalues}
    Let $\matH$ be Hermitian, $\matS$ Hermitian positive-definite, both with size $n$ and floating point elements, on a machine with precision $\umach$, $t=\log(1/\umach)$ bits for the significand, and $p=O(\log(\log(n)))$ bits for the exponent. Assume that $\|\matH\|,\|\matS^{-1}\|\leq 1$, $\kappa(\matS)\in\poly(n)$, and that we have access to $\widetilde\kappa\in[\kappa(\matS),Z\kappa(\matS)]$, where $Z>1$ might be a constant or a function of $n$. There exists an algorithm that returns a set of $n$ approximate eigenvalues $\widetilde\lambda_1,\ldots,\widetilde\lambda_n$ such that
    \begin{align*}
        \labs \widetilde\lambda_i - \lambda_i(\matH,\matS) \rabs
        \leq
        \epsilon
    \end{align*}
    using at most 
    \begin{align*}
        O\Big(
            n^{\omega}
            \flopcost(\log(n)\log(Z\kappa(\matS)) + \log(\tfrac{1}{\epsilon}))
            +
            n^2\polylog(\tfrac{n}{\epsilon})
        \Big)
    \end{align*}
    boolean operations. If $\widetilde\kappa$ is not given, it can be computed up to a factor $Z=3n$ with Corollary \ref{corollary:alg_cond}.
    \begin{proof}
         We can use \cite[Prop. C.3]{sobczyk2024invariant} with error $\epsilon/2$, which requires $O(n^{\omega})$ floating point operations and a total of $O(\log(n)\log(Z\kappa(\matS))+\log(\tfrac{1}{\epsilon}))$ bits, and it returns a Hermitian matrix $\matHtilde$ that satisfies
         $|\lambda_i(\matHtilde)-\lambda_i(\matH,\matS)|\leq \tfrac{\epsilon}{2}$.

         Then, Theorem \ref{theorem:hermitian_eigenvalues} is used with parameter $\epsilon/2$ to compute the eigenvalues of $\matHtilde$ up to additive error $\epsilon/2$. We can use it if $\kappa(\matS)\in\poly(n)$, otherwise the problem is ill-conditioned and we might need more than $O(\log(\log(n)))$ bits for the floating point exponent. Regarding the bits of precision $O(\log(n/\epsilon))$ bits are sufficient for Theorem \ref{theorem:hermitian_eigenvalues}. But this is already covered, since we assumed that we have at least $O(\log(n)\log(Z\kappa(\matS))+\log(1/\epsilon))$ bits of precision in the previous step. Then the complexity of Theorem \ref{theorem:hermitian_eigenvalues} for the increased number of bits becomes
         \begin{align*}
             O\Big(
                n^\matmulexponent
                \flopcost\left(
                \log(n)\log(Z\kappa(\matS))+\log(\tfrac{1}{\epsilon})
                \right)
                +
                n^2\polylog(\tfrac{n}{\epsilon})
            \Big)
         \end{align*}
         boolean operations. The returned eigenvalues $\widetilde\lambda_1,\ldots,\widetilde\lambda_n$ satisfy
         \begin{align*}
             \labs \widetilde\lambda_i-\lambda_i(\matH,\matS) \rabs
             \leq 
             \labs \widetilde\lambda_i-\lambda_i(\matHtilde) \rabs
             +
             \labs \lambda_i(\matHtilde) -\lambda_i(\matH,\matS)\rabs
             \leq 2\epsilon/2 = \epsilon.
         \end{align*}
    \end{proof}
\end{corollary}
\begin{remark}
    In Theorem \ref{corollary:hermitian_definite_pencil_eigenvalues} we assumed that $\|\matH\|,\|\matS^{-1}\|\leq 1$. This is not a limitation since we can approximate $\|\matS^{-1}\|$ with Proposition \ref{proposition:alg_sigmak}, and then scale accordingly. Formally, let $\eta \gtrsim \|\matH\|$ and $\sigma\gtrsim \|\matS^{-1}\|$. Then we can rewrite the generalized eigenproblem
\begin{align*}
\matH\matC=\matS\matC\geneigmatrix
\quad
\Leftrightarrow
\quad
(\tfrac{1}{\eta}\matH)\matC = (\sigma\matS) \matC (\geneigmatrix\tfrac{1}{\eta\sigma})
\quad
\Leftrightarrow
\quad
\matH'\matC = \matS'\matC\geneigmatrix',     
\end{align*}
i.e. it is the same generalized eigenproblem only with scaled eigenvalues. Assuming that the matrices $\matH$ and $\matS$ are ``well-conditioned,'' i.e. their norms and condition numbers $\in \poly(n)$, the eigenvalues are scaled by at most a $1/\poly(n)$ factor, and thus it suffices to scale $\epsilon$ by $1/\poly(n)$ as well. We can thus safely make the unit-norms assumption. 
\end{remark}

\subsection{Spectral gaps}
In a similar way we can compute the spectral gap between a pair of eigenvalues of Hermitian matrices and Hermitian-definite pencils.

\begin{corollary}
    \label{corollary:spectral_gap}
    Let $\matA$ be a banded Hermitian matrix of size $n$ with $1\leq d\leq n-1$ off-diagonals, $\|\matA\|\leq 1$, and its eigenvalues $\lambda_1\leq \lambda_2\leq \ldots \leq \lambda_n$. For some integer $k\in[n-1]$, let $\mu_k=\frac{\lambda_k+\lambda_{k+1}}{2}$ and $\gap_k=\lambda_{k+1}-\lambda_k$. Given an accuracy $\epsilon\in(0,1)$, we can compute two values $\widetilde\mu_k$ and $\widetilde\gap_k$ such that
    \begin{align*}
            \widetilde\mu_k\in \mu_k\pm \epsilon\gap_k,
            \quad
            \text{and}
            \quad
            \widetilde\gap_k \in(1\pm\epsilon)\gap_k,
    \end{align*}
    in
     \begin{align*}
            O\lpar
                n^2
                \lbrac
                    d^{\omega-2}
                    \cdot
                    \flopcost(\log(\tfrac{n}{\epsilon\gap_k}))
                    +
                    \polylog(\tfrac{n}{\epsilon\gap_k})
                \rbrac
                \log(\log(\tfrac{1}{\epsilon\gap_k}))
            \rpar
    \end{align*}
    boolean operations.
    \begin{proof}
        We start with an initial $\epsilon_0=1/2$, and we call the algorithm of Theorem \ref{theorem:hermitian_eigenvalues} to compute the eigenvalues of $\matA$ up to additive error $\epsilon_0$. We keep calling the algorithm iteratively, by squaring $\epsilon_t$ in every iteration $t$, i.e., $\epsilon_t=\epsilon_0^{2^t}$. In each iteration, the algorithm returns eigenvalues $\widetilde\lambda^{(t)}_i$ that satisfy
        \begin{align*}
            \labs \widetilde\lambda^{(t)}_i - \lambda_i(\matA) \rabs \leq \epsilon_t.
        \end{align*}
        We run the algorithm until $\epsilon_t \leq \frac{\epsilon}{2}\widetilde\gap_k^{(t)} = \frac{\epsilon}{2} \labs \widetilde\lambda_{k+1}^{(t)} - \widetilde\lambda_{k}^{(t)}\rabs.$ The latter condition ensures that
        \begin{align*}
            \epsilon_t 
            \leq
            \frac{\epsilon}{2} \labs\widetilde\lambda_{k+1}^{(t)} - \widetilde\lambda_{k}^{(t)}\rabs
            \leq
            \frac{\epsilon}{2}\lpar
                \labs
                \lambda_{k+1} -\lambda_{k}
                \rabs
                +
                2\epsilon_t
            \rpar
            \Rightarrow
            \epsilon_t\leq \epsilon\gap_k,
        \end{align*}
        where in the last we used the assumption that $\epsilon\in(0,1/2)$. Since at each iteration $\gap_k-2\epsilon_t\leq \widetilde\gap_k^{(t)}$, then the following condition is sufficient for termination:
        \begin{align*}
            \epsilon_t \leq \frac{\epsilon}{2}\gap_k-2\epsilon_t \Rightarrow \epsilon_t \leq \epsilon\frac{\gap_k}{6}.
        \end{align*}
        This is reached in at most $t=O(\log(\log(\tfrac{1}{\epsilon\gap_k})))$ iterations. From Theorem \ref{theorem:hermitian_eigenvalues}, each iteration costs
        \begin{align*}
            O\lpar
                n^2S_{\omega}
                (\log(d))
                \cdot
                \flopcost(\log(\tfrac{n}{\epsilon_t}))
                +
                n^2\polylog(\tfrac{n}{\epsilon_t})
            \rpar,
        \end{align*}
        boolean operations,
        which is maximized in the last iteration. This gives a total cost of at most
        \begin{align*}
            O\lpar
                n^2
                \log(\log(\tfrac{1}{\epsilon\gap_k}))
                \lbrac
                    S_{\omega}(\log(d))\cdot
                    \flopcost(\log(\tfrac{n}{\epsilon\gap_k}))
                    +
                    \polylog(\tfrac{n}{\epsilon\gap_k})
                \rbrac
            \rpar
        \end{align*}
        boolean operations, which is equal to
        \begin{align*}
            O\lpar
                n^2
                \log(\log(\tfrac{1}{\epsilon\gap_k}))
                \lbrac
                    d^{\omega-2}\log(n)\flopcost(\log(\tfrac{n}{\epsilon\gap_k}))
                    +
                    \polylog(\tfrac{n}{\epsilon\gap_k})
                \rbrac
            \rpar
        \end{align*}
        for constant $\omega\approx2.371$.
        We finally return $\widetilde\gap_k = \widetilde\lambda_{k+1}^{(t)} - \widetilde\lambda_{k}^{(t)}$ and $\widetilde\mu_k = \frac{\widetilde\lambda_{k+1}^{(t)} + \widetilde\lambda_{k}^{(t)}}{2} $ which satisfy the advertised bounds.
    
    \end{proof}
\end{corollary}

\begin{corollary}
    \label{corollary:alg_deterministic_spectral_gap}
    Let $\matH$ be Hermitian, $\matS$ Hermitian positive-definite, both with size $n$ and $\|\matH\|,\|\matS^{-1}\|\leq 1$, which define a Hermitian-definite pencil $(\matH,\matS)$. Given $k\in[n-1]$,  $\widetilde\kappa\in[\kappa(\matS),Z\kappa(\matS)]$, where $Z>1$ might be a constant or a function of $n$, and accuracy $\epsilon\in(0,1/2)$, we can compute $\widetilde\mu_k=\mu_k\pm\epsilon\gap_k$ and $\widetilde \gap_k=(1\pm\epsilon)\gap_k$, where $\mu_k=\tfrac{\lambda_k+\lambda_{k+1}}{2}$ and $\gap_k=\lambda_{k}-\lambda_{k+1}$. The algorithm requires 
    \begin{align*}
        O\lpar
            \lbrac
                n^{\omega}
                \flopcost\lpar
                    \log(n)\log(Z\kappa(\matS)) + \log(\tfrac{1}{\epsilon\gap_k})
                \rpar
                +
                n^2\polylog(\tfrac{n}{\epsilon\gap_k})
            \rbrac
            \log(\log(\tfrac{1}{\epsilon\gap_k}))
        \rpar
    \end{align*}
    boolean operations. If $\widetilde\kappa$ is not given, it can be computed up to a factor $Z=3n$ with Corollary \ref{corollary:alg_cond}.
    \begin{proof}
        We start with an initial $\epsilon_0=1/2$, and we call the algorithm of Theorem \ref{corollary:hermitian_definite_pencil_eigenvalues} to compute the eigenvalues of $(\matH,\matS)$ up to additive error $\epsilon_0$. We keep calling the algorithm iteratively, by squaring $\epsilon_t$ in every iteration $t$, i.e., $\epsilon_t=\epsilon_0^{2^t}$. In each iteration, the algorithm returns eigenvalues $\widetilde\lambda^{(t)}_i$ that satisfy
        \begin{align*}
            \labs \widetilde\lambda^{(t)}_i - \lambda_i(\matH,\matS) \rabs \leq \epsilon_t.
        \end{align*}
        We terminate when $\epsilon_t \leq \frac{\epsilon}{2}\widetilde\gap_k^{(t)} = \frac{\epsilon}{2} \labs \widetilde\lambda_{k+1}^{(t)} - \widetilde\lambda_{k}^{(t)}\rabs.$ This ensures that
        \begin{align*}
            \epsilon_t 
            \leq
            \frac{\epsilon}{2} \labs\widetilde\lambda_{k+1}^{(t)} - \widetilde\lambda_{k}^{(t)}\rabs
            \leq
            \frac{\epsilon}{2}\lpar
                \labs
                \lambda_{k+1} -\lambda_{k}
                \rabs
                +
                2\epsilon_t
            \rpar
            \Rightarrow
            \epsilon_t\leq \epsilon\gap_k,
        \end{align*}
        where in the last we used the assumption that $\epsilon\in(0,1/2)$. Since at each iteration $\gap_k-2\epsilon_t\leq \widetilde\gap_k^{(t)}$, then the following condition is sufficient for termination:
        \begin{align*}
            \epsilon_t \leq \frac{\epsilon}{2}\gap_k-2\epsilon_t \Rightarrow \epsilon_t \leq \epsilon\frac{\gap_k}{6}.
        \end{align*}
        This is reached in at most $t=O(\log(\log(\tfrac{1}{\epsilon\gap_k})))$ iterations. From Theorem \ref{corollary:hermitian_definite_pencil_eigenvalues}, each iteration costs
        \begin{align*}
            O\lpar
                n^{\omega}
                \flopcost\lpar
                    \log(n)\log(Z\kappa(\matS)) + \log(\tfrac{1}{\epsilon_t})
                \rpar
                +
                n^2\polylog(\tfrac{n}{\epsilon_t})
            \rpar
        \end{align*}
        boolean operations,
        which is maximized in the last iteration. This gives a total cost of at most
        \begin{align*}
            O\lpar
                \log(\log(\tfrac{1}{\epsilon\gap_k}))
                \lbrac
                    n^{\omega}
                    \flopcost\lpar
                        \log(n)\log(Z\kappa(\matS)) + \log(\tfrac{1}{\epsilon\gap_k})
                    \rpar
                    +
                    n^2\polylog(\tfrac{n}{\epsilon\gap_k})
                \rbrac
            \rpar
        \end{align*}
        boolean operations.
        We finally return $\widetilde\gap_k = \widetilde\lambda_{k+1}^{(t)} - \widetilde\lambda_{k}^{(t)}$ and $\widetilde\mu_k = \frac{\widetilde\lambda_{k+1}^{(t)} + \widetilde\lambda_{k}^{(t)}}{2} $ which satisfy the advertised bounds.
    
    \end{proof}
\end{corollary}

\subsection{Spectral projectors and invariant subspaces}
For a Hermitian matrix $\matA$, the algorithm called $\GAP$ in \cite{sobczyk2024invariant} computes the spectral gap and the midpoint in the spirit of Corollary \ref{corollary:spectral_gap} using
$O\lpar n^{\omega}\log(\tfrac{1}{\epsilon\gap_k})\log(\tfrac{1}{\delta\epsilon\gap_k})
\cdot 
\flopcost\Big( 
    \log^3(\tfrac{n}{\delta\epsilon\gap_k})\log(\tfrac{1}{\delta\epsilon\gap_k})\log(n)
\Big)
\rpar$ boolean operations and succeeds with probability $1-\delta$.
On the other hand, if $\omega$ is treated as a constant larger than two, the algorithm of Corollary \ref{corollary:spectral_gap} requires
\begin{align*}
    O\lpar
        n^{\omega}\flopcost(\log(n))
            +
        n^2\polylog(n/\epsilon)
    \rpar
\end{align*}
boolean operations, which is significantly faster than $\GAP$. 
Moreover, it is deterministic, and it has a lower complexity for banded matrices.
A key difference between the two algorithms is that the $\GAP$ algorithm of \cite{sobczyk2024invariant} is fully analyzed in floating point, while the algorithm of \ref{corollary:spectral_gap} requires \cite{bini1998computing} which is analyzed in Boolean RAM.

Note that originally the $\GAP$ algorithm was used as a preliminary step to locate the gap and the midpoint in order to compute spectral projectors on invariant subspaces. Corollary \ref{corollary:spectral_gap} can serve as a direct replacement, providing an end-to-end deterministic algorithm for computing spectral projectors.

\begin{corollary}
    \label{corollary:spectral_projector}
    Let $(\matH,\matS)$ be a Hermitian 
     definite pencil of size $n$, with $\|\matH\|,\|\matS^{-1}\|\leq 1$, and  $\lambda_1\leq\lambda_2\leq\ldots\leq \lambda_n$ its eigenvalues. Given as input $\matH$, $\matS$, an integer $1\leq k\leq n-1$, an error parameter $\epsilon\in(0,1)$, we can compute a matrix $\projectormatrixtilde_k$ such that
    \begin{align*}
            \lnorm \projectormatrixtilde_k - \projectormatrix_k \rnorm \leq \epsilon,
    \end{align*}
    where $\projectormatrix_k$ is the true spectral projector on the invariant subspace that is associated with the $k$ smallest eigenvalues, in 
    \begin{align*}
        O\lpar
            n^{\omega}
            \flopcost
            \lpar
                \log(n)
                \log^3(\tfrac{1}{\gap_k})\log(\tfrac{n\kappa}{\epsilon\gap_k})
            \rpar
            \lpar
            \log(\tfrac{1}{\gap_k})+\log(\log(\tfrac{n\kappa}{\epsilon\gap_k}))
            \rpar
            +
            n^{2}\polylog(\tfrac{n\kappa}{\gap_k})
        \rpar
    \end{align*}
    boolean operations, where $\kappa:=\kappa(\matS)=\|\matS\|\|\matS^{-1}\|$ and $\gap_k=\lambda_{k+1}-\lambda_k$.
\begin{proof}
We first use Corollary \ref{corollary:alg_cond} to compute $\widetilde\kappa\in[\kappa(\matS),3n\kappa(\matS)]$ in
\begin{align*}
    O\lpar
    \lbrac
        n^{\omega}\flopcost(\log(n\kappa))
        +
        n^2\polylog(n\kappa)
    \rbrac
    \log(\log(n\kappa))
\rpar
\end{align*}
boolean operations.
Then, given $\widetilde\kappa$, we call Corollary \ref{corollary:alg_deterministic_spectral_gap} to compute 
$\widetilde\mu_k$, $\widetilde\gap_k$,$\widetilde\kappa$
 in
       \begin{align*}
        O\lpar
            \log(\log(\tfrac{1}{\gap_k}))
            \lbrac
                n^{\omega}
                \flopcost\lpar
                    \log(n)\log(n\kappa) + \log(\tfrac{1}{\gap_k})
                \rpar
                +
                n^2\polylog(\tfrac{n}{\gap_k})
            \rbrac
        \rpar
    \end{align*}
    bit operations. 
    Afterwards we can use these values to call \cite[Prop. 2.1]{sobczyk2024invariant} to obtain the spectral projector $\projectormatrixtilde_k$ such that $\|\projectormatrixtilde_k-\projectormatrix_k\|\leq \epsilon$. 
    This requires
    \begin{align*}
        O\lpar
            n^{\omega}\lpar
                \log(\tfrac{1}{\gap_k})+\log(\log(\tfrac{n\kappa}{\epsilon\gap_k}))
            \rpar
            \cdot
            \flopcost
            \lpar
            \log(n)
            \log^3(\tfrac{1}{\gap_k})\log(\tfrac{n\kappa}{\epsilon\gap_k})
        \rpar
        \rpar
    \end{align*} bit operations.
    After some simplifications, the total boolean complexity is upper bounded by
    \begin{align*}
        O\lpar
            n^{\omega}
            \flopcost
            \lpar
                \log(n)
                \log^3(\tfrac{1}{\gap_k})\log(\tfrac{n\kappa}{\epsilon\gap_k})
            \rpar
            \lbrac
                \log(\tfrac{1}{\gap_k})+\log(\log(\tfrac{n\kappa}{\epsilon\gap_k}))
            \rbrac
            +
            n^{2}\polylog(\tfrac{n\kappa}{\gap_k})
        \rpar.
    \end{align*}
\end{proof}
\end{corollary}

\subsection{Stable inversion and factorizations}
Theorem \ref{theorem:stable_tridiagonal_reduction} can be used to invert a matrix stably in floating point, improving the bit requirement of the logarithmically-stable algorithm of \cite{demmel2007fastla}, however, at the cost of increasing the arithmetic complexity by a factor of $\log(n),$ which ultimately leads to a slower algorithm. To achieve this, we first form the matrix $\begin{pmatrix}
    0 & \matA^* \\
    \matA & 0
\end{pmatrix},$ and then reduce it to tridiagonal form. Afterwards we can solve $O(n)$ linear systems stably with QR factorization of the tridiagonal matrix, where the right hand sides are the columns of $\matQtilde$, in time $O(n)$ each. We perform one final multiplication to obtain the inverse. We do not expand the analysis further since the boolean complexity of the algorithm is slower than the one of \cite{demmel2007fastla} (even if we use a slow algorithm for basic arithmetic operations).
The stable inversion algorithm can be used to obtain stable factorization algorithms, such as LU \cite{demmel2007fastla} or Cholesky \cite{sobczyk2024invariant}.

\section{Conclusion}
\label{section:conclusion}
In this work we provided a deterministic complexity analysis for Hermitian eigenproblems. In the Real RAM model, we reported nearly-linear complexity upper bounds for the full diagonalization of arrowhead and tridiagonal matrices, and nearly matrix multiplication-type complexities for diagonalizing Hermitian matrices and for the SVD. This was achieved by analyzing the divide-and-conquer algorithm of \cite{gu1995divide}, when implemented with the Fast Multipole Method. We also showed that the tridiagonal reduction algorithm of \cite{schonhage1972unitare} is numerically stable in the floating point model. This paved the way to obtain improved deterministic boolean complexities for computing the eigenvalues, singular values,  spectral gaps, and spectral projectors, of Hermitian matrices and Hermitian-definite pencils.
Some interesting questions for future research are the following.
\begin{enumerate}
    \item \textbf{Stability of arrowhead diagonalization:} The FMM-accelerated arrowhead diagonalization algorithm was only analyzed in the Real RAM model. Several works \cite{gu1995divide,vogel2016superfast,ou2022superdc,cai2020stable} have provided stabilization techniques of related algorithms in floating point, albeit, without an end-to-end complexity analysis. Such an analysis will be insightful to better understand the boolean complexity of (Hermitian) diagonalization. 
    \item \textbf{Condition number in the SVD complexity:} The complexity of the SVD in Theorem \ref{theorem:svd} has a polylogarithmic dependence on the condition number. 
    Frustratingly, we were not able to remove it at the time of this writing.
    \item \textbf{Non-Hermitian diagonalization:} Sch\"onhage also proved that a non-Hermitian matrix can be reduced to upper Hessenberg form in matrix multiplication time \cite{schonhage1972unitare}. In this work we only provided the stability analysis for the Hermitian case. It would be interesting to investigate whether the Hessenberg reduction algorithm can be used to diagonalize non-Hermitian matrices in matrix multiplication time deterministically (e.g. in conjunction with \cite{banks2021global1,banks2022global2,banks2022global3}).
    \item \textbf{Deterministic pseudospectral shattering:} One of the main techniques of the seminal work of \cite{banks2022pseudospectral} is called ``pseudospectral shattering.'' The main idea is to add a tiny random perturbation to the original matrix to ensure that the minimum eigenvalue gap between any pair of eigenvalues will be at least polynomial in $1/n$. We highlight that the deflation preprocessing step in Proposition \ref{proposition:arrowhead_deflation} has this precise effect: the pseudospectrum is shattered with respect to a deterministic grid. Can this be generalized to obtain \textit{deterministic} pseudospectral shattering techniques, for Hermitian or even non-Hermitian matrices?
\end{enumerate}

\bibliographystyle{plain}

\newpage

\appendix

\section{Preliminaries for symmetric arrowhead diagonalization}
\label{section:arrowhead_preliminaries}

This section contains the required preliminaries to describe the algorithm for symmetric arrowhead diagonalization. Our analysis relies on the methodology of \cite{gu1995divide}, but we refer also to \cite{o1990computing,stor2015accurate} for related techniques. We first state the following lemma which describes the desired properties of the input matrix.
\begin{lemma}
\label{lemma:arrowhead_preliminaries}
Let $\matH=\begin{pmatrix}
        \alpha & \vecz^\top \\
        \vecz & \matD
\end{pmatrix}$, $\|\matH\|\leq 1$, be an arrowhead matrix in the form of Eq. \eqref{eq:arrowhead}.
Suppose that the elements of $\matH$ satisfy the following properties, for all $i=2,\ldots,n$:
\begin{align}
    \label{eq:arrowhead_desiderata}
    d_{i+1}-d_i \geq \tau, \quad \text{and} \quad |\vecz_i|\geq \tau,
\end{align}
for some $\tau\in(0,1)$, where $d_i$ is the $i$-th element of $\matD$. The eigenvalues $\lambda_i$ are the roots of the secular equation
\begin{align}
    f(\lambda)=\lambda-\alpha+\sum_{j=2}^n \tfrac{\vecz_j^2}{d_j-\lambda},
    \label{eq:arrowhead_secular_equation}
\end{align}
and they satisfy the following interlacing property:
\begin{align}
    \lambda_1 < d_2 < \lambda_2 < \ldots < d_n < \lambda_n.
    \label{eq:arrowhead_interlacing}
\end{align}
Moreover, it also holds that
\begin{align}
    \label{eq:arrowhead_eigenvalue_boundaries_distance}
    &\min\lcurly
        \labs d_i-\lambda_i \rabs,
        \labs d_{i+1}-\lambda_i \rabs
    \rcurly
    \geq
    \frac{\tau^3}{n+1},\quad \text{for all }i=2,\ldots, n-1,\nonumber
    \\
    &\min\lcurly
        \labs d_2-\lambda_1 \rabs,
        \labs d_n-\lambda_n \rabs
    \rcurly
    \geq
    \frac{\tau^3}{n},
\end{align}
i.e., there is a well defined gap between the eigenvalues and their boundaries.
\begin{proof}
    The fact that the eigenvalues are the roots of $f(\lambda)$ and that they satisfy the interlacing property is well-known (see e.g. \cite{o1990computing}). We now prove the lower bound, which will be useful for the complexity analysis.

    If $\lambda_i$ is an eigenvalue, for some $i\in\{1,\dots,n\}$, it holds that $f(\lambda_i)=0$. By expanding and rearranging the equation $f(\lambda_i)=0,$ we can write
        \begin{align}
            \label{eq:expanded_secular_equation_for_eigenvalue_bounds}
            \frac{\vecz_i^2}{d_i-\lambda_i} 
            = 
            \alpha 
            - 
            \lambda_i 
            -
            \sum_{j=2}^{i-1}
            \frac{\vecz_j^2}{d_j-\lambda_i}
            -
            \sum_{j=i+1}^{n}
            \frac{\vecz_j^2}{d_j-\lambda_i}.
        \end{align}
    For $i=n$, the right hand side of Eq. \eqref{eq:expanded_secular_equation_for_eigenvalue_bounds} can be bounded as
    \begin{align*}
            \labs
                \alpha 
                - 
                \lambda_n 
                -
                \sum_{j=2}^{n-1}
                \frac{\vecz_j^2}{d_j-\lambda_n}
            \rabs
            \leq
            |\alpha| + |\lambda_n| + \sum_{j=2}^{n-1}\frac{\vecz_j^2}{|d_j-\lambda_n|}
            \leq
            1 + 1 + (n-2)\frac{1}{\tau}
            \leq
            \frac{n}{\tau}
            .
        \end{align*}
        Then
        \begin{align*}
            \frac{\vecz_n^2}{|d_n-\lambda_n|} \leq \frac{n}{\tau}
            \Rightarrow
            |d_n-\lambda_n| \geq \frac{\tau\vecz_n^2}{n} \geq \frac{\tau^3}{n}.
        \end{align*}
    The corresponding bound for $i=1$ can be obtained with similar arguments after rearranging Eq. \eqref{eq:expanded_secular_equation_for_eigenvalue_bounds}.
    
    For $i=2,\ldots, n-1$, consider the following three cases:
    \begin{enumerate}
        \item $\lambda_i=\tfrac{d_{i+1}+d_i}{2}$: In this case $\lambda_i$ lies exactly in the center of the interval $[d_i,d_{i+1}]$, and since $|d_{i+1}-d_i|\geq \tau$, we have that $\min\lcurly
        \labs d_i-\lambda_i \rabs,
        \labs d_{i+1}-\lambda_i \rabs
        \rcurly
        \geq
        \frac{\tau}{2}\gg \frac{\tau^3}{n+1}$.
        \item $\lambda_i\in(d_i,\tfrac{d_{i+1}+d_i}{2})$: If $\lambda_i$ lies inside the left half of the interval, then clearly $|d_{i+1}-\lambda_i|> \frac{\tau}{2}$. 
        The magnitude of the right hand side of \eqref{eq:expanded_secular_equation_for_eigenvalue_bounds} can be bounded from above as follows
        \begin{align*}
            \labs
                \alpha 
                - 
                \lambda_i 
                -
                \sum_{j=2}^{i-1}
                \tfrac{\vecz_j^2}{d_j-\lambda_i}
                -
                \sum_{j=i+1}^{n}
                \tfrac{\vecz_j^2}{d_j-\lambda_i}
            \rabs
            &\leq
            |\alpha| + |\lambda_i| + \sum_{j=2,j\neq i}^n\tfrac{\vecz_j^2}{|d_j-\lambda_i|}\\
            &\leq
            1 + 1 + (n-3)\tfrac{1}{\tau} + \tfrac{1}{\tau/2}\\
            &\leq \tfrac{n+1}{\tau},
        \end{align*}
        which implies
        \begin{align*}
            \tfrac{\vecz_i^2}{|d_i-\lambda_i|} \leq \tfrac{n+1}{\tau}
            \Rightarrow
            |d_i-\lambda_i| \geq \tfrac{\tau\vecz_i^2}{n+1} \geq \tfrac{\tau^3}{n+1}.
        \end{align*}
        Thus $\min\lcurly
        \labs d_i-\lambda_i \rabs,
        \labs d_{i+1}-\lambda_i \rabs
        \rcurly
        \geq
        \min\lcurly
        \tfrac{\tau}{2},
        \tfrac{\tau^3}{n+1}
        \rcurly
        \geq
        \tfrac{\tau^3}{n+1}
        $.
        \item $\lambda_i\in(\tfrac{d_{i+1}+d_i}{2}, d_{i+1})$: With similar arguments we obtain that $\min\lcurly
        \labs d_i-\lambda_i \rabs,
        \labs d_{i+1}-\lambda_i \rabs
        \rcurly
        \geq
        \tfrac{\tau^3}{n+1}
        $ holds for this case as well, which concludes the proof.
    \end{enumerate}
\end{proof}
\end{lemma}

\subsection{Deflation}
\label{appendix:arrowhead_deflation}
In order to ensure that the given matrix satisfies the requirements of Equation \eqref{eq:arrowhead_desiderata} we will use deflation, specifically the methodology of section 4 in \cite{gu1995divide}.

\begin{proposition}[Arrowhead deflation]
    \label{proposition:arrowhead_deflation}
    Let $\matH=\begin{pmatrix}
        \alpha & \vecz^\top \\
        \vecz & \matD
    \end{pmatrix}$ be an arrowhead matrix in the form of Eq. \eqref{eq:arrowhead}. There exists an orthogonal matrix $\matG$ and a matrix $\matHtilde=\begin{pmatrix}
        \matHtilde' & 0 \\
        0 & \matDtilde
    \end{pmatrix}$ such that $\matDtilde$ is diagonal and $\matHtilde'=
    \begin{pmatrix}
        \widetilde\alpha' & \vecztilde'^\top \\
        \vecztilde' & \matDtilde'
    \end{pmatrix}
    $
    is an arrowhead matrix that satisfies the requirements of Equation \eqref{eq:arrowhead_desiderata}, specifically:
    \begin{align*}
    \widetilde d'_{j+1}-\widetilde d'_j \geq \tau, \quad
    |\vecztilde'_i|\geq \tau,
    \quad
    \text{and}
    \quad
    \|\matH-\matG\matHtilde\matG^\top\|\leq n\tau.    
    \end{align*}
      $\matHtilde$ can be computed in $O(n\log(n))$ arithmetic operations and comparisons, and the product $\matG^\top\matB$ for some matrix $\matB$ with $r$ columns can be computed in additional $O(nr)$ arithmetic operations on-the-fly.
    \begin{proof}
        This directly follows from the deflation strategy of Section 4.1 in \cite{gu1995divide}. We expand it's analysis here in three steps.

        \textbf{Step 1: Sorting.}
        First we sort the diagonal elements of $\matD=\begin{pmatrix}
            d_2 & & & \\
                & d_3 & & \\
                & & \ddots & \\
                & & & d_n
        \end{pmatrix}$ in ascending order, and then apply a permutation to ensure that
        \begin{align*}
            d_2 \leq d_3 \leq \ldots \leq d_n.
        \end{align*} Let $\matP_S$ be the corresponding permutation matrix. This procedure can be achieved in $O(n\log(n))$ comparisons, and it takes $O(n)$ to reorder $\matH_S = \matP_S\matH\matP_S^\top$ and $O(nr)$ to reorder $\matB_S = \matP_S^\top\matB$.
        
        \textbf{Step 2: Shaft deflation.} Next, we set all the elements in $\vecz$ that have magnitude smaller than $\tau$ to zero, by computing $\matH_Z = \matH_S-\matE_Z$, where $\|\matE_Z\|\leq (n-1)\tau$, since there are at most $n-1$ elements of $\vecz$ with magnitude less than $\tau$ that were set to zero. 
        We now choose another permutation matrix $\matP_Z$ such that the zeros in $\vecz$ go to the end. This way we obtain a matrix of the form
        \begin{align*} 
            \matP_Z\matH_Z\matP_Z^\top
            =
            \begin{pmatrix}
                \alpha & \vecz'^\top & 0 \\
                \vecz' & \matD' & 0 \\
                0 & 0 & \matD''
            \end{pmatrix}
            =
            \begin{pmatrix}
                \matH_Z' & 0 \\
                0 & \matD''
            \end{pmatrix}
            ,
        \end{align*}
        where $\matD'$ and $\matD''$ are diagonal, and  $\matD'$ remains sorted, while the elements of $\matD''$ are eigenvalues and can be deflated. The permutation costs $O(n)$ to apply on $\matH_Z$ and $O(nr)$ to apply on $\matB_S$. 

        \textbf{Step 3: Diagonal deflation.}
        We now focus on $\matH_Z'.$ First, observe that $\matH_Z'$ is in fact a principal submatrix of the original matrix $\matH$, i.e. it contains a subset of rows and columns of $\matH$. From the interlacing property, $\|\matH_Z'\|\leq \|\matH\|\leq 1$.
        It remains to ensure that $\matD_Z'$ satisfies the requirements of Equation \eqref{eq:arrowhead_desiderata}.
        Assume that $\matH'_Z$ has size $k\times k$, for some $0\leq k\leq n-1$. 
        We iterate over $j=k-1,\ldots,2$. If we meet a pair of entries such that $d'_{j+1}-d'_j \leq \tau$, then we apply the deflation technique of Equation $(15)$ of \cite{gu1995divide}. Specifically, we perform the following update:
            \begin{align*}
                \matH_Z' \leftarrow \matP_j (\matG_j \matH_Z' \matG_j^\top - \matE_j)\matP_j^\top,
            \end{align*}
        The matrices $\matP_j,\matG_j,$ and $\matE_j$ are described as follows. Let $j$ be the largest index such that $d'_{j+1}-d'_j\leq \tau$. The matrix $\matG_j$ encodes an elementary rotation and it has the same size $\matH'_Z$. They have the following form:
        \begin{align*}
        \matG_j &= \begin{pmatrix}
                1 &  &  &  &  &  & \\
                 & 1 &  &  &  & &\\
                 &  & \ddots &  &  & &\\
                 &  &  & c_j & -s_j & &\\
                 &  &  & s_j & c_j & & \\
                 &  &  &  && \ddots &\\
                 &  &  &  && &1
            \end{pmatrix},
            \quad
            \matH_Z' = \begin{pmatrix}
                \alpha & \vecz'_2 & \ldots & \vecz'_j & \vecz'_{j+1} & \ldots & \vecz'_k\\
                \vecz'_2 & d'_2 &  &  &  & &\\
                \vdots &  & \ddots &  &  & &\\
                \vecz'_j &  &  & d'_j &  & &\\
                \vecz'_{j+1} &  &  &  &d'_{j+1} & & \\
                \vdots &  &  &  && \ddots  &\\
                \vecz'_{k} &  &  &  && &d'_{k}
            \end{pmatrix},
            \\
            \matG_j\matH_Z'\matG_j^\top &= \begin{pmatrix}
                \alpha & \vecz'_2 & \ldots & \vecz''_j & 0 & \ldots & \vecz'_k\\
                \vecz'_2 & d'_2 &  &  &  & &\\
                \vdots &  & \ddots &  &  & &\\
                \vecz''_j &  &  & d''_j & \varepsilon_j & &\\
                0 &  &  & \varepsilon_j &d''_{j+1} & & \\
                \vdots &  &  &  && \ddots  &\\
                \vecz'_{k} &  &  &  && &d'_{k}
            \end{pmatrix},
            \quad
            \begin{array}{r l c r l}
                \vecz''_j=&\sqrt{\vecz'^2_j+\vecz'^2_{j+1}},
                & &
                s_j=&\vecz'_{j}/r_j,
                \\\\
                c_j=&\vecz'_{j+1}/r_j,
                & &
                d''_j=&d'_jc_j^2+d'_{j+1}s_j^2,
                \\\\
                d''_{j+1}=&d'_js_j^2+d'_{j+1}c_j^2,
                & &
                \varepsilon_j=&c_js_j(d'_{j+1}-d'_j).
            \end{array}
        \end{align*}
        $\matG_j$ sets $\vecz'_{j+1}$ to zero.
        Recall also that in $\matH_Z'$ for all $i=j+2,\ldots, k$ it holds that $d'_{i}-d'_{i-1}>\tau$, and for all $i=2,\dots,k$ it holds that $|\vecz'_i|\geq \tau$ and $d'_{i}\geq d'_{i-1}$.
        It is easy to see that $|\varepsilon_j|\leq \tau$. By setting $\matE_j=\begin{pmatrix}
            \ddots & & & \\
              & 0 & \varepsilon_j & \\
              &  \varepsilon_j & 0& \\
              & & & \ddots
        \end{pmatrix}$,
        we have that $\|\matE_j\|\leq \tau$ and $d''_{j+1}$ becomes an eigenvalue of the matrix $\matG_j\matH_Z'\matG_j^\top - \matE_j$. 
        Therefore, if we simply permute $d''_{j+1}$ to the bottom-right corner with a permutation matrix $\matP_j$, we obtain a matrix
        \begin{align*}
            \matP_{j}(\matG_j\matH'_Z\matG_j^\top - \matE_j)\matP_j^\top
            =
            \begin{pmatrix}
                \alpha & \vecz'_2 & \ldots & \vecz''_j & \ldots & \vecz'_k & 0
                \\
                \vecz'_2 & d'_2 &  &  &  & &
                \\
                \vdots &  & \ddots &  &  & &
                \\
                \vecz''_j &  &  & d''_j & & &
                \\
                \vdots &  &  &  &\ddots&   &
                \\
                \vecz'_{k} &  &  &  &&d'_{k} &
                \\
                0 &  &  &  & & & d''_{j+1}
            \end{pmatrix}
        \end{align*}
        For $d''_j$, recall that we can write $d'_{j+1}=d'_j+\gamma_j$, where $\gamma_j\in[0,\tau]$. Then $d''_j = c_j^2 d'_j + (1-c_j^2)d'_{j+1} = d'_j+(1-c_j^2)\gamma_j$. 
        Thus, $d''_j\geq d'_j\geq d'_{j-1}$. Also, by assumption $d'_{j+2}-d'_{j+1}>\tau$, which means that $d''_j =  d'_j+(1-c_j^2)\gamma_j \leq d'_{j+1}+\tau < d'_{j+2}$. Combining these two bounds, we conclude that \begin{align*}
            d'_{j-1}\leq d''_{j} \leq d'_{j+2}.
        \end{align*}
        Therefore, the diagonal part of the top-left $(k-1)\times(k-1)$ principal submatrix remains sorted after the deflation.
        
        After the deflation, if $j=k-1$ we update $j\leftarrow j-1$ and $k\leftarrow k-1$ and apply the same on the $k\times k$ leading principal submatrix of $\matP_{j}(\matG_j\matH'_Z\matG_j^\top - \matE_j)\matP_j^\top$. Otherwise  If the new diagonal elements do not satisfy $d'_{j+1}-d'_j>\tau$, we rotate and deflate once more. Otherwise, if $d'_{j+1}-d'_j>\tau$ or if $j=k$ after the deflation, we decrease $j\leftarrow j-1$ and continue. After every step, either the size of the arrowhead matrix (i.e. $k$) reduces by $1$ due to the deflation, or $j$ decreases by $1$. Therefore, after at most $n-1$ deflation steps the procedure will terminate. Each deflation step takes $O(1)$ arithmetic operations, and $O(r)$ operations to apply the corresponding rotation/permutation on $\matB$.

        Denoting by $\matH_0:=\matP_Z\matH_Z\matP_Z^\top
        =
        \matP_Z\lpar \matP_S\matH\matP_S^\top - \matE_Z\rpar \matP_Z^\top$,
        the final matrix $\matH_f$ is given by a sequence of operations
        \begin{align*}
        \matH_f
        &=
            \matP_1
            \bigg(
                \matG_1
                \bigg[
                    \ldots
                    \\
                    &\qquad\ldots
                    \lbrac
                        \matP_{k-2}
                        \lpar
                            \matG_{k-2}
                            \lbrac
                                \matP_{k-1}
                                \lpar \matG_{k-1}\matH_0\matG_{k-1} - \matE_{k-1} \rpar
                                \matP_{k-1}^\top
                            \rbrac
                            \matG_{k-2}^\top
                            -
                            \matE_{k-2}
                        \rpar
                        \matP_{k-2}^\top
                    \rbrac
                    \ldots
                    \\
                    &\quad\ldots
                \bigg]
                \matG_1^\top
                -
                \matE_1
            \bigg)
            \matP_1^\top
            \\
            &=
            \matP_1\matG_1\matP_2\matG_2\ldots\matP_{k-1}\matG_{k-1}
            \matH_0
            \matG_{k-1}^\top\matP_{k-1}^\top\ldots\matG_2^\top\matP_2^\top\matG_1^\top\matP_1^\top
            +
            \matE_1
            \\
            &=
            \matP_1\matG_1\matP_2\matG_2\ldots\matP_{k-1}\matG_{k-1}\matP_Z\matP_S\matH\matP_S^\top\matP_Z^\top
            \matG_{k-1}^\top\matP_{k-1}^\top\ldots\matG_2^\top\matP_2^\top\matG_1^\top\matP_1^\top
            +
            \matE_1
            +
            \matE_2
            \\
            &=
            \matG^\top \matH \matG
            +
            \matE_1
            +
            \matE_2,
        \end{align*}
        where $\matG^\top=\matP_1\matG_1\matP_2\matG_2\ldots\matP_{k-1}\matG_{k-1}\matP_Z\matP_S$ is orthogonal, the matrix $\matE_2$ is a rotated version of $\matE_Z$ and therefore $\|\matE_2\|=\|\matE_Z\|\leq (n-1)\tau$, and $\matE_1$ is the sum of at most $k-1$ rotated matrices $\matE_j$, in which case $\|\matE_1\|\leq (k-1)\tau$.

        We ultimately set the matrix to be returned $\matHtilde = \matG^\top\matH\matG = \matH_f-\matE_1-\matE_2$. It holds that
        \begin{align*}
            \lnorm \matH-\matG\matHtilde\matG^\top \rnorm
            &=
            \lnorm \matG\matE_1\matG^\top + \matG\matE_2\matG^\top \rnorm
            \leq
            n\tau.
        \end{align*}
        
        The total complexity is as follows. $O(n\log(n))$ comparisons are required for the sorting of the initial $\matD$. Step 2 needs $O(n)$ arithmetic operations to deflate and permute $\matH$, and $O(nr)$ are required to apply the transformations on $\matB$. Step 3 executes at most $n-1$ deflations, where each deflation requires $O(1)$ arithmetic operations and $O(r)$ operations to apply on $\matB$. Therefore the total complexity is $O(n\log(n))$ arithmetic operations and comparisons to obtain $\matHtilde$, plus additional $O(nr)$ operations to compute $\matG^\top\matB$ on-the-fly.
    \end{proof}
\end{proposition}

\subsection{Reconstruction from approximate eigenvalues}

If we have access to a set of approximate eigenvalues $\widehat\lambda_i$ of $\matH$ that also satisfy the same interlacing property, then we can construct a matrix $\matHhat$  that is close to $\matH$, and $\widehat\lambda_i$ are the eigenvalues of $\matHhat$. This is achieved with the following lemma from \cite{boley1977inverse}. We use its restatement from \cite{gu1995divide}.
\begin{lemma}[\cite{boley1977inverse,gu1995divide}]
    \label{lemma:arrowhead_reconstruction_from_shaft_and_eigenvalues}
    Given a set of $n$ numbers  $\widehat\lambda_1,\ldots,\widehat\lambda_n$ and a diagonal matrix $\matD=\diag(d_2,\ldots,d_n)$ such that
    \begin{align*}
        \widehat\lambda_1 < d_2 < \widehat\lambda_2 < \ldots < d_n <\widehat\lambda_n,
    \end{align*}
    there exists a symmetric arrowhead matrix $\matHhat=\begin{pmatrix}
        \widehat\alpha & \widehat\vecz^\top \\
        \widehat\vecz & \matD
    \end{pmatrix},$ whose eigenvalues are $\widehat\lambda_i$. In this case
    \begin{align}
        \label{eq:alpha_hat_and_zeta_hat}
        |\widehat\vecz_i| &= \sqrt{
            (d_i-\widehat\lambda_1)(\widehat\lambda_n-d_i)
            \prod_{j=2}^{i-1}
                \frac{\widehat\lambda_j-d_i}
                {d_j-d_i}
            \prod_{j=i}^{n-1}
                \frac{\widehat\lambda_j-d_i}
                {d_{j+1}-d_i}
        },
        \nonumber
        \\
        \widehat\alpha &= \widehat\lambda_1+\sum_{j=2}^n(\widehat\lambda_j-d_j),
    \end{align}
    where the sign of\ \  $\widehat\vecz_i$ can be chosen arbitrarily.
\end{lemma}

We shall use the spectral decomposition of $\matHhat$ as a backward approximate spectral decomposition of $\matH$. We write 
\begin{align*}
    \matH=\begin{pmatrix}
        \alpha & \vecz^\top \\
        \vecz & \matD
    \end{pmatrix},
    \quad \text{ and }\quad 
    \matHhat=\begin{pmatrix}
        \widehat\alpha & \widehat\vecz^\top \\
        \widehat\vecz & \matD
    \end{pmatrix},
\end{align*}
in which case $\|\matH-\matHhat\|\leq |\alpha-\widehat\alpha| + \|\vecz-\widehat\vecz\|$. It suffices to show that $\widehat\alpha \approx \alpha $ and $\widehat\vecz \approx \vecz$.

\begin{lemma}
    \label{lemma:h_backward_approximation}
    Let
    $\matH=\begin{pmatrix}
        \alpha & \vecz^\top \\
        \vecz & \matD
    \end{pmatrix}$
    be a symmetric arrowhead matrix with $\|\matH\|\leq 1$, satisfying the properties of Lemma \ref{lemma:arrowhead_preliminaries} with parameter $\tau$, and let $\widehat\lambda_1,\widehat\lambda_2,\ldots,\widehat\lambda_n$ be a set of approximate eigenvalues that satisfy
    \begin{align*}
        \widehat\lambda_1< d_2 < \widehat\lambda_2 < \ldots <d_n < \widehat\lambda_n, 
        \quad \text{and} \quad
        |\widehat\lambda_i -\lambda_i| \leq \epsilon\tfrac{\tau^3}{2(n+1)},
    \end{align*}
    for some $\epsilon\in(0,1/n)$. Then the quantities $\widehat\alpha$, $\veczhat$, and $\matHhat$ from Lemma \ref{lemma:arrowhead_reconstruction_from_shaft_and_eigenvalues} satisfy:
    \begin{align*}
        |\alpha-\widehat\alpha| &\leq \frac{\epsilon\tau^3}{2},\quad
        \|\vecz-\widehat\vecz\| \leq \frac{n\epsilon}{1-n\epsilon}, \quad
        \|\matH-\matHhat\| \leq \frac{\epsilon\tau^3}{2} + \frac{n\epsilon}{1-n\epsilon}.
    \end{align*}
    \begin{proof}
        From Eq. \eqref{eq:alpha_hat_and_zeta_hat} we have that $\alpha=\lambda_1+\sum_{j=2}^n(\lambda_j-d_j)$ and $\widehat\alpha=\widehat\lambda_1+\sum_{j-2}^n(\widehat\lambda_j-d_j)$. Then:
        \begin{align*}
            |\alpha-\widehat\alpha| = \labs 
                \sum_{j=1}^n (\lambda_j-\widehat\lambda_j)
            \rabs
            \leq
                \sum_{j=1}^n \labs \lambda_j-\widehat\lambda_j\rabs
            \leq \epsilon\tfrac{\tau^3}{2(n+1)} n
            \leq \tfrac{\epsilon\tau^3}{2}.
        \end{align*}

        We know that $|\widehat\lambda_i-\lambda_i| 
        \leq
        \epsilon\tfrac{\tau^3}{2(n+1)}.
        $
        Combined with Lemma \ref{lemma:arrowhead_preliminaries}, it implies that
        \begin{align*}
            |\lambda_i-\widehat \lambda_i| &\leq \epsilon |\lambda_i-d_i|, \text{ for all } i=2,\ldots,n,\\
            |\lambda_i-\widehat \lambda_i| &\leq \epsilon |\lambda_i-d_{i+1}|, \text{ for all } i=1,\ldots,n-1.
        \end{align*}
        Since $\max\{|\lambda_i-d_i|,|\lambda_i-d_{i+1}|\} \leq |\lambda_i-d_j|$, for all $j=1,\ldots,i-1,i+2,\ldots,n$, then we can write 
        \begin{align*}
            \widehat\lambda_i-d_j = \widehat\lambda_i-\lambda_i+\lambda_i-d_j=(1+\epsilon_{ij})(\lambda_i-d_j),
        \end{align*}
        for all $i,j\in[n]$, where $|\epsilon_{ij}|\leq \epsilon$.
        Then from Equation \eqref{eq:alpha_hat_and_zeta_hat} the elements of $\widehat\vecz$ are 
        \begin{align*}
            |\widehat\vecz_i| &= 
            \sqrt{
            (d_i-\widehat\lambda_1)(\widehat\lambda_n-d_i)
            \prod_{j=2}^{i-1}
                \frac{\widehat\lambda_j-d_i}
                {d_j-d_i}
            \prod_{j=i}^{n-1}
                \frac{\widehat\lambda_j-d_i}
                {d_{j+1}-d_i}
            }
            \\
            &=
            \sqrt{
            (d_i-\lambda_1)(1+\epsilon_{1i})
            (\lambda_n-d_i)(1+\epsilon_{ni})
            \prod_{j=2}^{i-1}
                \frac{\lambda_j-d_i}
                {d_j-d_i}(1+\epsilon_{ji})
            \prod_{j=i}^{n-1}
                \frac{\lambda_j-d_i}
                {d_{j+1}-d_i}(1+\epsilon_{ji})
            }
            \\
            &=
            |\vecz_i|
            \sqrt{
                \prod_{j=1}^{n}(1+\epsilon_{ji})
            }.
        \end{align*}
        By assumption, $\epsilon<1/n$. This allows us to use standard rounding error analysis (see e.g. \cite{higham2002accuracy}), from which we know that $\prod_{j=1}^n(1+\epsilon_{ij}) = 1+\theta_i,$ where $|\theta_i|\leq\frac{n\epsilon}{1-n\epsilon}$. 
        
        Finally, recall that from Lemma \ref{lemma:arrowhead_reconstruction_from_shaft_and_eigenvalues}, we can choose the sign of $\widehat\vecz_i$ arbitrarily. By choosing the sign of $\widehat\vecz_i$ to be the same as $\vecz_i$, we finally obtain
        \begin{align*}
            |\vecz_i-\widehat\vecz_i|  
            &= 
            |\vecz_i|
            \labs 
                \sqrt{1+\theta_i}-1
            \rabs
            =
            |\vecz_i|
            \labs
                \tfrac{\theta_i}
                {\sqrt{1+\theta_i}+1}
            \rabs
            \leq
            |\vecz_i|
            |\theta_i|
            \leq
            |\vecz_i|
            \tfrac{n\epsilon}{1-n\epsilon},
            \\
            \|\vecz-\widehat\vecz\|
            &\leq 
            \tfrac{n\epsilon}{1-n\epsilon}\|\vecz\| 
            \leq 
            \tfrac{n\epsilon}{1-n\epsilon}.
        \end{align*}
    \end{proof}
\end{lemma}

\section{Fast Arrowhead diagonalization}
\label{appendix:fast_arrowhead_diagonalization}
In this section we provide the full analysis of the algorithm of \cite{gu1995divide} when accelerated with the FMM.
The FMM was introduced in \cite{rokhlin1985rapid}, to accelerate the evaluation of integrals between interacting bodies, and it has been comprehensively analyzed in the literature \cite{darve2000fast,darve2000fast2,cai2020stable,sun2001matrix,martinsson2007accelerated,livne2002n,gu1993stable}. Consider a function of the form
\begin{align}
    f(x) = \sum_{j=1}^n c_j k(x-x_j),
    \label{equation:fmm_base}
\end{align}
where $x_j,c_j$ are constants and $k(x)$ is a suitably chosen kernel function, typically one of $\{\log(x), \tfrac{1}{x},\tfrac{1}{x^2}\}$. The FMM can be used to approximately evaluate $f(x)$ over $m$ different points $x_i$ in only $\widetilde O(m+n)$ arithmetic operations (suppressing logarithmic terms in the accuracy), instead of the naive $O(mn)$ evaluation. 

\subsection{Fast Multipole Method}
\label{appendix:fast_multipole_method}
For our analysis, we will need to use the FMM to evaluate the following functions \eqref{eq:fmm_1}-\eqref{eq:fmm_5}, each on $O(n)$ points. 
For each function, there is a guarantee that the evaluation points will satisfy certain criteria. More precisely, the magnitude of the denominators and the logarithm arguments  will have a well-defined lower bound. Here $\tau\in(0,1)$ is a parameter that is determined later.
\begingroup
\allowdisplaybreaks
\begin{align}
    &\text{Function:}
    &
    \text{Guaran}&\text{tees for FMM:}
    & 
    \text{  Used in:}\qquad
    \nonumber
    \\
    f(\lambda)&=\sum_{j=1}^n\frac{\vecz_j^2}{d_j-\lambda},
    &
    |d_j-\lambda|&\geq \Omega(\poly(\tfrac{\tau}{n})),
    & \text{Lemma \ref{lemma:fmm_approximate_eigenvalues}},
    \label{eq:fmm_1}
    \\
    f(d_i)&=\sum_{j=2}^{n}\log(|\widehat\lambda_j-d_i|),
    &
    |\widehat\lambda_j-d_i|&\geq \Omega(\poly(\tfrac{\tau}{n})),
    & \text{Lemma \ref{lemma:fmm_approximate_shaft}},
    \label{eq:fmm_2}
    \\
    f(d_i)&=\sum_{j=2}^{n}\log(|d_{j}-d_i|),
    &
    |d_j-d_i|&\geq \tau,
    & \text{Lemma \ref{lemma:fmm_approximate_shaft}},
    \label{eq:fmm_3}
    \\
    f(\lambda)&=\sum_{k=2}^n\frac{(1+\epsilon_k)\veczhat_k\vecq_k}{d_k-\lambda},
    &
    |d_k-\lambda|&\geq \Omega(\poly(\tfrac{\tau}{n})),
    & \text{Lemma \ref{lemma:fmm_approximate_inner_products}},
    \label{eq:fmm_4}
    \\
    f(\lambda)&=\sum_{k=2}^n\frac{(1+\epsilon_k)^2\veczhat^2_k}{(d_k-\lambda)^2},
    &
    |d_k-\lambda|&\geq \Omega(\poly(\tfrac{\tau}{n})),
    & \text{Lemma \ref{lemma:fmm_approximate_inner_products}}.
    \label{eq:fmm_5}
\end{align}
\endgroup

The fact that the magnitudes of the denonimators and the logarithm arguments are bounded from below allows us to use the seminal FMM analysis of \cite{gu1993stable,livne2002n,cai2020stable}. 
To the best of our knowledge, \cite{gu1993stable} is one of the first works to rigorously analyze the application of the FMM on such kernel functions with end-to-end approximation and complexity bounds. In Section 3.4 they achieved an error of $O\lpar
\epsilon \sum_i\tfrac{|x_i|}{|\omega^2-d_i^2|}
\rpar$
for the function $\Phi(\omega)=\sum_i\frac{x_i}{d_i^2-\omega^2}$, assuming that $d_i$ and $\omega$ satisfy similar interlacing properties to ours,
in $O(n\log^2(1/\epsilon))$ arithmetic operations. This complexity translates to $O(n\log^2(\tfrac{n}{\tau\epsilon}))$ arithmetic operations, if we rescale $\epsilon$ appropriately to obtain an absolute error $O(\epsilon)$, i.e., it satisfies the guarantees of Proposition \ref{proposition:fmm} with $\xi=2$. \cite{livne2002n} used a kernel-softening approach and report similar bounds in $O(n\log(1/\epsilon))$ operations. More recently, \cite{cai2020stable} developed a general framework based on the matrix-version of the FMM \cite{sun2001matrix}, and provided a thorough analysis and explicit bounds similar to \cite{gu1993stable}, which can be used to obtain guarantees for all the functions \eqref{eq:fmm_1}-\eqref{eq:fmm_5}. 

For completeness, we summarize the main ideas and results of the aforementioned works and provide a short proof for Proposition \ref{proposition:fmm} below.
We remind once more that we do not account for floating point errors, but rather focus only on the error FMM approximation errors. Obtaining full, end-to-end complexity analysis under floating point errors has its own merit, and it is left as future work.
\begin{proposition}[FMM]
    \label{proposition:fmm_appendix}
    There exists an algorithm, which we refer to as $(\epsilon,n)$-approximate FMM (or $(\epsilon,n)$-\fmmalgo, for short), which takes as input 
    \begin{itemize}
        \item a kernel function $k(x)=\lcurly \log(|x|), \frac{1}{x}, \frac{1}{x^2} \rcurly$,
        \item $2n+m$ real numbers: $\{x_1,\ldots,x_m\}\cup \{c_1,\ldots,c_n\}\cup\{y_1,\ldots,y_n\}$, and a constant $C$, such that $m\leq n$ and for all $i\in[m],j\in[n]$ it holds that
        \begin{align*}
            |x_i|,|c_j|,|y_j|<C
            \qquad
            \text{ and }
            \qquad
            |x_i-y_j|\geq \Omega(\poly(\tfrac{\epsilon}{n})).
        \end{align*}
    \end{itemize}
    It returns $m$ values $\widetilde f(x_1),\ldots,\widetilde f(x_m)$ such that
    $
        \labs
            \widetilde f(x_i)-f(x_i)
        \rabs
        \leq \epsilon,
    $
    for all $i\in[m]$, where $f(x) = \sum_{j=1}^n c_j k(x_i-y_j)$,
    in a total of $O\lpar 
        n\log^{\cfmm}(\tfrac{n}{\epsilon})
    \rpar$ arithmetic operations, where $\cfmm\geq 1$ is a small constant that is independent of $\epsilon,n$.
\begin{proof}
    All the evaluation points are always confined inside the interval $[-1,1]$.

    Following the seminal analysis of \cite[Sec. 2.1]{cai2020stable}, and simplifying it for the real case in exact arithmetic, let us initially focus on the kernel function $k(x-y)=\frac{1}{x-y}$. Consider  two sets of points: $X=\{x_1,\ldots,x_m\}$ and $Y=\{y_1,\ldots,y_n\}$, where  $x_i,y_j\in[-1,1]$. Moreover, assume that $X$ and $Y$ are \emph{$\beta$-separated}, with separation ratio $\beta$, which means that
    \begin{align*}
        \rho_X+\rho_Y \leq \beta|\mu_X-\mu_Y|, \quad 
        \beta\in(0,1),
    \end{align*}
    where $\mu_X=\frac{\sum x_i}{m}$ and $\mu_Y=\frac{\sum y_i}{n}$ are the centers of $X$ and $Y$, while $\rho_X=\max|x_i-\mu_X|$ and $\rho_Y=|y_i-\mu_Y|$ are the radii. Using a Taylor expansion we can write $k(x-y)$ in the so-called matrix form (\cite{sun2001matrix}) as follows:
    \begin{align}
        k(x-y)=\vecu^\top\matBbar\vecv + \epsilon_r,
        \label{eq:fmm_matvec_formulation}
    \end{align}
    where 
    \begingroup
    \allowdisplaybreaks
    \begin{align*}
        \vecu &= \begin{pmatrix}
            f_0(x-\mu_x) & f_1(x-\mu_x) & \ldots & f_{r-1}(x-\mu_x)
        \end{pmatrix}^\top
        \\
        \vecv &= \begin{pmatrix}
            f_0(y-\mu_Y) & f_1(y-\mu_Y) & \ldots & f_{r-1}(y-\mu_Y)
        \end{pmatrix}^\top
        \\
        \matBbar&=\begin{pmatrix}
            a_0 & a_1 &\cdots & a_{r-1}\\
            a_1 & \iddots & \iddots & \\
            \vdots & \iddots &  &\\
            a_{r-1} & & & 0\\
        \end{pmatrix}
        \begin{pmatrix}
            (-1)^0 & & & \\
            & (-1)^1 & &\\
            & & \ddots & \\
            & & & (-1)^{r-1}
        \end{pmatrix},
        \\
        f_j(x) &= \frac{x^j}{j!}, 
        \qquad 
        a_j=-\frac{j!}{(\mu_Y-\mu_X)^{j+1}}, 
        \qquad
        |\epsilon_r|\leq \beta^r\frac{1+\beta}{1-\beta}|k(x-y)|,
    \end{align*}
    \endgroup
    where the error bound holds due to the $\beta$-separation.
    
    Adapting the formulation for all $x\in X$ and $y\in Y$, we can write the entire interaction matrix as
    \begin{align*}
        \matK=\matUbar\matBbar\matVbar^\top + \matK\odot \matE,
    \end{align*}
    where $\odot$ is the element-wise product, $\matK_{i,j}=|k(x_i-y_j)|$, and $|\matE_{i,j}|\leq \beta^r\frac{1+\beta}{1-\beta}$.

    The goal of FMM is to efficiently approximate the matrix-vector product $\matK\vece$, where $\vece$ is the all-ones vector. This is achieved by truncating the $\matK\odot \matE$ term. In this case the total error for each evaluation point $x_i$ is bounded by 
    \begin{align}
        \label{eq:fmm_elementwise_error_bound}
        \epsilon_i := |\vece_i^\top(\matK\odot \matE)\vece| \leq \beta^r\frac{1+\beta}{1-\beta}\sum_{j=1}^{n}|k(x_i-y_j)|.
    \end{align}
    Assuming that the matrices have been pre-computed in $O((m+n)\poly(r))$ arithmetic operations, the product $\matUbar\matBbar\matVbar^\top \vece$ takes $O(r(r+m+n))$ operations.

    Next, we partition $[-1,1]$ in a constant number of subintervals with equal size, e.g., $I_1=[-1,-1/2], I_2=[-1/2,0], I_3=[0,1/2],$ and $I_4=[1/2,1]$. 
    Note that the sets $I_1$ and $I_3$ are $\frac{1}{2}$-separated. The same holds for $(I_2,I_4)$, while $(I_1,I_4)$ are $\frac{1}{3}$-separated. We can execute the following steps:
    \begin{enumerate}
        \item Split the sets $X$ and $Y$ into four subsets $X_1,X_2,X_3,X_4$, and  $Y_1,Y_2,Y_3,Y_4$, one subset for each interval $I_1,I_2,I_3,I_4$.
        \item Use a truncated Taylor expansion for each combination of ``long-range'' interactions, i.e., for the  set pairs $(X_1,Y_3),(X_1,Y_4), (X_2,Y_4), (X_3, Y_1), (X_4, Y_1)$, and $(X_4,Y_2)$.
        \item For the remaining ten pairs of ``short-range'' interactions, apply these steps recursively.
        \item Stop when  there are no more short-range interactions, i.e., when all the remaining 
        
        evaluations are between sets that are at least $\frac{1}{2}$-separated.
    \end{enumerate}

    Due to the deflation process, we have a guarantee that for \emph{all} $x\in X,y\in Y$, it holds that $|x-y|\geq \theta$, for some $\theta\in(0,1)$ to be specified later. Let $I^{(t)}_j$, $j=1,2,3,4$, be the four intervals at recursion depth $t\geq 1$. We know that $|I_j^{(t)}|=2\cdot4^{-t}$. Due to the lower bound on $|x-y|$, the final recursion depth is at most $t=O(\log(1/\theta))$, since at this depth every interval contains at most one element, either from $X$ or from $Y$. Moreover, all non-empty intervals are at least $\frac{1}{2}$-separated. 

    We now calculate the total number of arithmetic operations, starting at the bottom of the recursion tree, i.e., $t=C\log(1/\theta)$ for some constant $C$. The size of each interval at this depth is $2\cdot4^{-C\log(1/\theta)}=2\cdot \theta^{2C}$. Therefore, there are a total of $\frac{\labs[-1,1]\rabs}{2\cdot \theta^{2C}}=\frac{1}{\theta^{2C}}$ intervals, bundled in groups of four consecutive intervals. Of all those intervals, only $m+n$ are non-empty. Moreover, inside each group of four, all non-empty ones are at least $\frac{1}{2}$-separated, which implies that there are at most $2$ non-empty intervals in each group of four, each one containing a single element. 
    It is therefore sufficient to execute $O(m+n)$ evaluations using Eq. \eqref{eq:fmm_matvec_formulation} with a single vector each ($O(1)$ evaluations within each group of four). Each evaluation costs $O(r^2),$ which is independent of $m$ and $n$. Therefore, the total cost to calculate ``short-range'' interactions at the bottom of the recursion tree is at most $O((m+n)r^2)$.

    Next, we upper bound the cost for the recursion depth $t-1$, given the cost of depth $t$. Level $t$ prepares all the short-range interactions of level $t-1$, in time $T(t-1)$. At level $t-1$, there are four times fewer ``groups'' of intervals, and the intervals have four times larger size. Since the short-range interactions within each group were already prepared, it suffices to calculate the long-range interactions. We use again Eq. \eqref{eq:fmm_matvec_formulation}, but now potentially with more than one vectors within each group. 
    However, the total time for this calculation within each group scales linearly to the total number of points $X$ and $Y$ \emph{within each interval}.
    In particular, the cost within each (non-empty) group $l$ is $O(r^2(m_l+n_l))$. Summing the cost in all non-empty groups together, the total time to calculate the long-range interactions in \emph{all} groups scales as $O(r^2\sum_l (m_l+n_l))=O(r^2(m+n))$. 
    Since there are $O(\log(1/\theta))$ recursive steps, the total cost is bounded by $O(r^2(m+n)\log(1/\theta))$.
    
    It remains to bound the errors and the value of $r$. The final result for each evaluation point $x_i$ is a sum of applications of Eq. \eqref{eq:fmm_matvec_formulation}, and therefore the bound of Eq. \eqref{eq:fmm_elementwise_error_bound} applies for the total error, with $\beta\leq 1/2$. Due to the deflation, we have that $|k(x_i-y_j)|\leq \frac{1}{\theta}$, which means that Eq.  \eqref{eq:fmm_elementwise_error_bound} becomes
    \begin{align*}
        \epsilon_i \leq 3\left(\frac{1}{2}\right)^r\binom{n}{2}\frac{1}{\theta}.    
    \end{align*}
    If we set $r=O(\log(\frac{n}{\epsilon\theta}))$, then it holds that $\epsilon_i\leq \epsilon$, for all $i\in[n]$.
    Using the assumption of Proposition \ref{proposition:fmm} that $\theta \geq \Omega(\poly(\tfrac{\epsilon}{n}))$, we obtain the final bound for $r=O(\log(n/\epsilon))$.

    In a similar way we can analyze the other kernel functions of interest, namely, $\frac{1}{(x-y)^2}$ and $\log(|x-y|)$. Using their Taylor expansions we can write
    \begin{align*}
        \frac{1}{(x-y)^2} &= \frac{1}{(\rho_X-\rho_Y)^2}\sum_{j=0}^rj\frac{[(x-\rho_X)-(y-\rho_Y)]^{j-1}}{(\rho_Y-\rho_X)^{j-1}}+ \eta_r,
        \\
        \log(|x-y|) &= \log(|\rho_X-\rho_Y|)-\sum_{j=1}^r\frac{|(x-\rho_X)-(y-\rho_Y)|^j}{j|\rho_X-\rho_Y|^j} + \zeta_r.
    \end{align*}
    Further expanding the terms in the sums we can arrive to matrix formulations similar to Eq. \eqref{eq:fmm_matvec_formulation}. Moreover, assuming that the underlying sets $X$ and $Y$ are $\beta$-separated, the error terms $\eta_r$ and $\zeta_r$ are bounded in a similar way as $\epsilon_r$. 
    With further calculations we can obtain final bounds for $r$ to achieve a total additive error of at most $\epsilon$, similar to the kernel $1/(x-y)$.
    
    The analysis can be straightforwardly generalized when the summation coefficients of the kernel function are different than one, but  upper-bounded by some value $C$. Indeed, it suffices to replace the all-ones vector $\vece$ in Eq. \eqref{eq:fmm_elementwise_error_bound} with a vector that contains the numerators. If the upper bound $C$ is a constant the analysis is not affected.
\end{proof}
\end{proposition}

\subsection{Computing the eigenvalues with bisection}

Using the FMM as a black-box evaluation of the targeted kernel functions, we can compute all the eigenvalues of an arrowhead matrix which are given by the roots of the secular equation.
We highlight that \cite{livne2002n} provided a rigorous analysis of the Newton iteration, based on the results of \cite{melman1995numerical}, to compute all the roots of the secular equation. The reported complexity is $O(n\log(1/\epsilon))$, however, this does not include the number of Newton steps. Due to the well-known quadratic convergence properties, the number of Newton steps should be bounded by $O(\log(\log(1/\epsilon)))$. For the sake of simplicity and completeness, instead of repeating the analysis of \cite{livne2002n}, we will use the following standard bisection method to compute all the eigenvalues in a total of $O(n\log^2(1/\epsilon))$ operations. It might be slightly slower than the Newton iteration (up to a log factor), but it is significantly simpler and it does not require the evaluation of derivatives. It should therefore be both easier to implement and to analyze in finite precision. 

\begin{lemma}
    \label{lemma:fmm_approximate_eigenvalues}
    Let $\matH=\begin{pmatrix}
        \alpha & \vecz^\top\\
        \vecz & \matD
    \end{pmatrix}
    $
    be a symmetric arrowhead matrix 
    that satisfies the properties of Lemma \ref{lemma:arrowhead_preliminaries} for some parameter $\tau\in(0,1)$,
    with $\|\matH\|\leq 1$, 
    and assume that the diagonal elements of $\matD$ are sorted:
    $
        d_2 < d_3 <\ldots < d_n.
    $ 
    Given $\epsilon\in(0,1)$, and assuming an $(\epsilon,n)$-\fmmalgo\   implementation as in Proposition \ref{proposition:fmm}, we can compute approximate eigenvalues $\widehat\lambda_i$ such that
    \begin{align*}
        \widehat\lambda_1< d_2 < \widehat\lambda_2 < \ldots <d_n < \widehat\lambda_n, 
        \quad \text{and} \quad
        |\widehat\lambda_i -\lambda_i| \leq \epsilon,
    \end{align*}
    in $O\lpar
        n\log(\tfrac{1}{\epsilon})\log^{\cfmm}(\frac{n}{\tau\epsilon})
    \rpar$ arithmetic operations.
    \begin{proof}
        From Lemma \ref{lemma:arrowhead_preliminaries}, the eigenvalues $\lambda_i$ of $\matH$ are the roots of the secular equation $f(\lambda)=\lambda-\alpha+\sum_{j=2}^n \tfrac{\vecz_j^2}{d_j-\lambda}$ and they satisfy the interlacing property 
        \begin{align*}
            -1 \leq \lambda_1 < d_2 < \lambda_2 < d_2 <\ldots < d_n < \lambda_n \leq 1,
        \end{align*}
        and also that the lower bounds of Equation \eqref{eq:arrowhead_eigenvalue_boundaries_distance} hold.
        We thus need to search for the roots of $f(\lambda)$ inside the intervals $
        I_1=[-1,d_1-\tfrac{\tau^3}{n+1}]
        $, 
        $
        I_n=[d_n+\tfrac{\tau^3}{n+1},1]
        $, and, for all $i=1,\ldots, n-1$, $
        I_i=[d_i+\tfrac{\tau^3}{n+1},d_{i+1}-\tfrac{\tau^3}{n+1}]
        $. 
        
        Let $l_i$ and $r_i$ be the corresponding left and right boundaries of $I_i$, and initialize $\epsilon_{FMM}=\epsilon/2$, and $k=0$ (an iteration counter). 
        A binary search follows.
        \begin{enumerate}
            \item Set $\lambda_i^{(k)}=\frac{l_i+r_i}{2}$ and use the $(\epsilon_{FMM},\cfmm,n)$-\fmmalgo\   to compute $\widetilde f(\lambda_i^{(k)})$ such that $|f(\lambda_i^{(k)})-\widetilde f(\lambda_i^{(k)})|\leq \epsilon_{FMM}$.
            \item If $|l_i-r_i| \leq \epsilon$, stop and return $\widehat\lambda_i=\lambda_i^{(k)}$. (At every step $\lambda_i\in[l,r]$, and therefore $|\lambda_i^{(k)}-\lambda_i|\leq |l_i-r_i|\leq \epsilon$, which means that we can terminate).
            \item If $|\widetilde f(\lambda_i^{(k)})| \leq \epsilon_{FMM}$, stop and return $\widehat\lambda_i=\lambda_i^{(k)}$. In that case we can no longer determine the true sign of $f(\lambda_i^{(k)})$, and thus the binary search cannot continue further.
            \item If none of the above criteria are met, then we update the bounding interval $[l_i,r_i]$ as follows:
            \begin{itemize}
                \item If $\widetilde f(\lambda_i^{(k)})-\epsilon_{FMM}>0$, we set $r_i=\lambda_{i}^{(k)}$.
                \item Else, if $\widetilde f(\lambda_i^{(k)})+\epsilon_{FMM}<0$, we set $l_i=\lambda_i^{(k)}$.
            \end{itemize}
            \item Update $k\leftarrow k+1$, and go to Step 1.
        \end{enumerate}
        If the algorithm stops and returns at Step 2, then $|\lambda_i-\widehat\lambda_i|\leq \epsilon$ is already satisfied. 
        On the other hand, if the algorithm stops at Step 3,
        the returned solution $\widehat\lambda_i$ satisfies
        \begin{align*}
            |f(\widehat\lambda_i)| \leq |f(\widehat\lambda_i)-\widetilde f(\widehat\lambda_i)| + |\widetilde f(\widehat\lambda_i)|
            \leq \epsilon_{FMM} + \epsilon_{FMM} \leq 2\epsilon_{FMM} = \epsilon.
        \end{align*}
        Since $f(\lambda_i)=0$, we have that
        \begin{align*}
            f(\widehat\lambda_i) 
            &= 
            f(\widehat\lambda_i)-f(\lambda_i) 
            \\
            &= 
            \widehat\lambda_i
            -
            \alpha 
            + 
            \sum_{j=2}^n\frac{\vecz_j^2}{d_j-\widehat\lambda_i}
            -
            \lambda_i
            +
            \alpha
            -
            \sum_{j=2}^n\frac{\vecz_j^2}{d_j-\lambda_i}
            \\
            &=
            (\widehat\lambda_i-\lambda_i)
            \lbrac
                1+\sum_{j=2}^n \frac{
                    \vecz_j^2
                }
                {
                    (d_j-\widehat\lambda_i)(d_j-\lambda_i)
                }
            \rbrac.
        \end{align*}
        But $\widehat\lambda_i$ and $\lambda_i$ lie strictly inside the same interval $I_i\subset(d_{i-1},d_i)$, and therefore the denominator in the sum is always positive, which implies that
        \begin{align}
            |\widehat\lambda_i-\lambda_i| 
            &\leq
            |\widehat\lambda_i-\lambda_i| 
            \labs
                1+\sum_{j=2}^n \tfrac{
                    \vecz_i^2
                }
                {
                    (d_j-\widehat\lambda_i)(d_j-\lambda_i)
                }
            \rabs
            =
            \labs
                f(\widehat\lambda_i)-f(\lambda_i)
            \rabs
            =
            \labs
                f(\widehat\lambda_i)
            \rabs
            \leq
            \epsilon.
            \label{eq:lemma_fmm_approximate_eigenvalues}
        \end{align}
        Therefore, the returned solution satisfies the desired bound $|\lambda_i-\widehat\lambda_i|\leq \epsilon$ regardless of which stopping criterion is met first.
        The solutions $\widehat\lambda_i$ are inside the desired intervals $I_i$, since we only evaluate the secular equation on points that lie inside $I_i$. Therefore, the lower bounds of Equation \eqref{eq:arrowhead_eigenvalue_boundaries_distance} also hold for $\widehat\lambda_i$.

        At each iteration $k$, use the FMM to compute $\widetilde f(\widehat\lambda_i)$ for all $i$ simultaneously using in 
        $
        O(n\log^{\cfmm}(\tfrac{n}{\tau\epsilon_{FMM}}))
        =
        O(n\log^{\cfmm}(\tfrac{n}{\tau\epsilon}))
        $ arithmetic operations. We then need to iterate over all the returned values in $O(n)$ arithmetic operations to determine whether the corresponding eigenvalues have converged or if we need to perform additional iterations.

        It remains to bound the number of binary search iterations to reach any of the two termination criteria. Note that at every iteration the algorithm is halving the range of $[l_i,r_i]$ for every eigenvalue. This means that after $m$ iterations $\lambda_i$ is restricted inside an interval with size at most $2^{-m}$. Therefore, the termination criterion of Step 1 is achieved after at most $m=O(\log(1/\epsilon))$ iterations. Since the additional stopping criterion can only make the algorithm terminate faster, the total number of binary search iterations is at most $O(\log(1/\epsilon))$.
        The total runtime of the algorithm is therefore $O\lpar
            n\log(\tfrac{1}{\epsilon})\log^{\cfmm}(\tfrac{n}{\tau\epsilon})
        \rpar$.
    \end{proof}
\end{lemma}

\subsection{Approximating the elements of the shaft}
As a next step, we use the trick of \cite{gu1995divide} to approximate the elements of $\matHhat$ from Lemma \ref{lemma:h_backward_approximation}. 
\begin{lemma}
    \label{lemma:fmm_approximate_shaft}
    Let $\matH=\begin{pmatrix}
        \alpha & \vecz^\top\\
        \vecz & \matD
    \end{pmatrix}
    $
    be a symmetric arrowhead matrix with $\|\matH\|\leq 1$, that satisfies the requirements of Lemma \ref{lemma:arrowhead_preliminaries} with parameter $\tau$, and let $d_2 < \ldots < d_n$ be the diagonal elements of  $\matD$. From Lemma \ref{lemma:h_backward_approximation}, if we are given a set of approximate eigenvalues $\widehat\lambda_i$ that satisfy 
    \begin{align*}
        \widehat\lambda_1< d_2 < \widehat\lambda_2 < \ldots <d_n < \widehat\lambda_n, 
        \quad \text{and} \quad
        |\widehat\lambda_i -\lambda_i| \leq \epsilon\tfrac{\tau^3}{2(n+1)},
    \end{align*}
    for some $\epsilon\in(0,1/n)$, then there exists a matrix $\matHhat=\begin{pmatrix}
        \widehat\alpha & \widehat\vecz^\top \\
        \widehat\vecz & \matD
    \end{pmatrix},$ such that $\widehat\lambda_i$ are the exact eigenvalues of $\matHhat$ and 
    $\|\matH-\matHhat\| 
    \leq 
    \frac{\epsilon\tau^3}{2} + \frac{n\epsilon}{1-n\epsilon}
    $.
    Assuming an $(\epsilon,n)$-\fmmalgo, for any $\epsilon_{\vecz}\in(0,1/2)$, we can compute an approximate
    vector $\widehat\vecz'$ such that $|\widehat\vecz_i-\widehat\vecz_i'|\leq \epsilon_{\vecz}|\widehat\vecz_i|$ and  $\|\widehat\vecz'-\widehat\vecz\|\leq \epsilon_{\vecz} (1 + \frac{n\epsilon}{1-n\epsilon})$ in $
        O\lpar
            n\log(\tfrac{1}{\epsilon_{\vecz}})
            \log^{\cfmm}(\tfrac{n}{\tau\epsilon_{\vecz}}))
        \rpar
    $ arithmetic operations. The matrix $\matHhat'=\begin{pmatrix}
        \widehat\alpha & \widehat\vecz'^\top \\
        \widehat\vecz' & \matD
    \end{pmatrix}$ satisfies
    \begin{align*}
        \|\matH-\matHhat'\| \leq 
            \frac{\epsilon\tau^3}{2} 
            +
            \epsilon_{\vecz} + (1+\epsilon_{\vecz})\frac{n\epsilon}{1-n\epsilon}.
    \end{align*}
    \begin{proof}
        The observation is that $|\widehat\vecz_i|$ from Eq. \eqref{eq:alpha_hat_and_zeta_hat} can be written as
        \begin{align*}
            |\widehat\vecz_i|=\sqrt{
                (d_i-\widehat\lambda_1)(\widehat\lambda_n-d_i)
            }
            \exp(\Phi(d_i)),
        \end{align*}
        where 
        \begin{align*}
            \Phi(d_i)
            =
            \frac{1}{2}\lpar
                \sum_{j=2}^{i-1}
                    \log(d_i-\widehat\lambda_j)
                -
                \sum_{j=2}^{i-1}
                    \log(d_i-d_j)
                +
                \sum_{j=i+1}^{n}
                    \log(\widehat\lambda_j-d_i)
                -
                \sum_{j=i+1}^{n}
                    \log(d_{j}-d_i)
            \rpar.
        \end{align*}
        We thus need to evaluate the functions \eqref{eq:fmm_2}-\eqref{eq:fmm_5}, each at $n$ points, where all the points satisfy the corresponding guarantees. We can thus choose $\epsilon'=\epsilon_{\vecz}/2$, and call $(\epsilon',\cfmm,n)$-\fmmalgo\   to evaluate all the functions on all the points simultaneously in \begin{align*}
            O\lpar
                n\log(\tfrac{1}{\epsilon'})
                \log^{\cfmm}(\tfrac{n}{\tau\epsilon'}))
            \rpar
            =
            O\lpar
                n\log(\tfrac{1}{\epsilon_{\vecz}})
                \log^{\cfmm}(\tfrac{n}{\tau\epsilon_{\vecz}}))
            \rpar
        \end{align*}
        arithmetic operations, such that, for all $d_i$, $\widetilde\Phi(d_i)= \Phi(d_i)+\epsilon'_i$, where $|\epsilon'_i|\leq \epsilon'$. Then
        \begin{align*}
            |\widehat\vecz_i'| = \sqrt{
                (d_i-\widehat\lambda_1)(\widehat\lambda_n-d_i)
            }
            \exp(\widetilde\Phi(d_i))
            =
            \sqrt{
                (d_i-\widehat\lambda_1)(\widehat\lambda_n-d_i)
            }
            \exp(\Phi(d_i))
            \exp(\epsilon'_i)
            =
            |\widehat\vecz_i|
            \exp(\epsilon'_i).
        \end{align*}
        Recall that that for $1+2x\leq \exp(x)$ for $-1/4\leq x\leq 0$ and $\exp(x)\leq 1+2x$ for $0\leq x\leq 1/4$, which means that $\exp(\epsilon'_i)\in 1\pm 2\epsilon'$, 
        since $\epsilon'=\epsilon_{\vecz}/2\in(0,1/4)$.  This finally gives $\|\widehat\vecz-\widehat\vecz'\|\leq \epsilon_{\vecz} \|\widehat\vecz\|$.

        The matrix $\matHhat'=\begin{pmatrix}
        \widehat\alpha & \widehat\vecz'^\top \\
        \widehat\vecz' & \matD
        \end{pmatrix}$ satisfies
        \begin{align*}
            \|\matH-\matHhat'\| \leq \|\matH-\matHhat\| + \|\matHhat-\matHhat'\|
            \leq
            \frac{\epsilon\tau^3}{2} + \frac{n\epsilon}{1-n\epsilon}
            +
            \| \veczhat - \veczhat'\|
            \leq
            \frac{\epsilon\tau^3}{2} + \frac{n\epsilon}{1-n\epsilon}
            +
            \epsilon_{\vecz}\| \veczhat\|.
        \end{align*}
        From Lemma \ref{lemma:h_backward_approximation} we have that $\| \veczhat\| \leq \|\vecz\| + \|\vecz-\veczhat\| \leq 1 + \frac{n\epsilon}{1-n\epsilon}$ which gives the final bounds.
    \end{proof}
\end{lemma}

\subsection{Approximating inner products with the eigenvectors}
In the last part of the analysis we provide bounds on the errors for inner products of the form $\vecuhat^\top_i\vecq$, between the $i$-th eigenvector and some arbitrary vector $\vecq$. Following the procedure of Section 5 in \cite{gu1995divide}, we write
\begin{align*}
    \vecuhat^\top_i\vecq = \frac{-\vecq_1+\Phi(\widehat\lambda_i)}{\sqrt{1+\Psi(\widehat\lambda_i)}},
\end{align*}
where $\Phi(\lambda)=\sum_{k=2}^n\frac{\veczhat_k\vecq_k}{d_k-\lambda}$ and 
$\Psi(\lambda)=
    \sum_{k=2}^n
        \frac{\veczhat_k^2}{(d_k-\lambda)^2}.
$
To compute all the products for all $i$ with $\vecq$, we can use FMM to approximate $\Phi(\lambda)$ and $\Psi(\lambda)$ on $n$ points.

\begin{lemma}
    \label{lemma:fmm_approximate_inner_products}
        Let $\matH=\begin{pmatrix}
            \alpha & \vecz^\top\\
            \vecz & \matD
        \end{pmatrix}
        $
        be a symmetric arrowhead matrix with $\|\matH\|\leq 1$, that satisfies the requirements of Lemma \ref{lemma:arrowhead_preliminaries} with parameter $\tau$, and let $d_2 < d_3 < \ldots < d_n$ be the diagonal elements of  $\matD$. Let $\epsilon_{\vecz}$, $\widehat\lambda_i$, $\veczhat$, $\veczhat'$, $\matHhat$, and $\matHhat'$ be the same as in Lemma \ref{lemma:fmm_approximate_shaft}. 
        Let $\vecq$ be a fixed vector with $\|\vecq\|\leq 1$, and $\vecuhat_i, i\in[n]$, be the eigenvectors of $\matHhat$.
        Assuming an $(\epsilon,n)$-\fmmalgo, we can approximate all the inner products 
        $\vecuhat_i^\top\vecq$,  by some values $x_i$ such that 
        \begin{align*}
            |\vecuhat_i^\top\vecq - x_i| \leq 147\epsilon_{\vecz}
                \tfrac{(n+1)^2}{\tau^6},
        \end{align*}
        for all $i\in[n]$ simultaneously, in $O\lpar
            n\log(\tfrac{1}{\epsilon_{\vecz}})
            \log^{\cfmm}(\tfrac{n}{\tau\epsilon_{\vecz}}))
        \rpar$ arithmetic operations.
    \begin{proof}        
        First, let us rewrite the functions $\Phi$ and $\Psi$ to account for the errors in $\veczhat'$. Recall that from Lemma \ref{lemma:fmm_approximate_shaft} it holds that $|\veczhat_i-\veczhat'_i|\leq \epsilon_{\vecz}|\veczhat_i|$, in which case we can write 
        \begin{align*}
        \Phi'(\lambda)=\sum_{k=2}^n\frac{(1+\epsilon_k)\veczhat_k\vecq_k}{d_k-\lambda},
            \quad \text{ and } \quad
        \Psi'(\lambda)=
            \sum_{k=2}^n
                \frac{(1+\epsilon_k)^2\veczhat^2_k}{(d_k-\lambda)^2},
        \end{align*}
        where $|\epsilon_k|\leq\epsilon_{\vecz}$. Let $\vecuhat'_i$ be the $i$-th eigenvector of $\matHhat'$.
        Now we can consider the following bound:
        \begin{align}
            \labs
            \vecuhat_i^\top\vecq
            -
            \vecuhat_i'^\top\vecq
            \rabs
            &=
            \labs
            \frac{-\vecq_1+\Phi(\widehat\lambda_i)}{\sqrt{1+\Psi(\widehat\lambda_i)}}
            -
            \frac{-\vecq_1+\Phi'(\widehat\lambda_i)}{\sqrt{1+\Psi'(\widehat\lambda_i)}}
            \rabs\nonumber
            \\
            &=
            \labs
            \frac{
                (-\vecq_1+\Phi(\widehat\lambda_i))\sqrt{1+\Psi'(\widehat\lambda_i)}
                -
                (-\vecq_1+\Phi'(\widehat\lambda_i))\sqrt{1+\Psi(\widehat\lambda_i)}
            }{
                \sqrt{1+\Psi(\widehat\lambda_i)}\sqrt{1+\Psi'(\widehat\lambda_i)}
            }
            \rabs\nonumber
            \\
            &=
            \labs
            \frac{
                \Phi(\widehat\lambda_i)\sqrt{1+\Psi'(\widehat\lambda_i)}
                -
                \Phi'(\widehat\lambda_i)\sqrt{1+\Psi(\widehat\lambda_i)}
                +
                \vecq_1\lpar
                \sqrt{1+\Psi(\widehat\lambda_i)}
                -
                \sqrt{1+\Psi'(\widehat\lambda_i)}
                \rpar
            }{
                \sqrt{1+\Psi(\widehat\lambda_i)}\sqrt{1+\Psi'(\widehat\lambda_i)}
            }
            \rabs
            \nonumber
            \\
            &\leq
            \frac{
                \labs
                    \Phi(\widehat\lambda_i)\sqrt{1+\Psi'(\widehat\lambda_i)}
                -
                \Phi'(\widehat\lambda_i)\sqrt{1+\Psi(\widehat\lambda_i)}
                \rabs
                +
                \labs
                \vecq_1\lpar
                \sqrt{1+\Psi(\widehat\lambda_i)}
                -
                \sqrt{1+\Psi'(\widehat\lambda_i)}
                \rpar
                \rabs
            }{
                \labs
                    \sqrt{1+\Psi(\widehat\lambda_i)}\sqrt{1+\Psi'(\widehat\lambda_i)}
                \rabs
            }
            \label{eq:fmm_approximate_inner_product_1}
        \end{align}
        Let $\zeta_k=\frac{\veczhat_k}{d_k-\widehat\lambda_i}$. Assuming $\epsilon_{\vecz}\in(0,1/2)$, we get the following lower bound for the denominator
        \begin{align}
            \labs
                \sqrt{1+\Psi(\widehat\lambda_i)}\sqrt{1+\Psi'(\widehat\lambda_i)}
            \rabs
            &=
            \labs
                \sqrt{1+\sum_{k=2}^n\zeta_k^2}\sqrt{1+\sum_{k=2}^n(1+\epsilon_k)^2\zeta_k^2}
            \rabs
            \nonumber
            \\
            &\geq
            \labs
                \sqrt{1+\sum_{k=2}^n\zeta_k^2}\sqrt{1+\frac{1}{4}\sum_{k=2}^n\zeta_k^2}
            \rabs
            \nonumber
            \\
            &\geq
            \frac{1}{2}
            \labs
                1+\sum_{k=2}^n\zeta_k^2
            \rabs
            \nonumber
            \\
            &=
            \frac{1}{2}
            \labs
                1+\Psi(\widehat\lambda_i)
            \rabs.
            \label{eq:fmm_approximate_inner_product_2}
        \end{align}
        For the right part of the numerator, we have
        \begingroup
        \allowdisplaybreaks
        \begin{align}
            \label{eq:fmm_approximate_inner_product_2b}
            \labs
                \vecq_1\lpar
                \sqrt{1+\Psi(\widehat\lambda_i)}
                -
                \sqrt{1+\Psi'(\widehat\lambda_i)}
                \rpar
            \rabs
            &\leq
            \labs
                \sqrt{1+\sum_{k=2}^n\zeta_k^2}-\sqrt{1+\sum_{k=2}^n(1+\epsilon_k)^2\zeta_k^2}
            \rabs
            \nonumber
            \\
            &\leq
            \labs
                1+\sum_{k=2}^n\zeta_k^2
                - 
                1
                -
                \sum_{k=2}^n(1+\epsilon_k)^2\zeta_k^2
            \rabs        
            \nonumber
            \\
            &\leq
            \labs
                \sum_{k=2}^n\epsilon_k(2+\epsilon_k)\zeta_k^2
            \rabs 
            \nonumber
            \\
            &\leq
            3\epsilon_{\vecz}
            \sum_{k=2}^n\zeta_k^2
            \nonumber
            \\
            &=
            3\epsilon_{\vecz}\Psi(\widehat\lambda_i).
        \end{align}
        \endgroup
        For the left part of the numerator, we have
        \begingroup
        \allowdisplaybreaks
        \begin{align}
            &\labs
                \Phi(\widehat\lambda_i)\sqrt{1+\Psi'(\widehat\lambda_i)}
                -
                \Phi'(\widehat\lambda_i)\sqrt{1+\Psi(\widehat\lambda_i)}
            \rabs
            \nonumber
            \\
            &\quad=
            \labs
                \lpar
                    \sum_{k=2}^n\zeta_k\vecq_k
                \rpar
                \sqrt{1+\sum_{k=2}^n(1+\epsilon_k)^2\zeta_k^2}
                -
                \lpar
                    \sum_{k=2}^n(1+\epsilon_k)\zeta_k\vecq_k
                \rpar
                \sqrt{1+\sum_{k=2}^n\zeta_k^2}
            \rabs
            \nonumber
            \\
            &\quad=
            \labs
                \sum_{k=2}^n
                \lpar
                    \zeta_k\vecq_k
                    \lbrac
                        \sqrt{1+\sum_{l=2}^n(1+\epsilon_l)^2\zeta_l^2}
                        -
                        (1+\epsilon_k)
                        \sqrt{1+\sum_{l=2}^n\zeta_l^2}
                    \rbrac
                \rpar
            \rabs
            \nonumber
            \\
            &\quad=
            \labs
                \sum_{k=2}^n
                \lpar
                    \zeta_k\vecq_k
                    \frac{
                        1+\sum_{l=2}^n(1+\epsilon_l)^2\zeta_l^2
                        -
                        (1+\epsilon_k)^2
                        \lpar 1+\sum_{l=2}^n\zeta_l^2\rpar 
                    }
                    {
                        \sqrt{1+\sum_{l=2}^n(1+\epsilon_l)^2\zeta_l^2}
                        +
                        (1+\epsilon_k)
                        \sqrt{1+\sum_{l=2}^n\zeta_l^2}
                    }
                \rpar
            \rabs
            \nonumber
            \\
            &\quad=
            \labs
                \sum_{k=2}^n
                \lpar
                    \zeta_k\vecq_k
                    \frac{
                        1
                        -
                        (1+\epsilon_k)^2
                        +
                        \sum_{l=2}^n(1+\epsilon_l)^2\zeta_l^2
                        -
                        \sum_{l=2}^n(1+\epsilon_k)^2\zeta_l^2 
                    }
                    {
                        \sqrt{1+\sum_{l=2}^n(1+\epsilon_l)^2\zeta_l^2}
                        +
                        (1+\epsilon_k)
                        \sqrt{1+\sum_{l=2}^n\zeta_l^2}
                    }
                \rpar
            \rabs
            \nonumber
            \\
            &\quad=
            \labs
                \sum_{k=2}^n
                \lpar
                    \zeta_k\vecq_k
                    \frac{
                        1
                        -
                        (1+\epsilon_k)^2
                        +
                        \sum_{l=2}^n
                            \zeta_l^2\lbrac 
                                (1+\epsilon_l)^2-(1+\epsilon_k)^2
                            \rbrac
                    }
                    {
                        \sqrt{1+\sum_{l=2}^n(1+\epsilon_l)^2\zeta_l^2}
                        +
                        (1+\epsilon_k)
                        \sqrt{1+\sum_{l=2}^n\zeta_l^2}
                    }
                \rpar
            \rabs
            \nonumber
            \\
            &\quad=
            \labs
                \sum_{k=2}^n
                \lpar
                    \zeta_k\vecq_k
                    \frac{
                        -2\epsilon_k-\epsilon_k^2
                        +
                        \sum_{l=2}^n
                            \zeta_l^2\lbrac 
                                2\epsilon_l + \epsilon_l^2-2\epsilon_k+\epsilon_k^2
                            \rbrac
                    }
                    {
                        \sqrt{1+\sum_{l=2}^n(1+\epsilon_l)^2\zeta_l^2}
                        +
                        (1+\epsilon_k)
                        \sqrt{1+\sum_{l=2}^n\zeta_l^2}
                    }
                \rpar
            \rabs
            \nonumber
            \\
            &\quad\leq
            \sum_{k=2}^n
            \lpar
                |\zeta_k\vecq_k|
                \frac{
                    \labs
                        -2\epsilon_k-\epsilon_k^2
                        +
                        \sum_{l=2}^n
                            \zeta_l^2\lbrac 
                                2\epsilon_l + \epsilon_l^2-2\epsilon_k+\epsilon_k^2
                            \rbrac
                    \rabs
                }
                {
                    \labs
                    \sqrt{1+\sum_{l=2}^n(1+\epsilon_l)^2\zeta_l^2}
                    +
                    (1+\epsilon_k)
                    \sqrt{1+\sum_{l=2}^n\zeta_l^2}
                    \rabs
                }
            \rpar.
            \label{eq:fmm_approximate_inner_product_3}
        \end{align}
        \endgroup
        Once again, we lower bound the denominator and upper bound the numerator for Eq. \eqref{eq:fmm_approximate_inner_product_3}. For the denominator we have:
        \begin{align}
            \labs
                \sqrt{1+\sum_{l=2}^n(1+\epsilon_l)^2\zeta_l^2}
                +
                (1+\epsilon_k)
                \sqrt{1+\sum_{l=2}^n\zeta_l^2}
            \rabs
            \nonumber
            &\geq
            \labs
                \sqrt{1+\frac{1}{4}\sum_{l=2}^n\zeta_l^2}
                +
                \frac{1}{2}
                \sqrt{1+\sum_{l=2}^n\zeta_l^2}
            \rabs
            \nonumber\\
            &\geq
                \sqrt{1+\sum_{l=2}^n\zeta_l^2}
                \nonumber\\
            &=\sqrt{1+\Psi(\widehat\lambda_i)}.
            \label{eq:fmm_approximate_inner_product_4}
        \end{align}
        For the numerator it holds that:
        \begin{align}
            \labs
                -2\epsilon_k-\epsilon_k^2
                +
                \sum_{l=2}^n
                    \zeta_l^2\lbrac 
                        2\epsilon_l + \epsilon_l^2-2\epsilon_k+\epsilon_k^2
                    \rbrac
            \rabs
            \leq
            3\epsilon_{\vecz} + 6\epsilon_{\vecz}\sum_{l=2}^n\zeta_l^2
            =
            3\epsilon_{\vecz} + 6\epsilon_{\vecz}\Psi(\widehat\lambda_i)
            .
            \label{eq:fmm_approximate_inner_product_5}
        \end{align}
        Finally, recall that from Lemma \ref{lemma:arrowhead_preliminaries} it holds that
        \begin{align}
            |\zeta_k\vecq_k| 
            &=
            \labs
                \frac{\vecz_k\vecq_k}{d_k-\widehat\lambda_i}
            \rabs
            \leq
            \frac{2(n+1)}{\tau^3},
            \nonumber
            \\
            \Psi(\widehat\lambda_i) 
            &= 
            \sum_{k=2}^n \zeta_k^2 
            =
            \sum_{k=2}^n \frac{\vecz_k^2}{(d_k-\widehat\lambda_i)^2}
            \leq
            \sum_{k=2}^n \frac{4(n+1)^2}{\tau^6}
            \label{eq:fmm_approximate_inner_product_5b}
        \end{align}
        Combining \eqref{eq:fmm_approximate_inner_product_1}, \eqref{eq:fmm_approximate_inner_product_2}, \eqref{eq:fmm_approximate_inner_product_2b}, \eqref{eq:fmm_approximate_inner_product_3}, \eqref{eq:fmm_approximate_inner_product_4}, \eqref{eq:fmm_approximate_inner_product_5}, and
        \eqref{eq:fmm_approximate_inner_product_5b},
        we obtain
        \begin{align}
            \labs
            \vecuhat_i^\top\vecq
            -
            \vecuhat_i'^\top\vecq
            \rabs
            &\leq
            \frac{2}{
            \labs
                1+\Psi(\widehat\lambda_i)
            \rabs
            }
            \lbrac
                3\epsilon_{\vecz}\Psi(\widehat\lambda_i)
                +
                \sum_{k=2}^n
                \lpar
                    |\zeta_k\vecq_k|
                    \frac{
                        3\epsilon_{\vecz} + 6\epsilon_{\vecz}\Psi(\widehat\lambda_i)
                    }
                    {
                        \sqrt{1+\Psi(\widehat\lambda_i)}
                    }
                \rpar
            \rbrac
            \nonumber
            \\
            &\leq
                6\epsilon_{\vecz}\Psi(\widehat\lambda_i)
                +
                12\epsilon_{\vecz}
                \sum_{k=2}^n
                    |\zeta_k\vecq_k|
            \nonumber
            \\
            &\leq
                6\epsilon_{\vecz}\frac{4(n+1)^2}{\tau^6}
                +
                12\epsilon_{\vecz}
                (n-1)
                \frac{2(n+1)}{\tau^3}
                \nonumber
            \\
            &\leq
                48\epsilon_{\vecz}
                \frac{(n+1)^2}{\tau^6}.
            \label{eq:fmm_approximate_inner_product_main_1}
        \end{align}
        
        Next, we can choose some error parameter $\epsilon_{FMM}$ and include the errors introduced by using $(\epsilon_{FMM},\cfmm,n)$-\fmmalgo\   to approximate $\Phi'$ and $\Psi'$ (which are also listed in Equations \eqref{eq:fmm_4}, \eqref{eq:fmm_5}). The evaluation points satisfy the corresponding requirements.
        We shall denote by $\phi'_i$ the FMM error in $\Phi'(\widehat\lambda_i)$ and
        $\psi'_i$ the FMM error in $\Psi'(\widehat\lambda_i)$, such that $|\psi_i|,|\phi_i|\leq \epsilon_{FMM}$. In this case we will write
        \begin{align*}
            \vecuhat''^\top_i\vecq = 
            \frac{
                -\vecq_1+\Phi'(\widehat\lambda_i)+\phi_i
            }{
                \sqrt{1+\Psi'(\widehat\lambda_i)+\psi_i}
            }
            =
            \frac{
                -\vecq_1
                +
                \sum_{k=2}^n\frac{(1+\epsilon_k)\veczhat_k\vecq_k}{d_k-\widehat\lambda_i}
                +
                \phi_i
            }{
                \sqrt{1
                +
                \sum_{k=2}^n
                \frac{(1+\epsilon_k)^2\veczhat^2_k}{d_k-\widehat\lambda_i}
                +
                \psi_i}
            }.
        \end{align*}
        The error is bounded as follows.
        \begingroup
        \allowdisplaybreaks
        \begin{align}
            \labs \vecuhat_i''^\top \vecq - \vecuhat'^\top_i\vecq \rabs
            &=
            \labs
                \frac{
                -\vecq_1+\Phi'(\widehat\lambda_i)
                }{
                    \sqrt{1+\Psi'(\widehat\lambda_i)}
                }
                -
                \frac{
                -\vecq_1+\Phi'(\widehat\lambda_i)+\phi_i
                }{
                    \sqrt{1+\Psi'(\widehat\lambda_i)+\psi_i}
                }
            \rabs
            \nonumber
            \\
            &=
            \labs
                \frac{
                \lpar
                    -\vecq_1+\Phi'(\widehat\lambda_i)
                \rpar
                \sqrt{1+\Psi'(\widehat\lambda_i)+\psi_i}
                -
                \lpar
                    -\vecq_1+\Phi'(\widehat\lambda_i)+\phi_i
                \rpar
                \sqrt{1+\Psi'(\widehat\lambda_i)}
                }{
                    \sqrt{1+\Psi'(\widehat\lambda_i)}\sqrt{1+\Psi'(\widehat\lambda_i)+\psi_i}
                }
            \rabs
            \nonumber
            \\
            &=
            \labs
                \frac{
                    \lpar
                        -\vecq_1
                        +
                        \Phi'(\widehat\lambda_i)
                    \rpar
                    \lpar
                        \sqrt{1+\Psi'(\widehat\lambda_i)+\psi_i}
                        -
                        \sqrt{1+\Psi'(\widehat\lambda_i)}
                    \rpar
                    -
                    \phi_i
                    \sqrt{1+\Psi'(\widehat\lambda_i)}
                }{
                    \sqrt{1+\Psi'(\widehat\lambda_i)}\sqrt{1+\Psi'(\widehat\lambda_i)+\psi_i}
                }
            \rabs
            \nonumber
            \\
            &\leq
            \labs
                \frac{
                    \lpar
                        -\vecq_1
                        +
                        \Phi'(\widehat\lambda_i)
                    \rpar
                    \lpar
                        \sqrt{1+\Psi'(\widehat\lambda_i)+\psi_i}
                        -
                        \sqrt{1+\Psi'(\widehat\lambda_i)}
                    \rpar
                }{
                    \sqrt{1+\Psi'(\widehat\lambda_i)}\sqrt{1+\Psi'(\widehat\lambda_i)+\psi_i}
                }
            \rabs
            +
            \labs
                \frac{
                    \phi_i
                }{
                    \sqrt{1+\Psi'(\widehat\lambda_i)+\psi_i}
                }
            \rabs
            \nonumber
            \\
            &\leq
            \labs
                \frac{
                    \lpar
                        -\vecq_1
                        +
                        \Phi'(\widehat\lambda_i)
                    \rpar
                    \lpar
                        \sqrt{1+\Psi'(\widehat\lambda_i)+\psi_i}
                        -
                        \sqrt{1+\Psi'(\widehat\lambda_i)}
                    \rpar
                }{
                    \sqrt{1+\Psi'(\widehat\lambda_i)}\sqrt{1+\Psi'(\widehat\lambda_i)+\psi_i}
                }
            \rabs
            +
            \epsilon_{FMM}.
            \label{eq:fmm_approximate_inner_product_6}
        \end{align}
        \endgroup
        For the demoninator of the leftmost term we have
        \begin{align}
            \labs
                \sqrt{1+\Psi'(\widehat\lambda_i)}\sqrt{1+\Psi'(\widehat\lambda_i)+\psi_i}
            \rabs
            \geq
            \frac{1}{\sqrt{2}}
            \labs
                1+\Psi'(\widehat\lambda_i)
            \rabs.
            \label{eq:fmm_approximate_inner_product_7}
        \end{align}
        For the numerator we have
        \begin{align}
            &\labs
                    \lpar
                        -\vecq_1
                        +
                        \Phi'(\widehat\lambda_i)
                    \rpar
                    \lpar
                        \sqrt{1+\Psi'(\widehat\lambda_i)+\psi_i}
                        -
                        \sqrt{1+\Psi'(\widehat\lambda_i)}
                    \rpar
            \rabs
            \nonumber
            \\
            &\qquad=
            \labs
                \lpar
                    -\vecq_1
                    +
                    \Phi'(\widehat\lambda_i)
                \rpar
                \frac{
                    \lpar
                        1+\Psi'(\widehat\lambda_i)+\psi_i
                        -
                        1-\Psi'(\widehat\lambda_i)
                    \rpar
                }
                {
                    \lpar
                        \sqrt{1+\Psi'(\widehat\lambda_i)+\psi_i}
                        +
                        \sqrt{1+\Psi'(\widehat\lambda_i)}
                    \rpar
                }
            \rabs
            \nonumber
            \\
            &\qquad=
            \labs
                \lpar
                    -\vecq_1
                    +
                    \Phi'(\widehat\lambda_i)
                \rpar
                \frac{
                    \psi_i
                }
                {
                    \lpar
                        \sqrt{1+\Psi'(\widehat\lambda_i)+\psi_i}
                        +
                        \sqrt{1+\Psi'(\widehat\lambda_i)}
                    \rpar
                }
            \rabs
            \nonumber
            \\
            &\qquad\leq
            \labs
                \lpar
                    -\vecq_1
                    +
                    \Phi'(\widehat\lambda_i)
                \rpar
                \frac{
                    \psi_i
                }
                {
                        (\frac{1}{\sqrt{2}}+1)
                        \sqrt{1+\Psi'(\widehat\lambda_i)}
                }
            \rabs
            \nonumber
            \\
            &\qquad\leq
            \epsilon_{FMM}
            \labs
                \frac{
                    -\vecq_1
                    +
                    \Phi'(\widehat\lambda_i)
                }
                {
                \sqrt{1+\Psi'(\widehat\lambda_i)}
                }
            \rabs
            \nonumber
            \\
            &\qquad=
            \epsilon_{FMM}
            \labs
                \vecuhat'^\top_i\vecq
            \rabs.
            \label{eq:fmm_approximate_inner_product_8}
        \end{align}
        Combining the bounds of \eqref{eq:fmm_approximate_inner_product_6}, \eqref{eq:fmm_approximate_inner_product_7}, and \eqref{eq:fmm_approximate_inner_product_8}, we obtain
        \begin{align}
            \labs \vecuhat_i''^\top \vecq - \vecuhat'^\top_i\vecq \rabs
            \leq
            \frac{
                \epsilon_{FMM}
                \labs
                    \vecuhat'^\top_i\vecq
                \rabs
            }{
                \frac{1}{\sqrt{2}}
                \labs
                1+\Psi'(\widehat\lambda_i)
            \rabs}
            +\epsilon_{FMM}
            \leq
            \epsilon_{FMM}
            \lpar
                1
                +
                \sqrt{2}
                \labs
                    \vecuhat'^\top_i\vecq
                \rabs
            \rpar.
            \label{eq:fmm_approximate_inner_product_main_2}
        \end{align}
        Finally, the bounds of \eqref{eq:fmm_approximate_inner_product_main_1} and \eqref{eq:fmm_approximate_inner_product_main_2} are combined to provide
        \begin{align*}
            \labs \vecuhat_i^\top \vecq - \vecuhat''^\top_i\vecq\rabs
            &\leq
            \labs
                \vecuhat_i^\top \vecq - \vecuhat'^\top_i\vecq 
            \rabs
            +
            \labs
                \vecuhat_i'^\top \vecq - \vecuhat''^\top_i\vecq 
            \rabs
            \\
            &\leq
            48\epsilon_{\vecz}
                \tfrac{(n+1)^2}{\tau^6}
            +
            \epsilon_{FMM}
            \lpar
                1
                +
                \sqrt{2}
                \labs
                    \vecuhat'^\top_i\vecq
                \rabs
            \rpar
            \\
            &\leq
            48\epsilon_{\vecz}
                \tfrac{(n+1)^2}{\tau^6}
            +
            \epsilon_{FMM}
            \lpar
                1
                +
                \sqrt{2}
                (1+48\epsilon_{\vecz}
                \tfrac{(n+1)^2}{\tau^6})
            \rpar
            \\
            &\leq
            48\epsilon_{\vecz}
                \tfrac{(n+1)^2}{\tau^6}
            +
            3\epsilon_{\vecz}
            +
            96\epsilon_{\vecz}^2
                \tfrac{(n+1)^2}{\tau^6}
            \\
            &=
            147\epsilon_{\vecz}
                \tfrac{(n+1)^2}{\tau^6},
        \end{align*}
        where we set $\epsilon_{FMM}=\epsilon_{\vecz}$. The total cost is
        \begin{align*}
            O\lpar
                n\log(\tfrac{1}{\epsilon_{\epsilon_{FMM}}})
                \log^{\cfmm}(\tfrac{n}{\tau\epsilon_{FMM}})
            \rpar
            =
            O\lpar
                n\log(\tfrac{1}{\epsilon_{\vecz}})
                \log^{\cfmm}(\tfrac{n}{\tau\epsilon_{\vecz}})
            \rpar.
        \end{align*}
    \end{proof}
\end{lemma}

\subsection{Proof of Theorem \ref{theorem:arrowhead_diagonalization}}
We can finally combine all the results to prove Theorem \ref{theorem:arrowhead_diagonalization}, which we restate below for readability.
\begin{theorem}
    \label{theorem:arrowhead_diagonalization_appendix}
    Given a symmetric arrowhead matrix $\matH\in\mathbb{R}^{n\times n}$ as in Eq. \eqref{eq:arrowhead}, with $\|\matH\|\leq 1$, an accuracy parameter $\epsilon\in(0,1)$, a matrix $\matB$ with $r$ columns $\matB_i,i\in[r]$, where $\|\matB_i\|\leq 1$, and an $(\epsilon,n)$-\fmmalgo\   implementation (see Prop. \ref{proposition:fmm}), we can compute a diagonal matrix $\matLambdatilde$, and a matrix
    $\matQtilde_{\matB}$, such that 
    \begin{align*}
        \lnorm \matH-\matQ\matLambdatilde\matQ^\top \rnorm &\leq \epsilon,
        \quad
        \labs \lpar \matQ^\top\matB - \matQtilde_{\matB} \rpar_{i,j}\rabs \leq  \epsilon/n^2,
    \end{align*}
    where 
    $
        \matQ\in\mathbb{R}^{n\times n}
    $ is (exactly) orthogonal, in 
    $
        O\lpar nr\log^{\cfmm+1}(\tfrac{n}{\epsilon})\rpar
    $
    arithmetic operations and comparisons.

    Alternatively, if only want to compute a set of approximate values $\widetilde\lambda_1,\ldots,\widetilde\lambda_n$, such that $|\lambda_i(\matH)-\widetilde\lambda_i|\leq \epsilon$, the complexity reduces to $O\lpar n\log(\frac{1}{\epsilon})\log^{\cfmm}(\frac{n}{\epsilon})\rpar$ arithmetic operations.
    \begin{proof}
        \textbf{Step 1: Deflation.} The first step is to use deflation in order to ensure that the requirements of Lemma \ref{lemma:arrowhead_preliminaries} are satisfied. To that end we apply Proposition \ref{proposition:arrowhead_deflation} from the Appendix with parameter $\tau$. This way we obtain a symmetric  matrix $\matHtilde=\begin{pmatrix}
            \matHtilde' & 0\\
            0 & \matDtilde
        \end{pmatrix}$, such that $\matDtilde$ is diagonal, and the upper-left block $\matHtilde'=\begin{pmatrix}
            \widetilde\alpha' & \vecztilde'^\top \\
            \vecztilde' & \matDtilde'
        \end{pmatrix}$ is a symmetric arrowhead matrix that satisfies the requirements of Lemma \ref{lemma:arrowhead_preliminaries}, namely,
        \begin{align}
            \matDtilde'_{j+1,j+1}-\matDtilde'_{j,j} \geq \tau, \quad \text{and} \quad |\vecztilde'_i|\geq \tau.
        \end{align}
        The matrix $\matHtilde$ satisfies $\|\matH-\matG\matHtilde\matG^\top\|\leq n\tau$, where $\matG$ is orthogonal. 
        Since $\matG$ is orthogonal, $\matHtilde$ is similar to $\matG\matHtilde\matG^\top$, i.e., they have the same eigenvalues. Then Weyl's inequality gives 
        \begin{align*}
            |\lambda_i(\matH)-\lambda_i(\matHtilde)|
            =
            |\lambda_i(\matH)-\lambda_i(\matG\matHtilde\matG^\top)|
            \leq 
            \|\matH-\matG\matHtilde\matG^\top\|
            \leq
            n\tau.
        \end{align*}
        As described in Proposition \ref{proposition:arrowhead_deflation}, $\matHtilde'$ is a principal submatrix of $\matH$, and therefore due to the eigenvalue interlacing property it holds that $\|\matHtilde'\|\leq \|\matH\|\leq 1$.
        
        \textbf{Step 2: Diagonalization of $\matHtilde$.} 
        In the second step we focus on $\matHtilde$. There are two different cases, depending on whether we are interested only on the eigenvalues, or if we also want to multiply the eigenvector matrix with another matrix.
        
        \textbf{Case 1 - Eigenvalues only.} If we only care about the eigenvalues, we can ignore the computation of $\matG$, and therefore the deflation costs $O(n\log(n))$ operations. 
        It remains to approximate the eigenvalues of $\matHtilde'$. 
        For this we  use Lemma \ref{lemma:fmm_approximate_eigenvalues} with error parameter $\epsilon_{\lambda}$, which returns some values $\widetilde\lambda_1,\ldots,\widetilde\lambda_n$ in $O\lpar
            n\log(\tfrac{1}{\epsilon_{\lambda}})\log^{\cfmm}(\frac{n}{\tau\epsilon_{\lambda}})
        \rpar$ arithmetic operations such that $|\widetilde\lambda_i-\lambda_i(\matHtilde')|\leq \epsilon_{\lambda}$. 
        If we set:
        \begin{align*}
            \tau=\frac{\epsilon}{2n},
            \qquad
            \text{and}
            \qquad
            \epsilon_{\lambda}=\frac{\epsilon}{2},
        \end{align*}
        a triangle inequality gives that $|\widetilde\lambda_i-\lambda_i(\matH)|\leq n\tau+\epsilon_{\lambda}\leq \epsilon$.
        The complexity of Lemma \ref{lemma:fmm_approximate_eigenvalues} becomes $O\lpar
            n\log(\frac{1}{\epsilon})\log^{\cfmm}(\frac{n}{\epsilon})
        \rpar$, which gives a matching total number of arithmetic operations and comparisons.

        \textbf{Case 2 - Eigenvalues and eigenvectors.}
        In this case we do not only need the eigenvalues, but we are also given a matrix $\matB$ that needs to be multiplied from the left with the eigenvector matrix. We assume that the matrix product $\matG_{\matB}=\matG^\top\matB$ is already formed on-the-fly in a total of $O(nr)$ operations from Proposition \ref{proposition:arrowhead_deflation} in Step 1.
        
        We then proceed with the diagonalization of $\matHtilde'$. 
        The following steps are used to diagonalize $\matHtilde'$:
        \begin{enumerate}
            \item 
            We first set 
            $\epsilon_{\lambda}=\epsilon'\tfrac{\tau^3}{2(n+1)}$, for some $\epsilon'\in(0,1/n)$ that is specified later,
            and use Lemma \ref{lemma:fmm_approximate_eigenvalues}, to approximate all the eigenvalues of $\matHtilde'$  in 
            $O\lpar
                n\log(\tfrac{1}{\epsilon_{\lambda}})\log^{\cfmm}(\frac{n}{\tau\epsilon_{\lambda}})
            \rpar
            =
            O\lpar
                n\log^{\cfmm+1}(\frac{n}{\tau\epsilon'})
            \rpar$
            arithmetic operations. 
            The returned eigenvalues $\widehat\lambda_i$ satisfy $|\widehat\lambda_i-\lambda_i(\matHtilde')|\leq \epsilon_{\lambda}$.
            Note that this value of $\epsilon_{\lambda}$ satisfies the requirement of Lemmas \ref{lemma:h_backward_approximation} and \ref{lemma:fmm_approximate_shaft}, and therefore we can use them.
            
            From Lemma \ref{lemma:h_backward_approximation}, we know that $\widehat\lambda_i$ are the exact eigenvalues of the matrix $
            \matHhat = 
            \begin{pmatrix}
                \widehat\alpha & \veczhat^\top \\
                \veczhat & \matDtilde'
            \end{pmatrix}$, where:
            \begin{align*}
                |\widetilde\alpha'-\widehat\alpha| &\leq \frac{\epsilon'\tau^3}{2},
                \quad
                \|\widetilde\vecz'-\widehat\vecz\| \leq \frac{n\epsilon'}{1-n\epsilon'}, 
                \quad
                \|\matHtilde'-\matHhat\| \leq \frac{\epsilon'\tau^3}{2} + \frac{n\epsilon'}{1-n\epsilon'}.
            \end{align*}
            \item We now compute he elements of $\matHhat$. The value of $\widehat\alpha$ can be computed exactly in $O(n)$ operations from Lemma \ref{lemma:arrowhead_reconstruction_from_shaft_and_eigenvalues}. The vector $\veczhat$ can be approximated using Lemma \ref{lemma:fmm_approximate_shaft} in $O\lpar
                n\log(\tfrac{1}{\epsilon_{\vecz}})
                \log^{\cfmm}(\tfrac{n}{\tau\epsilon_{\vecz}})
            \rpar$ arithmetic operations, where $\epsilon_{\vecz}\in(0,1/2)$ is some chosen input parameter (specified below). The returned vector $\veczhat'$ satisfies 
            \begin{align*}
                \|\veczhat'-\veczhat\|\leq \epsilon_{\vecz}\|\veczhat\|.
            \end{align*}  
            \item Next, we partition the matrix $\matG_{\matB}=\matG^\top\matB$, which was computed in the previous steps, in the form $\matG^\top\matB=\begin{pmatrix}
                \matG_1^\top\matB \\
                \matG_2^\top\matB
            \end{pmatrix}$. 
            We then apply Lemma \ref{lemma:fmm_approximate_inner_products} on $(\matG_1^\top\matB).$ 
            Specifically, given the approximate eigenvalues $\widehat\lambda_i$ and the approximate vector $\veczhat'$, we use Lemma \ref{lemma:fmm_approximate_inner_products} to approximate the matrix product $\matUhat^\top(\matG_1^\top\matB)$ in  $O\lpar
                nr\log(\tfrac{1}{\epsilon_{\vecz}})
                \log^{\cfmm}(\tfrac{n}{\tau\epsilon_{\vecz}})
            \rpar$ operations, where $\matUhat$ denotes the orthogonal matrix whose columns are the eigenvectors of $\matHhat$ ($\matUhat$ is not computed explicitly). 
            Let $\matX_{\matB}=\matUhat^\top\matG_1^\top\matB$ be the true matrix product, and $\matX'_{\matB}$  be the matrix returned by Lemma \ref{lemma:fmm_approximate_inner_products}, respectively. Then for all $i,j$, $|(\matX_{\matB})_{i,j}-(\matX'_{\matB})_{i,j}|\leq 147\epsilon_{\vecz}
                \tfrac{(n+1)^2}{\tau^6}$. 
            \item We finally set $\matQtilde_{\matB}=\begin{pmatrix}
                \matX'_{\matB}\\
                \matG_2^\top\matB
            \end{pmatrix}$, and 
            $\matLambdatilde = \begin{pmatrix}
                    \matLambdahat & \\
                     & \matDtilde
                \end{pmatrix}.
            $
        \end{enumerate}
        Denote $\matQ:=
            \matG\begin{pmatrix}
                \matUhat & \\
                 & \matI
            \end{pmatrix}$.
        For the error $\lnorm\matQ^\top \matB - \matQtilde_{\matB}\rnorm$ we work as follows. Note that
        $
            \matQ^\top\matB = 
            \begin{pmatrix}
                \matUhat^\top & \\
                 & \matI
            \end{pmatrix}
            \begin{pmatrix}
                \matG_1^\top\matB \\
                \matG_2^\top\matB
            \end{pmatrix}
            = 
            \begin{pmatrix}
                \matUhat^\top\matG_1^\top\matB \\
                \matG_2^\top\matB
            \end{pmatrix}.
        $
        Then for all $i,j$:
        \begin{align*}
            \labs
            \lpar
            \matQ^\top\matB -\matQtilde_{\matB}
            \rpar_{i,j}
            \rabs
            =
            \labs
            \begin{pmatrix}
                \matUhat\matG_1^\top\matB \\
                \matG_2^\top\matB
            \end{pmatrix}_{i,j}
            -
            \begin{pmatrix}
                \matX'_{\matB} \\
                \matG_2^\top\matB
            \end{pmatrix}_{i,j}
            \rabs
            =
            \labs
                \begin{pmatrix}
                    \matUhat\matG_1^\top\matB- \matX'_{\matB} \\
                    0
                \end{pmatrix}_{i,j}
            \rabs
            =
            \labs
                \begin{pmatrix}
                    \matX_{\matB}- \matX'_{\matB} \\
                    0
                \end{pmatrix}_{i,j}
            \rabs.
        \end{align*}
        The bottom part is zero. The top part was already bounded above as
        $|(\matX_{\matB})_{i,j}-(\matX'_{\matB})_{i,j}|\leq 147\epsilon_{\vecz}
                \tfrac{(n+1)^2}{\tau^6}$. 
        
        We can now analyze the backward-error $\lnorm \matH-\matQ\matLambdatilde\matQ^\top\rnorm$. We have
        \begingroup
        \allowdisplaybreaks
        \begin{align*}
            \lnorm
                \matH-\matQ\matLambdatilde\matQ^\top
            \rnorm
            &=
            \lnorm
                \matH-\matG\begin{pmatrix}
                \matUhat & \\
                 & \matI
            \end{pmatrix}\matLambdatilde
            \begin{pmatrix}
                \matUhat & \\
                 & \matI
            \end{pmatrix}^\top \matG^\top
            \rnorm
            =
            \lnorm
                \matH-\matG\begin{pmatrix}
                \matHhat & \\
                 & \matDtilde            
            \end{pmatrix} \matG^\top
            \rnorm
            \\
            &=
            \lnorm
                \matH-\matG\begin{pmatrix}
                \matHtilde'+\matHhat-\matHtilde' & \\
                 & \matDtilde            
            \end{pmatrix} \matG^\top
            \rnorm
            =
            \lnorm
                \matH-\matG\matHtilde\matG^\top
                +
                \matG_1
                (\matHhat-\matHtilde')\matG_1^\top
            \rnorm
            \\
            &=
            \lnorm
                \matH-\matG\matHtilde\matG^\top
            \rnorm
            +
            \lnorm
                \matHhat-\matHtilde'
            \rnorm
            \leq
            n\tau 
            + 
            \frac{\epsilon'\tau^3}{2} + \frac{n\epsilon'}{1-n\epsilon'}.
        \end{align*}
        \endgroup

        The next task is to set the parameters $\tau,\epsilon_{\lambda}, \epsilon',$ and $\epsilon_{\vecz}$ and to upper bound the complexity. Given $\epsilon\in(0,1/2)$ the desired error parameter, the rest are set as follows:
        \begin{align*}
            \quad
            \tau=\tfrac{\epsilon}{2n},
            \quad
            \epsilon'=\tfrac{\epsilon}{4n},
            \quad
            \epsilon_{\lambda}
            =
            \epsilon'\tfrac{\tau^3}{2(n+1)}
            =
            \tfrac{\epsilon^4}{64n^4(n+1)}
            ,
            \quad
            \epsilon_{\vecz}=
            \epsilon\tfrac{\tau^6}{147n^2(n+1)^2}
            =
            \tfrac{\epsilon^7}{147\cdot 64\cdot n^8(n+1)^2}.
        \end{align*}
        Then the errors become:
        \begin{align*}
            \lnorm
                \matH-\matQ\matLambdatilde\matQ^\top
            \rnorm
            &\leq
            n\tau 
            + 
            \frac{\epsilon'\tau^3}{2} + \frac{n\epsilon'}{1-n\epsilon'}
            \leq
            \frac{\epsilon}{2} 
            + 
            \frac{\epsilon^4}{64n^4}
            + 
            \frac{\epsilon/4}{1-\epsilon/4}
            \leq
            \epsilon
            ,\\
            \labs
            \lpar
            \matQ^\top\matB -\matQtilde_{\matB}
            \rpar_{i,j}
            \rabs
            &=
            \labs
                \begin{pmatrix}
                    \matX_{\matB}- \matX'_{\matB} \\
                    0
                \end{pmatrix}_{i,j}
            \rabs
            \leq 
            147\epsilon_{\vecz}
                \frac{(n+1)^2}{\tau^6}
            = 
            \epsilon/n^2,
        \end{align*} 
        which also implies
        \begin{align}
            \label{eq:theorem_arrowhead_diagonalization_normwise_eigenvector_bound}
            \lnorm \matQ^\top\matB-\matQtilde^\top \rnorm \leq \epsilon/n.
        \end{align}
        The complexity is as follows:\\
        \begin{tabular}{r l}
            Deflation:
            &
            $O\lpar n\log(n) + nr\rpar$,
            \\
            Eigenvalues:
            &
            $O\lpar
                n\log(\tfrac{1}{\epsilon_{\lambda}})\log^{\cfmm}(\frac{n}{\tau\epsilon_{\lambda}})
            \rpar
            =
            O\lpar
                n\log^{\cfmm+1}(\frac{n}{\epsilon})
            \rpar$,
            \\
            $\veczhat'$:
            &
            $O\lpar
                n\log(\tfrac{1}{\epsilon_{\vecz}})\log^{\cfmm}(\frac{n}{\tau\epsilon_{\vecz}})
            \rpar
            =
            O\lpar
                n\log^{\cfmm+1}(\frac{n}{\epsilon})
            \rpar$,
            \\
            $\matQtilde_{\matB}$:
            &
            $O\lpar 
                nr\log(\tfrac{1}{\epsilon_{\vecz}})\log^{\cfmm}(\frac{n}{\tau\epsilon_{\vecz}})
            \rpar
            =
            O\lpar nr\log^{\cfmm+1}(\tfrac{n}{\epsilon})\rpar
            $,
            \\
            Total: 
            &
            $O\lpar nr\log^{\cfmm+1}(\tfrac{n}{\epsilon})\rpar$.
        \end{tabular}

    \end{proof}
\end{theorem}

\section{Tridiagonal diagonalization}
\label{appendix:tridiagonal_diagonalization}
\subsection{Omitted proofs}
The next lemma bounds the error of the reduction to arrowhead form when the spectral factorizations of the matrices $\matT_1$ and $\matT_2$ in Equation \eqref{eq:tridiagonal_to_arrowhead} are approximate rather than exact.
\begin{lemma}[Restatement of Lemma \ref{lemma:tridiagonal_assembly}]
\label{lemma:tridiagonal_assembly_appendix}
Let $\epsilon\in(0,1/2)$ be a given accuracy parameter and $\matT = \begin{pmatrix}
    \matT_1 & \beta_{k+1}\vece_k & \\
    \beta_{k+1}\vece_k^\top & \alpha_{k+1} & \beta_{k+2}\vece_1^\top \\
     & \beta_{k+2}\vece_1 & \matT_2
\end{pmatrix}$ be a tridiagonal matrix with size  $n\geq 3$ and $\|\matT\|\leq 1$, where $\matT_1=\matU_1\matD_1\matU_1^\top$ and $\matT_2=\matU_2\matD_2\matU_2^\top$ be the exact spectral factorizations of $\matT_1$ and $\matT_2$. Let $\matUtilde_1,\matDtilde_1,\matUtilde_2,\matDtilde_2$ be approximate spectral factorizations of $\matT_1,\matT_2$. If these factors satisfy
    \begin{align*}
        \lnorm
            \matT_{\{1,2\}} - \matUtilde_{\{1,2\}}\matDtilde_{\{1,2\}}\matUtilde_{\{1,2\}}^\top
        \rnorm 
        \leq \epsilon_1,
        \quad
        \lnorm \matUtilde_{\{1,2\}}\matUtilde_{\{1,2\}}^\top -\matI \rnorm &\leq \epsilon_1/n,
    \end{align*}
    for some $\epsilon_1\in(0,1/2)$, where $\matDtilde_{\{1,2\}}$ are both diagonal, then, assuming an $(\epsilon,n)$-\fmmalgo\   implementation as in Prop. \ref{proposition:fmm}, we can compute a diagonal matrix $\matDtilde$ and an approximately orthogonal matrix $\matUtilde$ such that
    \begin{align*}
        \lnorm\matUtilde^\top\matUtilde-\matI\rnorm\leq 3(\epsilon_1+\epsilon)/n,
        \quad \text{and} \quad
        \lnorm 
            \matT-\matUtilde\matDtilde\matUtilde^\top
        \rnorm \leq 2\epsilon_1+7\epsilon,
    \end{align*}
    in a total of $O\lpar n^2\log^{\cfmm+1}(\tfrac{n}{\epsilon})\rpar$ arithmetic operations and comparisons.
    \begin{proof}
        We first consider the matrix
        \begin{align*}
            \matTtilde
            =
            \begin{pmatrix}
                & \matUtilde_1 & \\
                1 & & \\
                & & \matUtilde_2
            \end{pmatrix}    
            \begin{pmatrix}
                \alpha_{k+1}& \beta_{k+1}\vecltilde_1^\top   & \beta_{k+2}\vecftilde_2^\top\\
                \beta_{k+1}\vecltilde_1 & \matDtilde_1 & \\
                 \beta_{k+2}\vecftilde_2& & \matDtilde_2
            \end{pmatrix}
            \begin{pmatrix}
                & 1 & \\
                \matUtilde_1^\top & & \\
                & & \matUtilde_2^\top
            \end{pmatrix}
            =
            \matW\matHtilde\matW^\top,
        \end{align*}
        where $\vecltilde_1^\top$ is the last row of $\matUtilde_1$ and $\vecftilde_2^\top$ is the first row of $\matUtilde_2$. $\matHtilde$ can be assembled in $O(n)$ operations, and $\matW$ in $O(n^2)$ operations. 
        Note that $\matW\matHtilde\matW^\top$ is not a similarity transformation, since $\matW$ is not orthogonal, but it can be seen as an approximate similarity transformation, which is sufficient for our analysis. 
        
        In the next step, we need to use Theorem \ref{theorem:arrowhead_diagonalization} to diagonalize $\matHtilde,$ and multiply its eigenvector matrix with $\matW$. 
        Since neither $\matHtilde$ nor $\matW$ are guaranteed to have $\|\matHtilde\|,\|\matW\|\leq 1$, we need to scale the appropriately so that we can apply Theorem \ref{theorem:arrowhead_diagonalization}.
        First note that the set of singular values of $\matW$, $\Sigma(\matW)$, is equal to
        \begin{align*}
            \Sigma(\matW) = \Sigma(\matUtilde_1) \cup \Sigma(\matUtilde_2) \cup \{1\}.
        \end{align*}
        Since the singular values of $\matUtilde_1$ and $\matUtilde_2$ lie inside $[\sqrt{1-\epsilon_1/n},\sqrt{1+\epsilon_1/n}]$ by assumption, then we have that $\sigma_i(\matW)\in[\sqrt{1-\epsilon_1/n},\sqrt{1+\epsilon_1/n}]$. This also implies that $\matW$ is invertible and that $\|\matW\|\leq \sqrt{1+\epsilon_1/n} <2 $. Next, consider the bound
        \begin{align*}
        \lnorm\matT-\matTtilde\rnorm
        &=
            \lnorm
            \begin{pmatrix}
                \matT_1 & \beta_{k+1}\vece_k & \\
                \beta_{k+1}\vece_k^\top & \alpha_{k+1} & \beta_{k+2}\vece_1^\top \\
                 & \beta_{k+2}\vece_1 & \matT_2
            \end{pmatrix}
            -
            \begin{pmatrix}
                \matUtilde_1\matDtilde_1\matUtilde_1^\top & \beta_{k+1}\matUtilde_1\vecltilde_1 & \\
                \beta_{k+1}\vecltilde_1^\top\matUtilde_1^\top & \alpha_{k+1} & \beta_{k+2}\vecftilde_2^\top\matUtilde_2^\top \\
                 & \beta_{k+2}\matUtilde_2\vecftilde_2 & \matUtilde_2\matDtilde_2\matUtilde_2^\top
            \end{pmatrix}
            \rnorm
            \\
            &\leq
            \lnorm
            \begin{pmatrix}
                \matT_1-\matUtilde_1\matDtilde_1\matUtilde_1^\top &  \\\
                 &  \matT_2 - \matUtilde_2\matDtilde_2\matUtilde_2^\top
            \end{pmatrix}
            \rnorm
            +
            \|\matT\|
            \lnorm
            \begin{pmatrix}
                 & \matUtilde_1\vecltilde_1-\vece_k & \\
                \vecltilde_1^\top\matUtilde_1^\top-\vece_k^\top &  & \vecftilde_2^\top\matUtilde_2^\top - \vece_1^\top \\
                 & \matUtilde_2\vecftilde_2 - \vece_1 & 
            \end{pmatrix}
            \rnorm
            \\
            &\leq
            \epsilon_1 + 2\epsilon_1/n.
        \end{align*}
        Then 
        \begin{align*}
            \lnorm \matHtilde \rnorm 
            &= 
            \lnorm \matW^{-1}\matW\matHtilde\matW^\top\matW^{-\top}\rnorm
            \\
            &\leq
            \lnorm \matW^{-1}\rnorm^2
            \lnorm \matW\matHtilde\matW^\top \rnorm
            \\
            &= 
            \lnorm \matW^{-1}\rnorm^2
            \lnorm \matTtilde \rnorm
            \\
            &\leq 
                \frac{1}{1-\epsilon_1/n}
                \lpar1+\epsilon_1+2\epsilon_1/n\rpar
                \\
            &<
            4.
        \end{align*}
        
        We can now apply Theorem \ref{theorem:arrowhead_diagonalization} for the matrices $\matHtilde/4$ and $\matB=\matW/4$, which satisfy the norm requirements, and for error  $\epsilon/4$. This way, in $O\lpar n^2\log^{\cfmm+1}(\tfrac{n}{\epsilon})\rpar$ operations, we can compute $\matUtilde,\matLambdatilde$, such that
        \begin{align*}
            \lnorm \matHtilde/4-\matQ\matLambdatilde\matQ^\top \rnorm &\leq \epsilon/4,
            \quad
            \lnorm \matQ^\top\matW^\top/4 - \matUtilde^\top\rnorm \leq \epsilon/4n,
        \end{align*}
        where $\matQ$ is an (unknown) orthogonal matrix, and the rightmost bound comes from Eq. \eqref{eq:theorem_arrowhead_diagonalization_normwise_eigenvector_bound}. If we rescale
        \begin{align*}
            \matLambdatilde \leftarrow 4\matLambdatilde, \quad \matUtilde\leftarrow 4 \matUtilde,
        \end{align*}
        this gives
        \begin{align*}
            \lnorm \matHtilde-\matQ\matLambdatilde\matQ^\top \rnorm &\leq \epsilon,
            \quad
            \lnorm \matQ^\top\matW^\top - \matUtilde^\top\rnorm \leq \epsilon/n.
        \end{align*}
        
        Then we consider the error
        \begin{align*}
            \lnorm\matT-\matUtilde\matLambdatilde\matUtilde^\top\rnorm
            \leq
            \lnorm\matT-\matTtilde\rnorm + \lnorm\matTtilde-\matUtilde\matLambdatilde\matUtilde^\top\rnorm.
        \end{align*}

        We can then bound the norm of $\matLambdatilde$ as follows:
        \begingroup
        \allowdisplaybreaks
        \begin{align*}
            \lnorm\matLambdatilde\rnorm
            &\leq 
            \lnorm\matQ\matLambdatilde\matQ^\top  - \matHtilde \rnorm
            + 
            \lnorm\matHtilde\rnorm
            \\
            &\leq 
            \epsilon + \lnorm\matW^{-1}\rnorm^2 \lnorm\matTtilde\rnorm
            \\
            &\leq 
            \epsilon + (\tfrac{1}{1-\epsilon_1/n})
            \lpar \lnorm\matT\rnorm + \lnorm\matTtilde-\matT\rnorm\rpar
            \\
            &\leq 
            \epsilon + (\tfrac{1}{1-\epsilon_1/n})\lpar1+\epsilon_1+2\epsilon_1/n\rpar
            \\
            &< 4.
        \end{align*}
        \endgroup
        Denoting $\matE_{\matU}:=\matUtilde-\matW\matQ$, we have that
        \begin{align*}
            \lnorm 
                \matTtilde-\matUtilde\matLambdatilde\matUtilde^\top
            \rnorm
            &=
            \lnorm
                \matW\matHtilde\matW^\top
                -
                (\matW\matQ+\matE_{\matU})\matLambdatilde(\matW\matQ+\matE_{\matU})^\top
            \rnorm
            \\
            &\leq
            \lnorm
                \matW\matHtilde\matW^\top
                -
                \matW\matQ\matLambdatilde\matQ^\top\matW^\top
            \rnorm
                +
            2
            \lnorm
                \matE_{\matU}\matLambdatilde\matQ^\top \matW^\top
            \rnorm
            +
            \lnorm
                \matE_{\matU}\matLambdatilde\matE_{\matU}^\top
            \rnorm
            \\
            &\leq
            \lnorm
                \matW
            \rnorm^2
            \lnorm 
            \matHtilde
                -
                \matQ\matLambdatilde\matQ^\top
            \rnorm
            +
            2
            \lnorm
                \matE_{\matU}
            \rnorm
            \lnorm
            \matW
            \rnorm
            \lnorm \matLambdatilde \rnorm
            +
            \lnorm
                \matE_{\matU}
            \rnorm^2
            \lnorm \matLambdatilde \rnorm
            \\
            &\leq
            (1+\epsilon_1/n)
            \epsilon
            +
            2\cdot 4
            (\epsilon/n)
            \sqrt{1+\epsilon_1/n}
            +
            4
            (\epsilon/n)^2
            \\
            &\leq 7\epsilon.
        \end{align*}
        This finally gives
        \begin{align*}
            \lnorm\matT-\matUtilde\matLambdatilde\matUtilde^\top\rnorm
            \leq
            \epsilon_1 + 2\epsilon_1/n + 7\epsilon \leq 2\epsilon_1 + 7\epsilon.
        \end{align*}

        We also need to prove that $\matUtilde$ is approximately orthogonal. From the stability of singular values (which is a consequence of Weyl's inequality), we know that $|\sigma_i(\matUtilde)-\sigma_i(\matW\matQ)| \leq \lnorm \matUtilde-\matW\matQ\rnorm \leq \epsilon/n$. The singular values of $\matW\matQ$ are the same as the singular values of $\matW$ since $\matQ$ is orthogonal. We already argued  that $\sigma_i(\matW)\in\lbrac
            \sqrt{1-\epsilon_1/n},\sqrt{1+\epsilon_1/n}
        \rbrac$, and thus
        $\sigma_i(\matUtilde)\in\lbrac
            \sqrt{1-\epsilon_1/n}-\epsilon/n,\sqrt{1+\epsilon_1/n}+\epsilon/n
        \rbrac\in\lbrac
            1-(\epsilon_1+\epsilon)/n,1+(\epsilon_1+\epsilon)/n
        \rbrac$, which gives roughly $\lnorm \matUtilde\matUtilde^\top-\matI\rnorm
        \leq 
        3(\epsilon_1+\epsilon)/n$.
    \end{proof}
\end{lemma}

\begin{lemma}
    \label{lemma:tridiagonal_assembly_eigenvalues_only}
Let $\epsilon\in(0,1/2)$ be a given accuracy parameter and $\matT = \begin{pmatrix}
    \matT_1 & \beta_{k+1}\vece_k & \\
    \beta_{k+1}\vece_k^\top & \alpha_{k+1} & \beta_{k+2}\vece_1^\top \\
     & \beta_{k+2}\vece_1 & \matT_2
\end{pmatrix}$ be a tridiagonal matrix with size $n\geq 2$, where $\matT_1=\matU_1\matD_1\matU_1^\top$ and $\matT_2=\matU_2\matD_2\matU_2^\top$ be the exact spectral factorizations of $\matT_1$ and $\matT_2$. Let $\matUtilde_1,\matDtilde_1,\matUtilde_2,\matDtilde_2$ be approximate spectral factorizations of $\matT_1,\matT_2$. Assume that these factors satisfy
    \begin{align*}
        \lnorm
            \matT_{\{1,2\}} - \matUtilde_{\{1,2\}}\matDtilde_{\{1,2\}}\matUtilde_{\{1,2\}}^\top
        \rnorm 
        \leq \epsilon_1,
        \quad
        \lnorm \matUtilde_{\{1,2\}}\matUtilde_{\{1,2\}}^\top -\matI \rnorm &\leq \epsilon_1/n,
    \end{align*}
    for some $\epsilon_1\in(0,1/2)$, where $\matDtilde_{\{1,2\}}$ are both diagonal. Assume also that $\matDtilde_1,\matDtilde_2$, as well as the last row $\vecltilde_1^\top$ of $\matUtilde_1$, and the first row $\vecftilde_2^\top$ of $\matUtilde_2$, are explicitly available. 
    
    Then we can compute a diagonal matrix $\matDtilde$, as well as the first row $\vecltilde$ and/or the last row $\vecftilde$ of an approximately orthogonal matrix $\matUtilde$, which satisfy
    \begin{align*}
        \lnorm\matUtilde^\top\matUtilde-\matI\rnorm\leq 3(\epsilon_1+\epsilon)/n,
        \quad \text{and} \quad
        \lnorm 
            \matT-\matUtilde\matDtilde\matUtilde^\top
        \rnorm \leq 2\epsilon_1+7\epsilon,
    \end{align*}
    in a total of $O\lpar n\log^{\cfmm+1}(\tfrac{n}{\epsilon})\rpar$ arithmetic operations.
    \begin{proof}
        The main observation is that, if we only care about $\matDtilde$ and the first and/or last row of $\matUtilde$, then the proof of Lemma \ref{lemma:tridiagonal_assembly} can be simplified as follows. Since we have explicit access to $\vecltilde_1$ and $\vecftilde_2$, we can construct $\matHtilde$ in $O(n)$ arithmetic operations.

        The next step is to apply Theorem \ref{theorem:arrowhead_diagonalization} on $\matHtilde$. The theorem returns $\matLambdatilde$, which is a diagonal matrix with approximate eigenvalues, and $\matUtilde^\top,$ which approximates the matrix product $\matQ^\top\matW^\top$. However, we do not need to compute the matrix product $\matQ^\top\matW^\top$. Instead, we only need to compute two matrix-vector products, $\matQ^\top\vecw_1$ and $\matQ^\top\vecw_n$, where $\vecw_1,\vecw_n$ are the first and last columns of $\matW^\top$.
    \end{proof}
\end{lemma}

\subsection{Approximating only the eigenvalues of a tridiagonal matrix}
The following result gives the complexity of approximating only the eigenvalues of $\matT$, instead of a full diagonalization. 
The main observation is that Lemma \ref{lemma:tridiagonal_assembly} can be simplified if we only need the eigenvalues. This simplification is listed in \ref{lemma:tridiagonal_assembly_eigenvalues_only} in the Appendix.
Using this as an inductive step, we obtain the following Corollary.
\begin{corollary}
    \label{corollary:tridiagonal_eigenvalues}
    Let $\matT$ be a symmetric unreduced tridiagonal matrix with $\|\matT\|\leq 1$ and $\epsilon\in(0,1/2)$ be a given accuracy parameter. Assuming access to an $(\epsilon,n)$-\fmmalgo, for $\tau\in\Theta(\poly(\tfrac{\epsilon}{n}))$, we can compute approximate eigenvalues $\widetilde\lambda_1,\ldots,\widetilde\lambda_n$ such that
    $|\widetilde\lambda_i-\lambda_i(\matT)|\leq \epsilon$ in $O\lpar n\log^{\cfmm+2}(\tfrac{n}{\epsilon})\rpar$ arithmetic operations.
    \begin{proof}
        The proof is similar to the one of Theorem \ref{theorem:alg_recursive_diagonalization}. During the recursion, instead of Lemma \ref{lemma:tridiagonal_assembly}, we use Lemma \ref{lemma:tridiagonal_assembly_eigenvalues_only}. 
        This way we reduce the arithmetic complexity of each recursive call from $O\lpar n^2\log^{\cfmm+1}(\tfrac{n}{\epsilon})\rpar$
        arithmetic operations 
        to
        $O\lpar n\log^{\cfmm+1}(\tfrac{n}{\epsilon})\rpar$, which becomes
        $O\lpar n\log^{\cfmm+2}(\tfrac{n}{\epsilon})\rpar$ after solving the recursion. The approximation guarantees remain the same, but in this case we only compute the eigenvalues, without the eigenvectors.
    \end{proof}
\end{corollary}

\subsection{Application: Hermitian diagonalization}
\label{appendix:hermitian_diagonalization_analysis}
Theorem \ref{theorem:alg_recursive_diagonalization} has a direct application to diagonalize Hermitian matrices. The following result is immediate.
\begin{corollary}
    \label{corollary:hermitian_diagonalization_appendix}
    Let $\matA$ be a Hermitian matrix of size $n$ with $\|\matA\|\leq 1$. Given accuracy $\epsilon\in(0,1/2)$, and an $(\epsilon,n)$-\fmmalgo\   implementation of Prop. \ref{proposition:fmm}, we can compute a matrix $\matQtilde$ and a diagonal matrix $\matLambdatilde$ such that
    \begin{align*}
        \lnorm \matA - \matQtilde\matLambdatilde\matQtilde^* \rnorm \leq \epsilon,
        \quad
        \lnorm \matQtilde^* \matQtilde - \matI \rnorm \leq \epsilon/n^2,
    \end{align*}
    or, stated alternatively,
    \begin{align*}
        \matQtilde=\matQ+\matE_{\matQ},
        \quad
        \matQ^\top\matQ=\matI,
        \quad
        \lnorm \matE_{\matQ} \rnorm \leq \epsilon/n^2,
        \quad
        \lnorm \matA - \matQ\matLambdatilde\matQ^\top \rnorm \leq \epsilon.
    \end{align*}
    The algorithm requires a total of $O\lpar n^\omega\log(n) + n^2\log^{\cfmm+1}(\tfrac{n}{\epsilon})\rpar$ arithmetic operations and comparisons.
    \begin{proof}
        Using the tridiagonal reduction algorithm of Schönhage \cite{schonhage1972unitare}, we can compute a unitary matrix $\matZ$ and a tridiagonal matrix $\matT$ in $O(n^\omega\log(n))$ arithmetic operations such that $\matA=\matZ\matT\matZ^*$ (see Theorem \ref{theorem:stable_tridiagonal_reduction} for the complexity analysis). The matrix $\matT$ can always be chosen to have real entries. We then use Theorem \ref{theorem:alg_recursive_diagonalization} to compute $\matUtilde$ and $\matLambdatilde$ such that \begin{align*}
            \lnorm \matT - \matUtilde\matLambdatilde\matUtilde^\top \rnorm \leq \epsilon,
            \quad
            \lnorm \matUtilde^\top \matUtilde - \matI \rnorm \leq \epsilon/n^2,
        \end{align*}
        in $O\lpar n^2\log^{\cfmm+1}(\tfrac{n}{\epsilon})\rpar$ arithmetic operations.
        We then form $\matQtilde=\matZ\matUtilde$ in $O(n^\omega)$ operations. It holds that
        \begin{align*}
            \lnorm \matA-\matQtilde\matLambdatilde\matQtilde^* \rnorm=
            \lnorm \matZ\matT\matZ^*-\matZ\matUtilde\matLambdatilde\matUtilde^\top\matZ^* \rnorm=
            \lnorm \matT-\matUtilde\matLambdatilde\matUtilde^\top \rnorm
            \leq \epsilon,
        \end{align*}
        and $\|\matQtilde^*\matQtilde-\matI\|=\|\matUtilde^\top\matZ^*\matZ\matUtilde-\matI\|\leq \epsilon/n^2$.
        The alternative statement can be obtained in a similar way as in the last part of the proof of Theorem \ref{theorem:alg_recursive_diagonalization}.
    \end{proof}
\end{corollary}

\subsection{Singular Value Decomposition}
\label{appendix:svd_analysis}

In this section we provide the proof of Theorem \ref{theorem:svd}. 
\begin{proof}[Proof of Theorem \ref{theorem:svd}]
        To ensure that $\|\matA\|$  satisfies the norm requirements, we first scale $\matA$ by $\frac{1}{\|\matA\|_F}=\frac{1}{\sqrt{\tr(\matA^*\matA)}}$. The trace can be computed in $O(mn)=O(n^{k+1})$. Using standard trace and norm inequalities,  it holds that
        \begin{align*}
            \tr(\matA^*\matA) \leq n\|\matA^\top\matA\| = n \|\matA\|^2 \leq n \|\matA\|_F^2 = n\tr(\matA^*\matA).
        \end{align*}
        which gives the desired $\frac{1}{\sqrt{n}} \leq \|\tfrac{\matA}{\sqrt{\tr(\matA^*\matA)}}\| \leq 1$.
    
        The next step is to compute the Gramian $\matA^*\matA$ in $O(n^{\omega(1,k,1)})$ and then reduce it to tridiagonal form $\matT$ using the algorithm of \cite{schonhage1972unitare} (see Theorem \ref{theorem:stable_tridiagonal_reduction}) in $O(n^\omega)$. We can now use Corollary \ref{corollary:tridiagonal_eigenvalues} to compute the condition number of $\matT$. We start with $\epsilon=1/2$, and call Corollary \ref{corollary:tridiagonal_eigenvalues} iteratively to compute the eigenvalues of $\matT$ up to error $\epsilon$. We keep halving $\epsilon$ and calling in each iteration until $\widetilde\lambda_{\min}>0$ and $\epsilon\leq \frac{\widetilde\lambda_{\min}}{4}$. This ensures that all eigenvalues are approximated up to additive error $\tfrac{1}{4}\lambda_{\min}(\matT)=\tfrac{1}{4}\sigma_{\min}(\matT)$ (the equality holds because $\matT$ is PSD). Due to the assumption $\frac{1}{n^c}\leq \|\matA\|\leq 1$, it holds that 
        $\kappa(\matA)\leq \frac{1}{\sigma_{\min}(\matA)}\leq n^c\kappa(\matA)$, 
        in which case we can use $\widetilde\lambda_{\min}$ to approximate the condition number of $\matT$ up to a factor $n^c$. In particular, if $\widetilde\sigma_{\min}
        =\sqrt{\widetilde\lambda_{\min}}\in\Theta(\sigma_{\min}(\matA))$ is the approximated smallest singular value of $\matT$,  we can set $
        \widetilde\kappa=\frac{1}{\widetilde\sigma}\in \Theta(\frac{1}{\sigma_{\min}(\matA)}) \in \lbrac \Omega(\kappa(\matA)), O(n^c\kappa(\matA)) \rbrac
        $. 
        
        We call Corollary \ref{corollary:tridiagonal_eigenvalues} $O(\log(n\kappa(\matA))$ times, and each call costs $O\lpar n\log^{\cfmm+2}(\tfrac{n}{\epsilon})\rpar$ arithmetic operations.
        The total cost is therefore at most $O\lpar n\log^{\cfmm+3}(n\kappa(\matA))\rpar$.
        
        Next, we set $\epsilon'=\epsilon/(n^c\widetilde\kappa)^2$ and use Corollary \ref{corollary:hermitian_diagonalization} 
        compute a nearly unitary $\matVtilde$ and diagonal $\matLambdatilde$ such that
        \begin{align*}
            \lnorm \matA^*\matA - \matVtilde\matLambdatilde\matVtilde^* \rnorm \leq \epsilon',
            \quad
            \lnorm \matVtilde^* \matVtilde - \matI \rnorm \leq \epsilon'/n^2.
        \end{align*}
        This costs $O\lpar 
        n^\omega\log(n) + n^2\log^{\cfmm+1}(\tfrac{n}{\epsilon'})
        \rpar
        =
        O\lpar 
            n^\omega\log(n) + n^2\log^{\cfmm+1}(\tfrac{n\kappa(\matA)}{\epsilon})
        \rpar
        $.
        Due to the upper bound of $\epsilon'$, the diagonal elements of $\matLambdatilde$ are positive, and we set $\matSigmatilde=\matLambdatilde^{1/2}$.

        The left singular vector matrix is set to
        $\matUtilde=\matA\matVtilde^{-*}\matSigmatilde^{-1}$, which implies that $\matA=\matUtilde\matSigmatilde\matV^*$.
        It remains to show that $\matUtilde$ has approximately orthonormal columns. We have that 
        \begin{align*}
            \matUtilde^*\matUtilde = \matSigmatilde^{-1}\matVtilde^{-1}\matA^*\matA\matVtilde^{-*}\matSigmatilde^{-1}
            = \matSigmatilde^{-1}\matVtilde^{-1}
                \lpar 
                    \matVtilde\matLambdatilde\matVtilde^* + \matE
                \rpar
            \matVtilde^{-*}\matSigmatilde^{-1}
            =
            \matI
            +
            \matSigmatilde^{-1}\matVtilde^{-1}\matE \matVtilde^{-*}\matSigmatilde^{-1},
        \end{align*}
        where $\|\matE\|\leq \epsilon'$. Then
        \begin{align*}
            \lnorm
                \matI-\matUtilde^*\matUtilde
            \rnorm
            =
            \lnorm
                \matSigmatilde^{-1}\matVtilde^{-1}\matE \matVtilde^{-*}\matSigmatilde^{-1}
            \rnorm
            \leq
            \lnorm
                \matSigmatilde^{-1}
            \rnorm^2
            \lnorm
                \matVtilde^{-1}
            \rnorm^2
            \lnorm
                \matE
            \rnorm
            \leq
            C(n^c\kappa(\matA))^2
            \epsilon'
            \leq
            C\epsilon,
        \end{align*}
        for some small constant $C$, which arises from $\|\matVtilde^{-1}\|\approx \tfrac{1}{1+\epsilon/\poly(n)}$ and the unspecified constant factor of the approximate condition number $\widetilde\kappa$, which should be also around $2$ in the worst case. Rescaling $\epsilon'$ gives the final result.

        For the alternative statement, we write $\matUtilde=\matU+\matE_{\matU}$, where $\matU^*\matU=\matI$ and $\|\matE_{\matU}\|\leq \epsilon$, and $\matVtilde=\matV+\matE_{\matV}$ where $\matV^*\matV=\matI$ and $\|\matE_{\matV}\|\ll \epsilon/n^2$.
        It is easy to see that $\lnorm \matSigmatilde \rnorm \leq 1+\epsilon$, $\lnorm \matUtilde\rnorm \leq 1+\epsilon$, and $\lnorm \matVtilde \rnorm \ll 1+\epsilon/n^2$.
        Then
        \begin{align*}
            \lnorm
                \matA - \matU\matSigmatilde\matV^*
            \rnorm
            &\leq
            \lnorm
                \matA - \matUtilde\matSigmatilde\matVtilde^*
            \rnorm
            +
            \lnorm \matSigmatilde \rnorm
            \lpar
                \lnorm
                    \matE_{\matU}
                \rnorm
                \lnorm \matVtilde \rnorm
                +
                \lnorm
                    \matE_{\matV}
                \rnorm
                \lnorm \matUtilde \rnorm
                +
                \lnorm
                    \matE_{\matU}
               \rnorm
                \lnorm \matE_{\matV} \rnorm
            \rpar
            \\
            &\leq
            \epsilon
            +
            (1+\epsilon)
            \lpar
                \epsilon(1+\epsilon/n^2) 
                +
                \epsilon(1+\epsilon)/n^2
                +
                \epsilon^2/n^2
            \rpar
            \\
            &< 4\epsilon,
        \end{align*}
        and thus it suffices to rescale $\epsilon$ by $1/4$.
    \end{proof}

\section{Floating point arithmetic}
\label{appendix:floating_point_arithmetic}

The sign $s$ is $+$ if the corresponding bit is one, and $-$ if the bit is zero. The exponent $e$ is stored as a binary number in the so-called \textit{biased form}, and its range is $e\in[-M,M]$, where $M=2^{p-1}$. The significand $m$ is an integer that satisfies $2^{t-1}\leq m \leq 2^t-1$, where the lower bound is enforced to ensure that the system is \textit{normalized}, i.e. the first bit of $m$ is always $1$.
We can therefore write $\fl(\alpha)$ in a more intuitive representation
\begin{align*}
    \fl(\alpha) = \pm 2^{e}\times \lpar \tfrac{m_1}{2} + \tfrac{m_2}{2^2} + \ldots + \tfrac{m_t}{2^t} \rpar,
\end{align*}
where the first bit $m_1$ of $m$ is always equal to one for normalized numbers. The range of normalized numbers is therefore $[2^{-M},2^{M}(2-2^{-t})]$. Numbers that are smaller than $2^{-M}$ are called \textit{subnormal} and they will be ignored for simplicity, since we can either add more bits in the exponent. Similarly, numbers that are larger than $2^{M}(2-2^{-t})$ are assumed to be numerically equal to infinity, denoted by $\FLINF$. 

From \cite[Theorem 2.2]{higham2002accuracy},
for all real numbers $\alpha$ in the normalized range it holds that \begin{align*}
    \fl(\alpha) = (1+\theta)\alpha,
\end{align*}
where $\theta\in\mathbb{R}$ satisfies $|\theta|\leq 2^{-t}:=\umach$, where $\umach$ is the \textit{machine precision}. Clearly, $t=O(\log(1/\umach))$, in which case we can always obtain a bound for the number of required bits of a numerical algorithm if we have an upper bound for the precision $\umach$. We will write the same for complex numbers which are represented as a pair of normalized floating point numbers.

The floating point implementation of each arithmetic operation $\odot \in\{+,-,\times,/\}$ also satisfies
\begin{align}
    \fl(\alpha\odot\beta) = (1+\theta)(\alpha\odot\beta),\quad |\theta|\leq\umach.
\end{align}
Divisions and multiplications with $1$ and $2$ do not introduce errors (for the latter we simply increase/decrease the exponent).
We assume that we also have an implementation of $\sqrt{\cdot}$ such that $\fl(\sqrt{\alpha})=(1+\theta)\sqrt{\alpha}$ where $|\theta|\leq \umach$.
From \cite[Lemma 3.1]{higham2002accuracy}, we can bound products of errors as
\begin{align*}
    \prod_{i=1}^n (1+\theta_i)^{\rho_i} = 1+\eta_n,
\end{align*}
where $\rho_i=\pm 1$ and $|\eta_n|\leq \tfrac{n\umach}{1-n\umach}$.

The above can be extended also for complex arithmetic (see \cite[Lemma 3.5]{higham2002accuracy}), where the bound becomes $|\theta|\leq O(\umach)$, but we will ignore the constant prefactor for simplicity. 

Operations on matrices can be analyzed in a similar manner. Let $\otimes$ denote the element-wise multiplication between two matrices and $\oslash$ the element-wise division. The floating point representation of a matrix $\matA$ satisfies
\begin{align*}
    \fl(\matA) = \matA+\matDelta \otimes \matA,\quad |\matDelta_{i,j}|\leq \umach.
\end{align*}
It can be shown that $\|\matDelta\|\leq \umach\sqrt{n}\|\matA\|$. 
For any operation $\odot\in\{+,-,\otimes,\oslash\}$ and matrices $\matA$ and $\matB$ it holds that
\begin{align}
    \label{eq:matrix_flop_errors}
    \fl(\matA\odot \matB) = \matA \odot \matB +\matDelta\otimes(\matA \odot \matB), 
    \quad 
    |\matDelta_{i,j}|\leq \umach, 
    \quad 
    \|\matDelta\otimes(\matA \odot \matB)\|\leq \umach\sqrt{n}\|\matA\odot\matB\|.
\end{align}

\section{Reduction to tridiagonal form - omitted proofs and definitions}
\label{appendix:tridiagonal_reduction}

\subsection{Imported subroutines}
\label{appendix:tridiagonal_reduction_subroutines}
In this appendix we mention some preliminary results that we use in the analysis.
\begin{theorem}[$ \MM$, stable fast matrix multiplication \cite{demmel2007fastmm,demmel2007fastla}]
\label{theorem:fast_mm}
For every $\eta>0$, there exists a fast matrix multiplication algorithm $ \MM$ which takes as input two matrices $\matA,\matB\in\mathbb{C}^{n\times n}$ and returns $\matC\leftarrow  \MM(\matA,\matB)$ such that
\begin{align*}
    \|\matC-\matA\matB\| \leq n^{c_{\eta}}\cdot\umach\|\matA\|\|\matB\|,
\end{align*}
on floating point machine with precision $\umach$, for some constant $c_{\eta}$ independent of $n$. It requires $O(n^{\omega+\eta})$ floating point operations.
\end{theorem}
We can also assume that if the result $\matA\matB$ is Hermitian, then $\MM(\matA,\matB)$ will be Hermitian as well (see e.g. \cite{sobczyk2024invariant}).

\begin{theorem}[Cf. \cite{demmel2007fastla}]
\label{theorem:alg_qr}
Given a matrix $\matA\in\mathbb{C}^{m\times n}$, $m\geq n$, there exists an algorithm $[\matQ,\matR]\leftarrow\QR(\matA)$ which returns an upper triangular matrix $\matR\in\mathbb{C}^{m\times n}$ and a matrix $\matQ\in\mathbb{C}^{m\times m}$, such that:
\begin{align*}
    (\matQ+\matE_{\matQ})^*(\matA+\matE_{\matA}) = \matR, 
    \quad
    \|\matE_{\matQ}\|\leq n^{c_{\QR}}\umach,
    \quad
    \|\matE_{\matA}\|\leq n^{c_{\QR}}\umach\|\matA\|,
\end{align*}
for some constant $c_{\QR}$, where the matrix $\matQ+\matE_{\matQ}$ is unitary. The algorithm executes $O(mn^{\omega-1})$ floating point operations on a machine with precision $\umach$.
\end{theorem}
In our analysis it will be useful the following re-statement of the aforementioned results.
\begin{corollary}
\label{corollary:alg_qr}
There exists a global constant $\cmm\geq \max\{c_{\QR},c_{\eta},1\}$, such that the matrices from Theorems \ref{theorem:fast_mm} and \ref{theorem:alg_qr} satisfy the following properties
    \begin{align*}
        \lnorm  \MM(\matA,\matB)-\matA\matB \rnorm &\leq n^{\cmm} \umach \|\matA\|\|\matB\|,
        \\
        \lnorm \matA - \matQ\matR \rnorm &\leq  n^{\cmm}\umach\|\matA\|, 
        \\
        \|\matR\|&\leq (1+n^{\cmm}\umach)\|\matA\|,
        \\
        \|\matQ\|&\leq \sqrt{1+n^{\cmm}\umach},
        \\
        \|\matQ^{-1}\|&\leq \frac{1}{\sqrt{1-n^{\cmm}\umach}},
        \\
        \max\{
            \|\matI-\matQ\matQ^*\|,\|\matI-\matQ^*\matQ\|
        \}
        &\leq
        n^{\cmm}\umach.
    \end{align*}
\begin{proof}
    $\|\matR\|=\|\matA+\matE_{\matA}\|\leq \|\matE_{\matA}\|+\|\matA\|\leq (1+n^{c_{\QR}}\umach)\|\matA\|$.     Moreover, $\|\matQ\|\leq \|\matQ+\matE_{\matQ}\|+\|\matE_{\matQ}\|\leq 1+n^{c_{\QR}}\umach$.
    Then from Theorem \ref{theorem:alg_qr} we can rewrite $\matA+\matE_{\matA}=(\matQ+\matE_{\matQ})\matR$. This gives
    \begingroup
    \allowdisplaybreaks
    \begin{align*}
        \lnorm \matA-\matQ\matR\rnorm 
        &= 
        \lnorm
        \matE_{\matQ}\matR-\matE_{\matA}
        \rnorm\\
        &\leq 
        \|\matE_{\matQ}\| \|\matA+\matE_{\matA}\| + \|\matE_{\matA}\|\\
        &\leq
        n^{c_{\QR}}\umach (1+n^{c_{\QR}}\umach)\|\matA\| + n^{c_{\QR}}\umach \|\matA\|\\
        &=
        n^{c_{\QR}}\umach (2+n^{c_{\QR}}\umach)\|\matA\|
        \\
        &\leq
        n^{2c_{\QR}+1}\umach\|\matA\|,
    \end{align*}
    \endgroup
    where in the last we assumed $n\geq 2$ and $\umach\leq 1$.
\end{proof}
\end{corollary}

Using square fast matrix multiplication we can obtain an algorithm to compute two banded matrices with the same bandwidth
\begin{corollary}
    \label{corollary:block_tridiagonal_mm}
    Let $\matA$,$\matB$ in $\mathbb{C}^{n\times n}$, where, without loss of generality, $n$ is a power of two. Assume that $\matA$ and $\matB$ are block-tridiagonal, with block size $n_k=2^k$ for some $k\in \{0,1,\ldots,\log(n)-2\}$. We can compute a matrix $\matC'$ such that $\|\matC'-\matA\matB\|\leq \umach n_k^{\beta}\|\matA\|\|\matB\|$ in $O(nn_k^{\omega-1})$ floating point operations.
    \begin{proof}
        By definition, the true matrix $\matC=\matA\matB$ is block-pentadiagonal with the same block size as $\matA$ and $\matB$. Each block of $\matC$ can be computed using at most three square matrix multiplications between blocks of $\matA$ and $\matB$, and two square matrix additions. Therefore, each block $\matC_{i,j}$ of $\matC$ can be computed in $O(n_k^{\omega})$ floating point operations, and, by Theorem \ref{theorem:fast_mm} and Corollary \ref{corollary:alg_qr} the result will satisfy
        $\|\matC'_{i,j}-\matC_{i,j}\| \leq O(n_k^\beta)\umach\|\matA\|\|\matB\|$. Splitting the matrix $\matC$ in three block-diagonals, $\matC_{-1},\matC_0,$ and $\matC_1$, we have that
        \begin{align*}
            \|\matC'-\matC\| \leq \|\matC_{-1}-\matC'_{-1}\| + \|\matC_0-\matC'_0\| + \|\matC_{1}-\matC'_1\|
            \leq 3\cdot O(n_k^\beta)\umach\|\matA\|\|\matB\|.
        \end{align*}
        There are $3n/n_k-2$ blocks in $\matC$, which gives the total complexity to be $O(n_k^{\omega}n/n_k)= O(nn_k^{\omega-1})$.
    \end{proof}
\end{corollary}

\subsection{Rotations}
An example of the rotations is illustrated in Equations \eqref{eq:rotation_r_i} and \eqref{eq:rotation_r_i_prime}. In Equation \eqref{eq:rotation_r_i}, the matrices $\matA'_{1,2}$ and $\matA'_{2,1}$ are lower and upper triangular, respectively. In Equation \eqref{eq:rotation_r_i_prime}, the matrices $\matA'_{4,2}$ and $\matA'_{5,3}$ are upper triangular, while $\matA'_{2,4}$ and $\matA'_{3,5}$ are lower triangular. 
\begingroup
\setlength\arraycolsep{1.5pt}
\begin{align}
    \label{eq:rotation_r_i}
    \overbrace{
    \begin{pmatrix}
        \matA_{1,1}  & \matA_{1,2}  & \boxed{\matA_{1,3}}   & 0             & 0           & 0           & 0           & 0           \\
        \matA_{2,1}  & \matA_{2,2}  &\matA_{2,3}   & \matA_{2,4}   & 0           & 0           & 0           & 0           \\
        \boxed{\matA_{3,1}}  & \matA_{3,2}  &\matA_{3,3}   & \matA_{3,4}   & \matA_{3,5} & 0           & 0           & 0           \\
        0            & \matA_{4,2}  &\matA_{4,3}   & \matA_{4,4}   & \matA_{4,5} & \matA_{4,6} & 0           & 0           \\
        0            & 0            &\matA_{5,3}   & \matA_{5,4}   & \matA_{5,5} & \matA_{5,6} & \matA_{5,7} & 0           \\
        0            & 0            & 0            & \matA_{6,4}   & \matA_{6,5} & \matA_{6,6} & \matA_{6,7} & \matA_{6,8} \\
        0            & 0            & 0            & 0             & \matA_{7,5} & \matA_{7,6} & \matA_{7,7} & \matA_{7,8} \\
        0            & 0            & 0            & 0             & 0           & \matA_{8,6} & \matA_{8,7} & \matA_{8,8} \\
    \end{pmatrix}
    }^{
        \matA^{(k,2,0)},\ i=2
    }
    \stackrel{R_2}{\longrightarrow}
    \overbrace{
    \begin{pmatrix}
        \matA_{1,1}  & \matA'_{1,2}  & 0   & 0             & 0           & 0           & 0           & 0           \\
        \matA'_{2,1}  & \matA_{2,2}  &\matA_{2,3}   & \matA_{2,4}   & \boxed{\matA_{2,5}}           & 0           & 0           & 0           \\
        0  & \matA_{3,2}  &\matA_{3,3}   & \matA_{3,4}   & \matA_{3,5} & 0           & 0           & 0           \\
        0            & \matA_{4,2}  &\matA_{4,3}   & \matA_{4,4}   & \matA_{4,5} & \matA_{4,6} & 0           & 0           \\
        0            & \boxed{\matA_{5,2}} &\matA_{5,3}   & \matA_{5,4}   & \matA_{5,5} & \matA_{5,6} & \matA_{5,7} & 0           \\
        0            & 0            & 0            & \matA_{6,4}   & \matA_{6,5} & \matA_{6,6} & \matA_{6,7} & \matA_{6,8} \\
        0            & 0            & 0            & 0             & \matA_{7,5} & \matA_{7,6} & \matA_{7,7} & \matA_{7,8} \\
        0            & 0            & 0            & 0             & 0           & \matA_{8,6} & \matA_{8,7} & \matA_{8,8} \\
    \end{pmatrix}
    }^{
        \matA^{(k,3,5)}
    }
\end{align}
\endgroup

\begingroup
\setlength\arraycolsep{1.5pt}
\begin{align}
    \label{eq:rotation_r_i_prime}
    \overbrace{
    \begin{pmatrix}
        \matA_{1,1}  & \matA'_{1,2}  & 0   & 0             & 0           & 0           & 0           & 0           \\
        \matA'_{2,1}  & \matA_{2,2}  &\matA_{2,3}   & \matA_{2,4}   & \boxed{\matA_{2,5}}           & 0           & 0           & 0           \\
        0  & \matA_{3,2}  &\matA_{3,3}   & \matA_{3,4}   & \matA_{3,5} & 0           & 0           & 0           \\
        0            & \matA_{4,2}  &\matA_{4,3}   & \matA_{4,4}   & \matA_{4,5} & \matA_{4,6} & 0           & 0           \\
        0            & \boxed{\matA_{5,2}} &\matA_{5,3}   & \matA_{5,4}   & \matA_{5,5} & \matA_{5,6} & \matA_{5,7} & 0           \\
        0            & 0            & 0            & \matA_{6,4}   & \matA_{6,5} & \matA_{6,6} & \matA_{6,7} & \matA_{6,8} \\
        0            & 0            & 0            & 0             & \matA_{7,5} & \matA_{7,6} & \matA_{7,7} & \matA_{7,8} \\
        0            & 0            & 0            & 0             & 0           & \matA_{8,6} & \matA_{8,7} & \matA_{8,8} \\
    \end{pmatrix}
    }^{
        \matA^{(k,3,5)}
    }
    \stackrel{R'_4}{\longrightarrow}
    \overbrace{
        \begin{pmatrix}
            \matA_{1,1}  & \matA'_{1,2} & 0            & 0                   & 0           & 0           & 0           & 0           \\
            \matA'_{2,1} & \matA_{2,2}  &\matA_{2,3}   & \matA'_{2,4}        & 0           & 0           & 0           & 0           \\
            0            & \matA_{3,2}  &\matA_{3,3}   & \matA_{3,4}         & \matA'_{3,5} & 0           & 0           & 0           \\
            0            & \matA'_{4,2} &\matA_{4,3}   & \matA_{4,4}         & \matA_{4,5} & \matA_{4,6} & \boxed{\matA_{4,7}} & 0           \\
            0            & 0            &\matA'_{5,3}  & \matA_{5,4}         & \matA_{5,5} & \matA_{5,6} & \matA_{5,7} & 0           \\
            0            & 0            & 0            & \matA_{6,4}         & \matA_{6,5} & \matA_{6,6} & \matA_{6,7} & \matA_{6,8} \\
            0            & 0            & 0            & \boxed{\matA_{4,7}} & \matA_{7,5} & \matA_{7,6} & \matA_{7,7} & \matA_{7,8} \\
            0            & 0            & 0            & 0                   & 0           & \matA_{8,6} & \matA_{8,7} & \matA_{8,8} \\
        \end{pmatrix}
    }^{
        \matA^{(k,3,7)}
    }
\end{align}
\endgroup

The rotations can be conveniently implemented in floating point using fast QR factorizations.

\begin{lemma}
    \label{lemma:rotation_r_i_floating_point}
    Let $\matA^{(k,i,0)}$ be a block-pentadiagonal matrix with no bulge of the form of Equation \eqref{eq:block_pentadiagonal}, consisting of $b_k\times b_k$ blocks of size $n_k\times n_k=2^{k}\times 2^{k}$ each. The rotation $[\matA^{(k,i+1,i+3)},\matQ] \leftarrow R_i(\matA^{(k,i,0)})$ can be implemented in $O(n_k^{\omega})$ floating point operations. The resulting matrix $\matA^{(k,i+1,i+3)}$ has a bulge at positions $(i,i+3)$ and $(i+3,i)$, and $\matQ$ is approximately unitary. Moreover, if $\umach\leq 1/n^{\cmm}$, then
    \begin{align*}
        \lnorm 
            \matA^{(k,i,0)} - \matQ\matA^{(k,i+1,i+3)} \matQ^*
        \rnorm
        \leq
        C\umach n^{\cmm}\|\matA^{(k,i,0)}\|,
    \end{align*}
    where $C$ is a constant independent of $n$.
    \begin{proof}
        The first step is to compute $\matQ_i,\matR_i\leftarrow \QR(\matA_i)$, where $\matA_i=\begin{pmatrix}
            \matA_{i,i-1}\\
            \matA_{i+1,i-1}
        \end{pmatrix}$. From Theorem \ref{theorem:alg_qr}, $\matQ_i$ has size $2n_k\times 2n_k$, and $\matR_i$ has size $2n_k\times n_k$ and it is upper triangular.
        
        We then isolate the matrices $\matB_{i}=\begin{pmatrix}
            \matA_{i,i} & \matA_{i,i+1} \\
            \matA_{i+1,i} & \matA_{i+1,i+1} \\
        \end{pmatrix}$,
        as well as $\matC_{i}=\begin{pmatrix}
            \matA_{i,i+2} & 0 \\
            \matA_{i+1,i+2} & \matA_{i+1,i+3} \\
        \end{pmatrix}$, and compute the two matrix products $\matB_i'= \MM(\matQ_i^*,\matB_i)$ and $\matC_i'= \MM(\matQ_i^*,\matC_i)$. We finally compute the product $\matB_i''= \MM(\matB_i',\matQ_i)$. The matrix $\matB_i''$ can be forced to be Hermitian by copying the lower triangular part to the upper triangular part. This increases the error norm by a factor of $O(\log(n_k))$ (see eg \cite{sobczyk2024invariant}). Now consider the matrix $\matA^{(k,i+1,i+3)}$, which is exactly the same as the original $\matA^{(k,i,0)}$, having only replaced the following blocks:
        \begin{align*}
        \begin{pmatrix}
        \ldots            &\matA_i^*        & 0       \\
        \matA_{i}         & \matB_{i}       & \matC_i \\
        0                 & \matC_{i}^*     & \ldots
        \end{pmatrix}
        \longrightarrow
        \begin{pmatrix}
        \ldots            &\matR_i^*        & 0       \\
        \matR_{i}         & \matB''_{i}     & \matC'_i \\
        0                 & \matC_{i}'^*    & \ldots
        \end{pmatrix}.
        \end{align*}
        If we set $\matQ=\begin{pmatrix}
            \matI_{(i-1)n_k} & 0       & 0 \\
               0             & \matQ_i &  0\\
               0             & 0       & \matI_{n-(i+1)n_k}
        \end{pmatrix}$, then
        \begin{align}
            \label{eq:lemma_rotation_r_i_backward_error}
            \lnorm \matA^{(k,i,0)} - \matQ\matA^{(k,i+1,i+3)} \matQ^* \rnorm
            &=
            \lnorm
                \begin{pmatrix}
                    \ldots            &\matA_i^*        & 0       \\
                    \matA_{i}         & \matB_{i}       & \matC_i \\
                    0                 & \matC_{i}^*     & \ldots
                \end{pmatrix}
                -
                \begin{pmatrix}
                \ldots            &\matR_i^*\matQ_i^*        & 0       \\
                \matQ_i\matR_{i}         & \matQ_i\matB''_{i}\matQ_i^*     & \matQ_i\matC'_i \\
                0                 & \matC_{i}'^*\matQ_i^*    & \ldots
                \end{pmatrix}
            \rnorm
            \nonumber
            \\
            &\leq
            \lnorm
                \matA_i-\matQ_i\matR_i
            \rnorm
            +
            \lnorm
                \matB_i-\matQ_i\matB''_i\matQ_i^*
            \rnorm
            +
            \lnorm
                \matC_i-\matQ_i\matC'_i
            \rnorm.
        \end{align}
        From Corollary \ref{corollary:alg_qr} we have that $\|\matA_i-\matQ_i\matR_i\| \leq (2n_k)^{\cmm}\umach\|\matA_i\|$.
        For the middle term, note that we can write $\matB''_i=(\matQ_i^*\matB_i+\matE_1)\matQ_i+\matE_2$. From Corollary \ref{corollary:alg_qr}, $\|\matE_1\|\leq (2n_k)^{\cmm}\umach\|\matQ_i\|\|\matB_i\| 
        \leq (2n_k)^{\cmm}\umach (1+(2n_k)^{\cmm}\umach) \|\matB_i\|
        $, and
        \begin{align*}
            &\|\matE_2\| \leq (2n_k)^{\cmm}\umach \|\matQ_i^*\matB_i+\matE_1\|\|\matQ_i\| 
            \\
            &\quad\leq 
            (2n_k)^{\cmm}\umach
            \lpar
                (1+(2n_k)^{\cmm}\umach)\|\matB_i\|+(2n_k)^{\cmm}\umach (1+(2n_k)^{\cmm}\umach) \|\matB_i\|
            \rpar
            (1+(2n_k)^{\cmm}\umach)
            \\
            &\quad\leq 
            (2n_k)^{\cmm}\umach
            \lpar
                1+(2n_k)^{\cmm}\umach
            \rpar^3
            \|\matB_i\|.
        \end{align*}
        From Corollary \ref{corollary:alg_qr}, we can write $\matQ\matQ^*=\matI+\matE_3$ where $\|\matE_3\|\leq (2n_k)^{\cmm}\umach$. The matrix $\matQ_i\matB''_i\matQ_i^*$ can be written as
        \begin{align*}
            \matQ_i\matB''_i\matQ_i^* 
            &=
            \matQ_i\lpar  (\matQ_i^*\matB_i+\matE_1)\matQ_i+\matE_2 \rpar \matQ_i^*
            \\
            &=
            \matQ_i\matQ_i^*\matB_i\matQ_i\matQ_i^*+\matQ_i\matE_1\matQ_i\matQ_i^*+\matQ_i\matE_2\matQ_i^*
            \\
            &=
            (\matI+\matE_3)\matB_i(\matI+\matE_3) + \matQ_i\matE_1(\matI+\matE_3) + \matQ_i\matE_2\matQ_i^*
            \\
            &=
            \matB_i+\matE_3\matB_i + \matB_i\matE_3 + \matE_3\matB_i\matE_3 + \matQ_i\matE_1+\matQ_i\matE_1\matE_3 + \matQ_i\matE_2\matQ_i^*.
        \end{align*}
        Then
        \begin{align*}
            \lnorm 
                \matB_i - \matQ_i\matB''_i\matQ_i^*
            \rnorm
            &=
            \lnorm 
                \matE_3\matB_i + \matB_i\matE_3 + \matE_3^2 + \matQ_i\matE_1+\matQ_i\matE_1\matE_3 + \matQ_i\matE_2\matQ_i^*
            \rnorm
            \\
            &\leq
            2\|\matE_3\matB_i\|
            + 
            \|\matE_3\matB_i\matE_3\|
            + 
            \|\matQ_i\matE_1\|
            +
            \|\matQ_i\matE_1\matE_3\|
            + 
            \|\matQ_i\matE_2\matQ_i^*\|
            \\
            &\leq
            2(2n_k)^{\cmm}\umach\|\matB_i\|
            + 
            ((2n_k)^{\cmm}\umach)^2\|\matB_i\|
            + 
            (2n_k)^{\cmm}\umach (1+(2n_k)^{\cmm}\umach)^2 \|\matB_i\|
            +\ldots
            \\
            &\quad\quad
            \ldots
            ((2n_k)^{\cmm}\umach)^2 (1+(2n_k)^{\cmm}\umach)^2 \|\matB_i\|
            + 
            (2n_k)^{\cmm}\umach
            \lpar
                1+(2n_k)^{\cmm}\umach
            \rpar^5
            \|\matB_i\|
            \\
            &\leq
            (2n_k)^{\cmm}\umach
            \|\matB_i\|
            \lpar
                2
                + 
                (2n_k)^{\cmm}\umach
                + 
                (1+(2n_k)^{\cmm}\umach)^2
                +
                (2n_k)^{\cmm}\umach (1+(2n_k)^{\cmm}\umach)^2
                + 
                \lpar
                    1+(2n_k)^{\cmm}\umach
                \rpar^5
            \rpar
            \\
            &\leq C_1(2n_k)^{\cmm}\umach \|\matB_i\|,
        \end{align*}
        for some constant $C_1$, where $\umach \leq 1/n^{\cmm}$ is a sufficient condition for the last inequality.
    
        Finally, note that from Corollary \ref{corollary:alg_qr} $\matC'_i$ can be written as $\matC'_i=(\matQ_i^*\matC_i+\matE_4)$ where $\|\matE_{4}\|\leq (2n_k)^{\cmm}\umach\|\matQ_i\|\|\matC_i\|\leq (2n_k)^{\cmm}\umach(1+(2n_k)^{\cmm}\umach)\|\matC_i\|,$ in which case
        \begin{align*}
            \|\matC_i-\matQ_i\matC_i'\|
            &=
            \|\matC_i-\matQ_i(\matQ_i^*\matC_i+\matE_4)\|
            \\
            &\leq
            \|\matC_i-(\matI+\matE_3)\matC_i\|+\|\matQ_i\matE_4\|
            \\
            &\leq
            \|\matE_3\matC_i\|+\|\matQ_i\matE_4\|
            \\
            &\leq
            (2n_k)^{\cmm}\umach\|\matC_i\|+(2n_k)^{\cmm}\umach(1+(2n_k)^{\cmm}\umach)^2\|\matC_i\|
            \\
            &=
            (2n_k)^{\cmm}\umach\|\matC_i\|(1+(2n_k)^{\cmm}\umach(1+(2n_k)^{\cmm}\umach)^2)
            \\
            &\leq
            C_2(2n_k)^{\cmm}\umach\|\matC_i\|,
        \end{align*}
        where $C_2$ is a constant which arises if we assume that $\umach\leq 1/(2n_k)^{\cmm}$.
        Replacing all the derived bounds in Eq. \eqref{eq:lemma_rotation_r_i_backward_error} it follows that
        \begin{align*}
            \lnorm \matA^{(k,i,0)} - \matQ\matA^{(k,i+1,i+3)} \matQ^* \rnorm
            &\leq
            (2n_k)^{\cmm}\umach\lpar \|\matA_i\| + C_1\|\matB_i\| + C_2 \|\matC_i\|\rpar
            \leq
            C_3(2n_k)^{\cmm}\umach \|\matA^{(k,i,0)}\|,
        \end{align*}
        where we bounded the norms of the submatrices by the norm of the original matrix and $C_3=1+C_1+C_2$.
    \end{proof}
\end{lemma}

\begin{lemma}
    \label{lemma:rotation_r_i_prime_floating_point}
    Let $\matA^{(k,s,j+1)}$, $j\geq s+1$, be a block-pentadiagonal matrix of the form of Equation \eqref{eq:block_pentadiagonal}, with a bulge at position $(j+1,j-2)$, consisting of $b_k\times b_k$ blocks of size $n_k\times n_k=2^{k}\times 2^{k}$ each. The rotation $[\matA^{(k,s,j+3)},\matQ] \leftarrow R_j'(\matA^{(k,s,j+1)})$ can be implemented in $O(n_k^{\omega})$ floating point operations. The resulting matrix $\matA^{(k,s,j+3)}$ has a bulge at positions $(j+3,j)$ and $(j,j+3)$, and $\matQ$ is approximately unitary. Moreover, if $\umach\leq 1/n^{\cmm}$, then
    \begin{align*}
        \lnorm 
            \matA^{(k,s,j+1)} - \matQ\matA^{(k,s,j+3)} \matQ^*
        \rnorm
        \leq
        C'\umach (2n_k)^{\cmm}\|\matA^{(k,s,j+1)}\|,
    \end{align*}
    where $C'$ is a constant independent of $n$.
    \begin{proof}
        The proof is almost identical to the one of Lemma \ref{lemma:rotation_r_i_floating_point} and therefore it is omitted. The only change is that instead of computing the QR of a $2n_k\times n_k$ submatrix, we compute the QR of a $2n_k\times 2n_k$ one.
    \end{proof}
\end{lemma}

\subsection{Bandwidth halving}

\begin{lemma}
    \label{lemma:bandwidth_halving_floating_point_appendix}
    Let $\matA^{(k,2,0)}$ be a full block-pentadiagonal matrix of the form of Equation \eqref{eq:block_pentadiagonal}, with no bulge, consisting of $b_k\times b_k$ blocks of size $n_k\times n_k=2^{k}\times 2^{k}$ each. For any $k\in [\log(n)-2]$, if $\umach\leq \tfrac{1}{C_1n^{\cmm+3}}$, where $C_1$ is a constant, then Algorithm \ref{algorithm:halve} returns a matrix $\matQhat^{(k)}$ that is approximately orthogonal, and a matrix $\matA^{(k-1,2,0)}$ such that
    \begin{align*}
        \lnorm 
            \matA^{(k,2,0)} - \matQhat^{(k)}\matA^{(k-1,2,0)} \matQhat^{(k)*}
        \rnorm
        &\leq
        C_2\umach n^{\cmm+3}\|\matA^{(k,2,0)}\|,
        \\
        \lnorm 
            \matQhat^{(k)}\matQhat^{(k)*} - \matI
        \rnorm
        &\leq
        C_3\umach n^{\cmm+3}
    \end{align*}
    where $C_2,C_3$ are also constants independent of $n$. It requires 
    $O\lpar
        n^2
        \lpar 
            S_{\omega}(\log(n)-1)
            -
            S_{\omega}(k)
        \rpar
    \rpar$
    floating point operations, where $S_{x}(m):=\sum_{l=1}^{m} \lpar 2^{x-2} \rpar^{l}$. If $\matQhat^{(k)}$ is not required, the complexity reduces to 
    $O(n^2n_k^{\omega-2})=O(n^2(2^{\omega-2})^k)$.
    \begin{proof}
        The first observation is that, for fixed $i$ (i.e. for every outer iteration), the rotation matrices  $\matQ_{i,j}$ for $j=i+1,\ldots,b_k$ are all independent. This means that the  product can be formed without any multiplications. Specifically, for even $i$, the product $\matQ_i=\matQ_{i,b_k}\cdot\matQ_{i,b_k-2}\cdot\ldots\cdot\matQ_{i,i+2}\cdot\matQ_{i,i}$ has the form
        \begin{align}
            \label{eq:lemma_bandwidth_halving_q_i_block_diagonal_form}
            \matQ_i=\begin{pmatrix}
            \matI &       &      &                 &                 &       &                  \\
                 & \ddots &      &                 &                 &       &                  \\
                 &       & \matI &                 &                 &       &                  \\
                 &       &      & \matZ_{i,i} &                 &       &                  \\
                 &       &      &                 & \matZ_{i,i+2} &       &                  \\
                 &       &      &                 &                 & \ddots &                  \\
                 &       &      &                 &                 &       & \matZ_{i,b_k}
            \end{pmatrix},
        \end{align}
        where the matrices $\matZ$ have size $2n_k\times 2n_k$ and originate from the QR factorizations inside the rotations $R_i$ and $R_j'$. For odd $i$, the matrix $\matQ_i=\matQ_{i,b_k-1}\cdot\matQ_{i,b_k-3}\cdot\ldots\cdot\matQ_{i,i+2}\cdot\matQ_{i,i}$ has a similar form.

        Using Lemmas \ref{lemma:rotation_r_i_floating_point} and \ref{lemma:rotation_r_i_prime_floating_point}, for some fixed even $i$, we can write
        \begin{align*}
            \matA^{(k,i,0)}
            &=
            \matQ_{i,i}\matA^{(k,i+1,i+3)}\matQ_{i,i}^*+\matE_{i,i}
            \\
            &=
            \matQ_{i,i}\lpar
                \matQ_{i,i+2}\matA^{(k,i+1,i+5)}\matQ_{i,i+2}^* + \matE_{i,i+2}
            \rpar
            \matQ_{i,i}^*+\matE_{i,i}
            \\
            &=
            \matQ_i\matA^{(k,i+1,0)}\matQ_i^*
            +
            \matE_{i,i}
            +
            \sum_{l=1}^{b_k/2}
            \lbrac
                \lpar
                    \prod_{m=1}^{l}
                    \matQ_{i,i+2m}
                \rpar
                \matE_{i,i+2l}
                \lpar
                    \prod_{m=1}^{l}
                    \matQ_{i,i+2m}
                \rpar^*
            \rbrac.
        \end{align*}
        Here error matrix $\matE_{i,i}$ originates from the first rotation $R_i$ and is equal to $\matE_{i,i}=\matA^{(k,i,0)}-\matQ_{i,i}\matA^{(k,i+1,i+3)}\matQ_{i,i}^*$, while for $l=1,\ldots,b_k/2$ the error matrices originate from the rotations $R'$ and are equal to $\matE_{i,i+2l}=\matA^{(k,i+1,i+1+2l)}-\matQ_{i,i+2l}\matA^{(k,i+1,i+1+4l)}\matQ_{i,i+2l}^*$. It holds that
        \begin{align*}
            \lnorm
                \matE_{i,i+2l}
            \rnorm
            \leq
            C'\umach (2n_k)^{\cmm} \lnorm \matA^{(k,i+1,i+1+2l)} \rnorm
            \leq C'\umach n^{\cmm} \lnorm \matA^{(k,i+1,i+1+2l)} \rnorm.
        \end{align*}
        From Lemma \ref{lemma:rotation_r_i_floating_point} it also holds that $\|\matE_{i,i}\|\leq C\umach (2n_k)^{\cmm}\|\matA^{(k,i,0)}\|\leq C\umach n^{\cmm}\|\matA^{(k,i,0)}\|$, and
        Corollary \ref{corollary:alg_qr} implies that the matrices $\matQ_{i,i+2l}$ are invertible and 
        $\|\matQ_{i,i+2l}^{-1}\|=\frac{1}{\sigma_{\min}(\matQ_{i,i+2l})}\geq \frac{1}{\sqrt{1-(2n_k)^{\cmm}\umach}}\geq \frac{1}{\sqrt{1-n^{\cmm}\umach}}$. 
        This provides that
        \begin{align*}
            \lnorm \matA^{(k,i+1,i+1+4l)} \rnorm 
            =
            \lnorm 
            \matQ_{i,i+2l}^{-1} (\matA^{(k,i+1,i+1+2l)} -  \matE_{i,i+2l})\matQ_{i,i+2l}^{-*}
            \rnorm 
            \leq
            \frac{1+C'n^{\cmm}\umach }{1-n^{\cmm}\umach}
            \lnorm 
            \matA^{(k,i+1,i+1+2l)}
            \rnorm,
        \end{align*}
        and accordingly $\lnorm \matA^{(k,i+1,i+3)}\rnorm \leq \frac{1+Cn^{\cmm}\umach}{1-n^{\cmm}\umach}
        \lnorm
            \matA^{(k,i,0)}
        \rnorm.
        $
        Since, as already mentioned, the product $\prod_{m=1}^{l}\matQ_{i,i+2m}$ involves independent diagonal blocks, then for any $l$:
        \begin{align*}
            \lnorm
                \prod_{m=1}^{l}\matQ_{i,i+2m}
            \rnorm
            =
            \max_{m=1,\ldots,l}\lcurly
                \lnorm \matQ_{i,i+2m}\rnorm
            \rcurly 
            \leq
            \sqrt{1+n^{\cmm}\umach}.
        \end{align*}
        where the last comes from Corollary \ref{corollary:alg_qr}. Similarly,
        \begin{align*}
            \lnorm
                \lpar \prod_{m=1}^{l}\matQ_{i,i+2m} \rpar^{-1}
            \rnorm
            =
            \max_{m=1,\ldots,l}\lcurly
                \lnorm \lpar \matQ_{i,i+2m} \rpar ^{-1}\rnorm
            \rcurly 
            \leq
            \tfrac{1}{
            \sqrt{1-n^{\cmm}\umach}}.
        \end{align*}
        Combining the aforementioned observations we obtain
        \begin{align*}
            \lnorm
                \matA^{(k,i,0)}
                -
                \matQ_i\matA^{(k,i+1,0)}\matQ_i^*
            \rnorm
            &=
            \lnorm
                \matE_{i,i}
                +
                \sum_{l=1}^{b_k/2}
                \lbrac
                    \lpar
                        \prod_{m=1}^{l}
                        \matQ_{i,i+2m}
                    \rpar
                    \matE_{i,i+2l}
                    \lpar
                        \prod_{m=1}^{l}
                        \matQ_{i,i+2m}
                    \rpar^*
                \rbrac
            \rnorm
            \\
            &\leq
            \lnorm
                \matE_{i,i}
            \rnorm
            +
            \sum_{l=1}^{b_k/2}
                \lbrac
                    \lnorm
                        \matE_{i,i+2l}
                    \rnorm
                    \lnorm
                    \prod_{m=1}^{l}
                        \matQ_{i,i+2m}
                    \rnorm^2
                \rbrac
            \\
            &\leq
            \lnorm
                \matE_{i,i}
            \rnorm
            +
            \lpar \sqrt{1+n^{\cmm}\umach} \rpar^2
            \sum_{l=1}^{b_k/2}
                \lnorm
                    \matE_{i,i+2l}
                \rnorm.
            \\
            &\leq
            C\umach n^{\cmm}\|\matA^{(k,i,0)}\|
            +
            \lpar 1+n^{\cmm}\umach \rpar
            \sum_{l=1}^{b_k/2}
                C'\umach n^{\cmm} \lnorm \matA^{(k,i+1,i+1+2l)} \rnorm.
            \\
            &\leq
            \umach n^{\cmm}
            \lbrac
                C\|\matA^{(k,i,0)}\|
                +
                C'\lpar 1+n^{\cmm}\umach \rpar
                \lnorm 
                    \matA^{(k,i+1,i+3)}
                \rnorm
                \sum_{l=1}^{b_k/2}
                    \lpar \tfrac{1+C'n^{\cmm}\umach }{1-n^{\cmm}\umach} \rpar^l
            \rbrac
            \\
            &\leq
            \umach n^{\cmm}
            \lbrac
                C\|\matA^{(k,i,0)}\|
                +
                C'\lpar 1+n^{\cmm}\umach \rpar
                \tfrac{1+Cn^{\cmm}\umach }{1-n^{\cmm}\umach}
                \lnorm 
                    \matA^{(k,i,0)}
                \rnorm
                \sum_{l=1}^{b_k/2}
                    \lpar \tfrac{1+C'n^{\cmm}\umach }{1-n^{\cmm}\umach} \rpar^l
            \rbrac.
            \\
            &=
            \umach n^{\cmm}
            \|\matA^{(k,i,0)}\|
            \lbrac
                C
                +
                C'\lpar 1+n^{\cmm}\umach \rpar
                \tfrac{1+Cn^{\cmm}\umach }{1-n^{\cmm}\umach}
                \sum_{l=1}^{b_k/2}
                    \lpar \tfrac{1+C'n^{\cmm}\umach }{1-n^{\cmm}\umach} \rpar^l
            \rbrac.
        \end{align*}
        If we further assume that $\umach \leq \frac{1}{\max\{1,C,C'\}n^{\cmm+1}}$, then all the summands are bounded by constants, and $b_k\leq n$, in which case we conclude
        \begin{align}
            \label{eq:lemma_bandwidth_halving_single_i}
            \lnorm
                \matA^{(k,i,0)}
                -
                \matQ_i\matA^{(k,i+1,0)}\matQ_i^*
            \rnorm
            \leq
            \umach n^{\cmm}
            \cdot
                C_1n
                \|\matA^{(k,i,0)}\|,
        \end{align}
        for some constant $C_1$. Moreover, the matrix $\matQ_i$ is approximately orthogonal, since,
        from Eq. \eqref{eq:lemma_bandwidth_halving_q_i_block_diagonal_form}, it consists of approximately orthogonal diagonal blocks. In particular,
        \begin{align}
            \label{eq:lemma_bandwidth_halving_q_i_approximately_orthogonal}
            \lnorm \matQ_i\matQ_i^* - \matI \rnorm
            \leq
            \max_{j=i+2,\ldots,b_k-1}
            \lcurly 
            \lnorm 
                \matZ_{i,j}\matZ_{i,j}^* - \matI
            \rnorm
            \rcurly
            \leq n^\cmm \umach.
        \end{align}

        We can now analyze the total error after applying the previous steps for all $i\in 2,\ldots,b_k-1$.
        From Eq. \eqref{eq:lemma_bandwidth_halving_single_i}, if we denote by $\matE_i=\matA^{(k,i,0)}
                -
                \matQ_i\matA^{(k,i+1,0)}\matQ_i^*
        $, we can directly obtain
        \begin{align*}
            \lnorm
                \matA^{(k,i+1,0)}
            \rnorm
            &=
            \lnorm
                \matQ_i^{-1}
                \lpar
                    \matA^{(k,i,0)}
                    -
                    \matE_i
                \rpar
                \matQ_i^{-*}
            \rnorm
            \leq
            \lnorm
                \matQ_i^{-1}
            \rnorm^2
            \lpar
                \lnorm
                    \matA^{(k,i,0)}
                \rnorm
                +
                \lnorm
                    \matE_i
                \rnorm
            \rpar
            \leq
            \tfrac{1+\umach n^{\cmm}
                C_1n}{1-n^\cmm \umach}
                \lnorm
                    \matA^{(k,i,0)}
                \rnorm.
        \end{align*}

        We now denote $\matQ^{(k)}=\matQ_{b_k}\cdot\matQ_{b_k-1}\cdot\ldots\cdot\matQ_3\cdot\matQ_2$. For now this product is assumed in exact arithmetic. We will later show how to efficiently compute it in floating point. 
        In exact arithmetic, for any $l\in\{2,\ldots,b_k\}$:
        \begin{align*}
            \lnorm
                \prod_{m=2}^{l}\matQ_{m}
            \rnorm
            \leq
            \lpar \sqrt{1+n^{\cmm}\umach}\rpar^{l-1},
            \quad
            \text{and}
            \quad
            \lnorm
                \lpar
                \prod_{m=2}^{l}\matQ_{m}
                \rpar^{-1}
            \rnorm
            \leq
            \lpar \frac{1}{\sqrt{1-n^{\cmm}\umach}}\rpar^{l-1}.
        \end{align*}
        From Bernoulli's inequalities, since $n^{\cmm}\umach\leq 1$, it follows that 
        \begin{align*}
            \sigma_{\max}(\matQ^{(k)})^2
            =
            \lnorm \matQ^{(k)} \rnorm^2 &\leq (1+n^{\cmm}\umach)^{\log(n)-1} \leq 1+n^{\cmm+1}\umach,
            \\
            \sigma_{\min}(\matQ^{(k)})^2
            =
            \lnorm 
                \lpar\matQ^{(k)}\rpar^{-1} 
            \rnorm^{-2} 
            &\geq (1-n^{\cmm}\umach)^{(\log(n)-1)} 
            \geq 1-(\log(n)-1)n^{\cmm}\umach
            \geq 1-n^{\cmm+1}\umach,
        \end{align*}
        and
        \begin{align}
            \label{eq:lemma_bandwidth_halving_q_approximately_orthogonal}
            \lnorm \matQ^{(k)}\matQ^{(k)*}-\matI \rnorm \leq n^{\cmm+1}\umach.
        \end{align}
        With similar arguments as above, we write
        \begin{align*}
            \matA^{(k,2,0)}
            &=
            \matQ_2\matA^{(k,3,0)}\matQ_2^* + \matE_2
            \\
            &=
            \matQ_2\lpar
                \matQ_3\matA^{(k,4,0)}\matQ_3^*+\matE_3
            \rpar
            \matQ_2^*
            +\matE_2
            \\
            &=
            \matE_2
            +
            \sum_{l=3}^{b_k}
            \lbrac
                \lpar
                \prod_{m=3}^l
                    \matQ_{m-1}
                \rpar
                \matE_l
                \lpar
                \prod_{m=3}^l
                    \matQ_{m-1}
                \rpar^*
            \rbrac
            +
            \matQ^{(k)}\matA^{(k-1,2,0)}
            \matQ^{(k)*}.
        \end{align*}    
        Rearranging the terms, we can bound the error norm by
        \begin{align*}
            \lnorm
                \matA^{(k,2,0)}
                -
                \matQ\matA^{(k-1,2,0)}\matQ^*
            \rnorm
            &=
            \lnorm
                \matE_2
                +
                \sum_{l=3}^{b_k}
                \lbrac
                    \lpar
                    \prod_{m=3}^l
                        \matQ_{m-1}
                    \rpar
                    \matE_l
                    \lpar
                    \prod_{m=3}^l
                        \matQ_{m-1}
                    \rpar^*
                \rbrac
            \rnorm
            \\
            &\leq
            \lnorm
                \matE_2
            \rnorm
            +
            \sum_{l=3}^{b_k}
                \lbrac
                    \lnorm
                        \matE_l
                    \rnorm
                    \lnorm
                        \prod_{m=3}^l
                            \matQ_{m-1}
                    \rnorm^2
                \rbrac
            \\
            &\leq
            \umach n^{\cmm}
            \cdot
                C_1n
                \|\matA^{(k,2,0)}\|
            +
            \sum_{l=3}^{b_k}
                \lbrac
                    \umach n^{\cmm}
                        C_1n
                        \lnorm\matA^{(k,l,0)}\rnorm
                    \lpar 1+n^{\cmm}\umach\rpar^{l-1}
                \rbrac
            \\
            &\leq
            \umach n^{\cmm}
            \cdot
                C_1n
                \lnorm\matA^{(k,2,0)}\rnorm
            \lpar
                1
                +
                \sum_{l=3}^{b_k}
                    \lbrac
                        \lpar 
                            1+n^{\cmm}\umach
                        \rpar^{l-1}
                        \lpar
                            \tfrac{1+\umach n^{\cmm}
                            C_1n}{1-n^\cmm \umach}
                        \rpar^{l-1}
                    \rbrac
            \rpar
        \end{align*}
        Reducing $\umach$ further down to $\umach\leq \frac{1}{\max\{1,C,C',C_1\}n^{\cmm+2}}$ then
        $\lpar 
            1+n^{\cmm}\umach
        \rpar^{l-1}
        \leq
        \lpar 
            1+1/n^2
        \rpar^{b_k}
        \leq
        \lpar 
            1+1/n^2
        \rpar^{n}
        \leq c_1,
        $
        and
        \begin{align*}
            \lpar
            \tfrac{1+\umach n^{\cmm}
                C_1n}{1-n^\cmm \umach}
            \rpar^{l-1}
            \leq
            \lpar
            \tfrac{1+1/n}{1-1/n^2 }
            \rpar^{b_k}
            \leq
            \lpar
            \tfrac{n}{n-1}
            \rpar^{n}
            \leq c_2,
        \end{align*}
        where both $c_1$ and $c_2$ are constants for all $n\geq 2$. This gives
        \begin{align}
            \lnorm
                \matA^{(k,2,0)}
                -
                \matQ\matA^{(k-1,2,0)}\matQ^*
            \rnorm
            &\leq
            \umach n^{\cmm}
            \cdot
                C_1n
                \lnorm\matA^{(k,2,0)}\rnorm
            \lpar
                1
                +
                b_k(1+c_1+c_2)
            \rpar
            \nonumber
            \\
            &\leq
            \umach n^{\cmm}
            \cdot
                C_2n^2
                \lnorm\matA^{(k,2,0)}\rnorm.
            \label{eq:lemma_bandwidth_halving_base_backward_error}
        \end{align}
        A bound for the norm of $\matA^{(k-1,2,0)}$ is also directly implied
        \begin{align*}
            \lnorm
                \matA^{(k-1,2,0)}
            \rnorm
            &\leq
            \lnorm \lpar \matQ^{(k)} \rpar^{-1}\rnorm^2
            \lnorm
                \matQ^{(k)}
                \matA^{(k-1,2,0)}
                \matQ^{(k)*}
            \rnorm
            \\
            &\leq
            \lnorm \lpar \matQ^{(k)} \rpar^{-1}\rnorm^2
            \lpar
                \lnorm
                    \matQ^{(k)}
                    \matA^{(k-1,2,0)}
                    \matQ^{(k)*}
                    -
                    \matA^{(k,2,0)}
                \rnorm
                +
                \lnorm                
                    \matA^{(k,2,0)}
                \rnorm
            \rpar
            \\
            &\leq
            \lpar\frac{1}{1-n^{\cmm}\umach}\rpar^{b_k-2} 
            \lpar 
            1+
            C_2\umach n^{\cmm+2} 
            \rpar
            \lnorm\matA^{(k,2,0)}\rnorm
            \\
            &\leq
            C_3\lnorm\matA^{(k,2,0)}\rnorm.
        \end{align*}

        Each rotation (Lemmas \ref{lemma:rotation_r_i_floating_point} and \ref{lemma:rotation_r_i_prime_floating_point}) requires $O(n_k^{\omega})$ floating point operations. There are at most $O(b_k^2)$ rotations, which gives a total of $O(n^2n_k^{\omega-2})$ floating point operations for the rotations. If the final similarity transformation matrix $\matQhat^{(k)}$ is not required, the algorithm can already terminate. Otherwise, some additional work is required to compute it.
        
        We now describe how to compute the matrix $\matQhat^{(k)}$ which approximates $ \matQ^{(k)}$  from the matrices $\matQ_i$ in floating point. Each matrix $\matQ_i$ is block-diagonal with block size $2n_k$. They can also be seen as block-tridiagonal matrices with block-size $n_k$. To construct each matrix $\matQ_i$ we need to perform $O(b_kn_k^{\omega})$ floating point operations. This means that all the matrices $\matQ_i$ for $i=2,\ldots,b_k$ can be constructed in $O(b_k^2n_k^{\omega})=O(n^2n_k^{\omega-2})$ floating point operations.
        
        For any $i\in2,4,6,\ldots ,b_k-2$, we can multiply $\matQ_i$ with $\matQ_{i+1}$ using Corollary \ref{corollary:block_tridiagonal_mm}. 
        Each such multiplication returns a matrix which is again block-tridiagonal, but with half the number of blocks $(b_k/2)$ and with double the block-size $(2n_k)$. 
        The cost of the multiplication is $C_{ \MM}n\cdot n_k^{\omega-1}$ operations. There are $b_k-2=n/n_k-2$ such matrices $\matQ_i$. Therefore, we can perform $(b_k-2)/2$ block-tridiagonal multiplications to compute all the matrices
        $\matQ_{3:2},\matQ_{5:4},\ldots \matQ_{b_k-1:b_k-2}$, in a total of $\tfrac{b_k}{2} C_{ \MM}n\cdot n_k^{\omega-1}$ operations, where each $\matQ_{i+1:i}$ is block-tridiagonal with double the block-size $2n_k$. We keep performing the multiplications in a binary-tree structure. At each tree level $l=2,\ldots,\log(b_k)-1$, we have half the number of block-tridiagonal matrices than in the level $l-1$, with double the block size.  The total complexity becomes
        \begin{align*}
            T_k
            &= 
            \sum_{l=1}^{\log(b_k)-1} 
            \overbrace{
                \lpar \tfrac{b_k}{2^l} \rpar
            }^{
                \text{number of multiplications at level } l
            }
            \cdot
            \overbrace{
                C_{ \MM}n(2^{l-1}n_k)^{\omega-1}
            }^{
                \text{cost per multiplication at level } l
            }
            \\
            &=
            2^{1-(\omega)}
            C_{ \MM}
            n
            \sum_{l=1}^{\log(n)-k-1}
            \tfrac{n}{2^l2^k}
            \lpar
                2^l2^k
            \rpar^{\omega-1}
            \\
            &=
            2^{1-(\omega)}
            C_{ \MM}
            n
            \sum_{l=k+1}^{\log(n)-1}
            (2^{\omega-2})^l
            \\
            &=
            O\lpar
                n^2
                \lpar
                    S_{\omega}(\log(n)-1) - S_{\omega}(k)
                \rpar 
            \rpar,
        \end{align*}
        where $S_{x}(m):=\sum_{l=1}^{m} \lpar 2^{x-2} \rpar^{l}.$ In floating point arithmetic, since $\eta>0$, for every $k=2,\ldots,\log(n)-2$ the geometric series gives
        \begin{align*}
            S_{\omega}(\log(n)-1)-S_{\omega}(k) = \frac{
                (2^{\omega-2})^{\log(n)-1}
                -
                (2^{\omega-2})^k
            }{
                2^{\omega-2}-1
            }.
        \end{align*}
        However, if one day it turns out that $\omega=2$, in exact arithmetic the term becomes $S_2(\log(n)-1)-S_2(k) = \log(n)-1-k$.
        
        In terms of the floating point errors, recall first that the matrices in the lowest level of the tree, $\matQ_i$, are approximately orthogonal, i.e. $\matQ_i=\matW_i+\matE_i$ where $\matW_i$ is exactly orthogonal and $\|\matE_i\|\leq n^\cmm\umach:=\err(0)$. Then we can write $\|\matQ_i\|\leq 1+\err(0)$, and $\err(l)$ will be used to denote the total error at the tree level $l=1,\ldots,\log(b_k)$. 
        Assume that we want to multiply $\matQ_2$ with $\matQ_3$, to form \begin{align*}
            \matQhat^{1}_{1}=\BTMM(\matQ_2,\matQ_3) = 
            \matQ_2\matQ_3 + \matE_{\BTMM}
            = 
            \matW_2\matW_3 + \matW_2\matE_3+\matE_2\matW_3+\matE_2\matE_3+\matE_{\BTMM}.
        \end{align*}
        Let $\matQ_1^1=\matW_2\matW_3$ be the true product between the unitary matrices $\matW_2$ and $\matW_3$.
        We have the following:
        \begin{align*}
            \|\matE_{\{2,3\}}\| &\leq \err(0),
            \\
            \|\matE_{\BTMM}\| &\leq \umach n^{\cmm}\|\matQ_2\|\|\matQ_3\|\leq \err(0)(1+\err(0))^2,
            \\
            \|\matE_1^1\| &= \|\matQ_1^1-\matQhat_1^1\|
            \leq 2\err(0)+\err(0)^2+\err(0)(1+\err(0))^2:=\err(1),
            \\
            \|\matQhat_1^1\| &\leq 1+\err(1),
        \end{align*}
        Now if we multiply two matrices from level $l$, e.g. $\matQhat_1^l$ and $\matQhat_2^l$, we can write similarly
        \begin{align*}
            \matQhat^{l+1}_{1}=\BTMM(\matQhat_1^l,\matQhat_2^l) = 
            \matQhat_1^l\matQhat_2^l + \matE_{\BTMM}
            = 
            \matQ_1^l\matQ_2^l + \matQ_1^l\matE_2^l+\matE_1^l\matQ_2^l+\matE_1^l\matE_2^l+\matE_{\BTMM},
        \end{align*}
        where $\matQ_{\{1,2\}}^l$ are unitary. Denoting $\matQ_1^{l+1}=\matQ_1^l\matQ_2^l$
        we have
        \begin{align*}
            \|\matE_{\{1,2\}}^l\| &\leq \err(l),
            \\
            \|\matE_{\BTMM}\| &\leq \umach n^{\cmm}\|\matQhat_1^l\|\|\matQhat_2^l\|\leq \err(0)(1+\err(l))^2,
            \\
            \err(l+1) = \|\matE_1^{l+1}\| &= \|\matQhat_1^{l+1}-\matQhat_1^{l+1}\|
            \leq 2\err(l)+\err(l)^2+\err(0)(1+\err(l))^2.
        \end{align*}
        A crude solution for the error at level $\log(n)$ is the following. 
        The errors are increasing, i.e. $\err(l+1)\geq \err(l)$. 
        For all $l$, if $\err(l)\leq 1$, then $\err(l+1)\leq 8\err(l)$. Assume that $\umach$ satisfies $\umach \leq \frac{1}{n^{\cmm+3}}$. Then $\err(0)\leq 1/n^3 \leq 1$, which means that $\err(1)\leq 8\err(0)\leq 8/n^3\leq 1$. Continuing until level $\log(n-1)$ we obtain that $\err(\log(n)-1)\leq 1/8\leq 1$. Thus, for every $\umach \leq \frac{1}{n^{\cmm+3}}$ we can upper bound the error up to level $\log(n)$ with the function
        \begin{align*}
            \err'(l) = 8\err'(l-1).
        \end{align*}
        Then $\err(\log(n))\leq \err'(\log(n)) \leq  8^{\log(n)}\err(0)=n^3\err(0)=n^{\cmm+3}\umach.$
        
        Denoting by $\matQhat^{(k)}$ the final matrix returned by the floating point multiplications at the last level, and by $\matQ^{(k)}=\matQ_2\matQ_3\ldots\matQ_{b_k-1}$ as above,  we obtain
        \begin{align*}
            \lnorm
                \matQhat^{(k)} - \matQ^{(k)}
            \rnorm
            =
            \lnorm
                \matE^{(k)}_{\matQ}
            \rnorm
            \leq
            \err(\log(b_k))\leq \err(\log(n)) \leq n^{\cmm+3}\umach,
        \end{align*}
        and from above we also already have that $\lnorm \matQ^{(k)} \rnorm \leq \sqrt{1+n^{\cmm+1}\umach}$. 
        It is straightforward to see that $\matQhat^{(k)}$ is approximately unitary, since
        \begin{align}
        \lnorm \matQhat^{(k)}\matQhat^{(k)*}-\matI \rnorm
        &=
            \lnorm 
                \lpar \matQ^{(k)}+\matE^{(k)}_{\matQ}\rpar
                \lpar \matQ^{(k)}+\matE^{(k)}_{\matQ}\rpar^*
                -
                \matI 
            \rnorm
            \nonumber
        \\
            &\leq
            \lnorm
            \matQ^{(k)}\matQ^{(k)*}
            -\matI
            \rnorm
            +
            2\lnorm \matQ^{(k)}\matE^{(k)}_{\matQ} \rnorm
            +
            \lnorm \matE^{(k)}_{\matQ} \rnorm ^2
            \nonumber
        \\
            &\leq
            n^{\beta+1}\umach  +
            2 \sqrt{1+n^{\cmm+1}\umach} n^{\cmm+3}\umach
            +
            (n^{\cmm+3}\umach)^2
            \nonumber
        \\
            &\leq
            6n^{\beta+3}\umach.
        \label{eq:bandwidth_halving_lemma_error_q_k}
        \end{align}

        Denoting by $\matE=\matQhat^{(k)} - \matQ^{(k)}$, and combining the aforementioned bound with the bound in \ref{eq:lemma_bandwidth_halving_base_backward_error}, we obtain
        \begin{align*}
            &\lnorm
                \matA^{(k,2,0)} - \matQhat^{(k)}\matA^{(k-1,2,0)}\matQhat^{(k)*}
            \rnorm
            \\
            &\qquad\leq
            \lnorm
                \matA^{(k,2,0)} - \matQ^{(k)}\matA^{(k-1,2,0)}\matQ^{(k)*}
            \rnorm
            +
            2
            \lnorm
            \matE\matA^{(k-1,2,0)}\matQ^{(k)*}
            \rnorm
            +
            \lnorm
            \matE\matA^{(k-1,2,0)}\matE^*
            \rnorm
            \\
            &\qquad\leq
            \umach n^{\cmm}
            \cdot
                C_2n^2
                \lnorm\matA^{(k,2,0)}\rnorm
            +
            2
            \cdot
            \umach n^3
            \zeta
            \lnorm
            \matA^{(k-1,2,0)}
            \rnorm            
            +
            (\umach n^3)^2
            \lnorm
            \matA^{(k-1,2,0)}
            \rnorm
            \\
            &\qquad\leq
            \lpar
                \umach n^{\cmm+2}
                    C_2
                +
                2
                \cdot
                \umach n^3
                \zeta
                C_3            
                +
                (\umach n^3)^2
                C_3
            \rpar
            \lnorm\matA^{(k,2,0)}\rnorm
            \\
            &\qquad\leq
            \umach C_4n^{\cmm+3}
            \lnorm\matA^{(k,2,0)}\rnorm,
        \end{align*}
        where $C_4$ is a constant. We conclude that $\matQhat^{(k)}$ satisfies all the advertised properties.

        The requirement for $\umach$ is
        \begin{align*}
            \umach \leq \min \lcurly
                \frac{1}{\max\{1,C,C'\}n^{\cmm+1}},
                \frac{1}{\max\{1,C,C',C_1\}n^{\cmm+2}},
                \frac{1}{n^{\cmm+3}}
            \rcurly,
        \end{align*}
        which are all satisfied by setting
        $
            \umach \leq \frac{1}{C_5n^{\cmm+3}},
        $
        where $C_5=\max\{1,C,C',C_1\}$.
    \end{proof}
\end{lemma}

\begin{theorem}[Restatement of Theorem \ref{theorem:stable_tridiagonal_reduction}]
    \label{theorem:stable_tridiagonal_reduction_appendix}
    There exists a floating point implementation of the tridiagonal reduction algorithm of \cite{schonhage1972unitare}, which takes as input a Hermitian matrix $\matA$, and returns a tridiagonal matrix $\matTtilde$, and (optionally) an approximately unitary matrix $\matQtilde$. If the machine precision $\umach$ satisfies
    $
        \umach \leq \epsilon\frac{1}{cn^{\beta+4}},
    $
    where $\epsilon\in(0,1)$, $c$ is a constant, and $\cmm$ is the same as in Corollary \ref{corollary:alg_qr}, which translates to $O(\log(n)+\log(1/\epsilon))$ bits of precision, then the following hold:
    \begin{align*}
        \lnorm \matQtilde\matQtilde^*-\matI\rnorm \leq \epsilon,
        \quad
        \text{and}
        \quad
        \lnorm \matA - \matQtilde\matTtilde\matQtilde^* \rnorm \leq \epsilon \lnorm \matA \rnorm.
    \end{align*}
    The algorithm executes at most $O\lpar
                    n^2 S_{\omega}(\log(n))
    \rpar$ floating point operations to return only $\matTtilde$, where $S_x(m)=\sum_{l=1}^m (2^{x-2})^l$.
    If $\matA$ is banded with $1\leq d\leq n$ bands, the floating point operations reduce to $O(n^2 S_{\omega}(\log(d))$.
    If $\matQtilde$ is also returned, the complexity increases to $O(n^2C_{\omega}(\log(n)))$, 
    where $C_{x}(n) := 
    \sum_{k=2}^{\log(n)-2}
    \lpar
        S_{x}(\log(n)-1) - S_{x}(k)
    \rpar$. If $\omega$ is treated as a constant $\omega\approx 2.371$ the corresponding complexities are $O(n^{\omega}), O(n^2d^{\omega-2}),$ and $O(n^{\omega}\log(n))$, respectively.
    \begin{proof}
        We apply the bandwidth halving algorithm recursively. Starting with the original matrix $\matA=\matA^{(\log(n)-2,2,0)}$, call Algorithm \ref{algorithm:halve} to produce a series of matrices $\matQhat^{(k)},\matA^{(k,2,0)}$ for all $k=\log(n)-2,\ldots,0$. For $k=0$, the notation of the final matrix $\matA^{(-1,2,0)}$ is symbolic and it denotes that the matrix is tridiagonal.

        From Lemma \ref{lemma:bandwidth_halving_floating_point_appendix}, for all $k=0,\ldots,\log(n)-2$, it holds that
        \begin{align*}
            \matA^{(k,2,0)}= \matQhat^{(k)}\matA^{(k-1,2,0)}\matQhat^{(k)*} + \matE^{(k)}, \quad \lnorm 
            \matE^{(k)}
            \rnorm
            \leq
            C_2\umach n^{\cmm+3}\lnorm\matA^{(k,2,0)}\rnorm.
        \end{align*}
        Combining this with Inequality \eqref{eq:lemma_bandwidth_halving_q_approximately_orthogonal} we have
        \begin{align*}
            \lnorm
                \matA^{(k-1,2,0)}
            \rnorm
            &=
            \lnorm
                \lpar \matQhat^{(k)} \rpar^{-1}
                \lpar
                    \matA^{(k,2,0)}
                    -
                    \matE^{(k)}
                \rpar
                \lpar\matQhat^{(k)}\rpar^{-*}
            \rnorm
            \\
            &\leq
            \lnorm
                \lpar \matQhat^{(k)}\rpar^{-1}
            \rnorm^2
            \lpar
                \lnorm
                    \matA^{(k,2,0)}
                \rnorm
                +
                \lnorm
                    \matE^{(k)}
                \rnorm
            \rpar
            \\
            &\leq
            \tfrac{1+C_2\umach n^{\cmm+3}}{1-n^{\cmm+1} \umach}
                \lnorm
                    \matA^{(k,2,0)}
                \rnorm.
        \end{align*}
        Note that for the first step $k=\log(n)-2$, the initial matrix is not block pentiadiagonal, but it rather has three block off-diagonals, in which case we need to simply modify the first rotation of Lemma \ref{lemma:bandwidth_halving_floating_point_appendix} to include also the additional block. 
        Then
        \begin{align*}
            &\matA^{(\log(n)-2,2,0)}
            \\
            &\quad=
            \matQhat^{(\log(n)-2)}\matA^{(\log(n)-3,2,0)}\matQhat^{(\log(n)-2)*} + \matE^{(\log(n)-2)}
            \\
            &\quad=
            \matQhat^{(\log(n)-2)}\lpar
            \matQhat^{(\log(n)-3)}
            \matA^{(\log(n)-4,2,0)}
            \matQhat^{(\log(n)-3)*} 
            + 
            \matE^{(\log(n)-3)}
            \rpar\matQhat^{(\log(n)-2)*} + \matE^{(\log(n)-2)}
            \\
            &\quad=
            \matE^{(\log(n)-2)}
            +
            \sum_{k=0}^{\log(n)-3}
            \lbrac
                \matQhat^{(\log(n)-2)}
                \ldots
                \matQhat^{(k+1)}
                \matE^{(k)}
                \matQhat^{(k+1)*}
                \ldots
                \matQhat^{(\log(n)-2)*}
            \rbrac
            +
            \matQhat\matA^{(-1,2,0)}
            \matQhat^*,
        \end{align*}
        Imposing a bound on $\umach\leq\frac{\epsilon}{cn^{\cmm+4}}$, this gives
        \begingroup
        \allowdisplaybreaks
        \begin{align*}
            \lnorm \matA-\matQhat\matTtilde\matQhat^* \rnorm
            &=
            \lnorm 
                \matA^{(\log(n)-2,2,0)}
                -
                \matQhat\matA^{(-1,2,0)}\matQhat^*
            \rnorm
            \\
            &=
            \lnorm
                \matE^{(\log(n)-2)}
                +
                \sum_{k=0}^{\log(n)-3}
                \lbrac
                    \matQhat^{(\log(n)-2)}
                    \ldots
                    \matQhat^{(k+1)}
                    \matE^{(k)}
                    \matQhat^{(k+1)*}
                    \ldots
                    \matQhat^{(\log(n)-2)*}
                \rbrac
            \rnorm
            \\
            &\leq
            \lnorm
                \matE^{(\log(n)-2)}
            \rnorm
            +
            \sum_{k=0}^{\log(n)-3}
            \lbrac
                \lnorm
                    \matE^{(k)}
                \rnorm
                \prod_{m=k+1}^{\log(n)-2}
                \lnorm
                    \matQhat^{(m)}
                \rnorm^2
            \rbrac
            \\
            &\leq
            C_2\umach n^{\cmm+3}\lnorm\matA\rnorm
            +
            \sum_{k=0}^{\log(n)-2}
            \lbrac
                C_2\umach n^{\cmm+3}\lnorm\matA^{(k,2,0)}\rnorm
                \lpar
                \sqrt{1+6n^{\beta+3}\umach}
                \rpar^{\log(n)-k-1}
            \rbrac
            \\
            &\leq
            C_2\umach n^{\cmm+3}
            \lpar
                \lnorm\matA\rnorm
                +
                \sum_{k=0}^{\log(n)-2}
                \lbrac
                    \lpar
                        \tfrac{1+C_2\umach n^{\cmm+3}}{1-n^{\cmm+1} \umach}
                    \rpar^{\log(n)-k-1}
                    \lnorm
                        \matA
                    \rnorm
                    \lpar
                    1+n^{\cmm+1}\umach
                    \rpar^{\log(n)-k-1}
                \rbrac
            \rpar
            \\
            &\leq
            C_2\umach n^{\cmm+3}
                \lnorm\matA\rnorm
            \lpar
                1
                +
                \log(n)
                \lbrac
                    \lpar
                        \tfrac{1+C_2\umach n^{\cmm+3}}{1-n^{\cmm+1} \umach}
                    \rpar^{\log(n)}
                    \lpar
                    1+n^{\cmm+1}\umach
                    \rpar^{\log(n)}
                \rbrac
            \rpar
            \\
            &\leq
            C_2\umach n^{\cmm+3}
                \lnorm\matA\rnorm
            \lpar
                1
                +
                \log(n)
                \lbrac
                    \lpar
                        \tfrac{1+C_2/n}{1-1/n^3}
                    \rpar^{n}
                    \lpar
                    1+1/n^3
                    \rpar^{n}
                \rbrac
            \rpar
            \\
            &\leq
            C_2'\umach n^{\cmm+4}
                \lnorm\matA\rnorm,
        \end{align*}
        \endgroup
        where we used the assumed bound on $\umach$. $C_2'\geq C_2$ is a constant.

        We finally bound the floating point errors that arise by multiplying the matrices $\matQhat^{(k)}$ to form $\matQtilde$, which approximates $\matQhat=\matQhat^{(\log(n)-2)}\matQhat^{(\log(n)-3)}\ldots \matQhat^{(1)}\matQhat^{(0)}$. Assume the following algorithm:
        \begin{enumerate}
            \item $\matQtilde^{(0)}=\matQhat^{(0)}$
            \item For $k=1,\ldots,\log(n)-2$:
            \item $\quad \matQtilde^{(k)}\leftarrow  \MM(\matQhat^{(k)},\matQtilde^{(k-1)})$.
            \item $\matQtilde=\matQtilde^{(\log(n)-2)}$.
        \end{enumerate}
        Each multiplication introduces an error $\matE^{ \MM}_{k}$. Specifically:
        \begin{align*}
            \matQtilde^{(k)} = 
             \MM(\matQhat^{(k)},\matQtilde^{(k-1)}) 
            = 
            \matQhat^{(k)}\matQtilde^{(k-1)} + \matE^{ \MM}_{k},
        \end{align*}
        where, from Corollary \ref{corollary:alg_qr}, $\lnorm \matE_k^{ \MM}\rnorm 
        \leq 
        \umach n^{\cmm} \lnorm  \matQtilde^{(k-1)} \rnorm \lnorm \matQhat^{(k)} \rnorm 
        \leq
        \umach n^{\cmm} \lpar \sqrt{1+6n^{\cmm+3}\umach} \rpar \lnorm  \matQtilde^{(k-1)} \rnorm 
        $.
        Then
        \begin{align*}
            \lnorm 
                 \matQtilde^{(k)}
            \rnorm 
            &\leq
            \lpar
                1+\umach n^{\beta}
            \rpar
            \lpar
                \sqrt{1+6n^{\cmm+3}\umach} 
            \rpar
            \lnorm  
                \matQtilde^{(k-1)} 
            \rnorm
            \\
            &\leq
            \lpar
                1+\umach n^{\beta}
            \rpar^{k}
            \lpar
                \sqrt{1+6n^{\cmm+3}\umach}
            \rpar^{k} 
            \lnorm  
                \matQtilde^{(0)} 
            \rnorm
            \\
            &\leq
            \lpar
                1+\umach n^{\beta}
            \rpar^{k}
            \lpar
                \sqrt{1+6n^{\cmm+3}\umach}
            \rpar^{k+1}. 
        \end{align*}
        Now take
        \begin{align*}
            \matQtilde^{(\log(n)-2)} 
            &=
            \matQhat^{(\log(n)-2)}
            \matQtilde^{(\log(n)-3)}
            + 
            \matE^{ \MM}_{\log(n)-2}
            \\
            &= 
            \matQhat^{(\log(n)-2)}
            \lpar 
               \matQhat^{(\log(n)-3)}
                \matQtilde^{(\log(n)-4)}
                + 
                \matE^{ \MM}_{\log(n)-3}
            \rpar
            + 
            \matE^{ \MM}_{\log(n)-2}
            \\
            &=
            \matQhat^{(\log(n)-2)}\matQhat^{(\log(n)-3)}\ldots \matQhat^{(0)} 
            + 
            \matE^{ \MM}_{\log(n)-2}
            + 
            \sum_{k=0}^{\log(n)-3}
                \matQhat^{(\log(n)-2)}
                \ldots
                \matQhat^{(k+1)}
                \matE^{ \MM}_{k}
                ,
        \end{align*}
        which means that
        \begingroup
        \allowdisplaybreaks
        \begin{align*}
            &\lnorm 
                \matQtilde^{(\log(n)-2)} 
                -
                \matQhat^{(\log(n)-2)}\ldots \matQhat^{(1)}\matQhat^{(0)}  
            \rnorm
            \\
            &\quad\leq
            \lnorm \matE^{ \MM}_{\log(n)-2}\rnorm
            + 
            \sum_{k=1}^{\log(n)-3}
                \lnorm 
                    \matE^{ \MM}_{k}
                \rnorm
                \prod_{l=k+1}^{\log(n)-2}        
                \lnorm
                    \matQhat^{(l)}
                \rnorm
            \\
            &\quad\leq
            \sum_{k=1}^{\log(n)-2}
                \lnorm
                    \matE^{ \MM}_{k}
                \rnorm
                \lpar
                    \sqrt{1+6n^{\cmm+3}\umach}
                \rpar^{\log(n)-2-k}  
             \\
            &\quad\leq
            \sum_{k=1}^{\log(n)-2}
                \umach n^{\cmm} \lpar \sqrt{1+6n^{\cmm+3}\umach} \rpar \lnorm  \matQtilde^{(k-1)} \rnorm
                \lpar
                    \sqrt{1+6n^{\cmm+3}\umach}
                \rpar^{\log(n)-2-k}  
            \\
            &\quad\leq
            \umach n^{\cmm}  
            \lpar
                \sqrt{1+6n^{\cmm+3}\umach}
            \rpar^{\log(n)-2}
            \sum_{k=1}^{\log(n)-2}
                \lpar
                1+\umach n^{\beta}
            \rpar^{k-1}
            \lpar
                \sqrt{1+6n^{\cmm+3}\umach}
            \rpar^{k}
            \\
            &\quad\leq
                \umach n^{\cmm}  
                \lpar
                    \sqrt{1+6n^{\cmm+3}\umach}
                \rpar^{\log(n)-2}
                (\log(n)-2)
                \lpar
                    1+\umach n^{\beta}
                \rpar^{\log(n)-3}
                \lpar
                    \sqrt{1+6n^{\cmm+3}\umach}
                \rpar^{\log(n)-2}
            \\
            &\quad\leq
                \umach n^{\cmm+1}  
                \lpar
                    1+6n^{\cmm+3}\umach
                \rpar^{\log(n)-2}
                \lpar
                    1+\umach n^{\beta}
                \rpar^{\log(n)-3}
            \\
            &\quad\leq
            \umach n^{\cmm+1} 
            \lpar
                1+1/n
            \rpar^{n}
            \\
            &\quad\leq
            C_Q\umach n^{\cmm+1},
        \end{align*}
        \endgroup
        where $C_Q$ is a small constant (the bounds are quite loose, but it does not affect the asymptotic complexity analysis as long as they stay polynomial in $n$).

        Next, we recall that the matrix $\matQhat$ is approximately orthogonal. From Lemma \ref{lemma:bandwidth_halving_floating_point_appendix}, for each $k=0,\ldots,\log(n)-2$:
        \begin{align*}
            \lnorm
                \matQhat^{(k)}\ldots \matQhat^{(1)}\matQhat^{(0)}
            \rnorm
            \leq
            \lpar \sqrt{1+n^{\cmm+1}\umach}\rpar^{k+1},
            \quad
            \text{and}
            \quad
            \lnorm
                \lpar
                \matQhat^{(k)}\ldots \matQhat^{(1)}\matQhat^{(0)} 
                \rpar^{-1}
            \rnorm
            \leq
            \lpar \frac{1}{\sqrt{1-n^{\cmm+1}\umach}}\rpar^{k+1}.
        \end{align*}
        Since $n^{\cmm+1}\umach\leq 1/n^2\ll 1$, using again Bernoulli's inequalities it follows that 
        \begin{align*}
            \lnorm \matQhat \rnorm^2 &\leq 
                (1+n^{\cmm+1}\umach)^{\log(n)-1} 
                \leq 
                (1+n^{\cmm+1}\umach)^{n} 
                \leq
                1+n^{\cmm+2}\umach,
            \\
            \lnorm \matQhat^{-1} \rnorm^{-2} &\geq (1-n^{\cmm+1}\umach)^{\log(n)-1} \geq 1-(\log(n)-1)n^{\cmm+1}\umach\geq1-n^{\cmm+2}\umach,
        \end{align*}
        and since all the (squared) singular values of $\matQhat$ are inside $[1-n^{\cmm+2}\umach,1+n^{\cmm+2}\umach]$ then
        \begin{align}
            \lnorm \matQhat\matQhat^*-\matI \rnorm \leq n^{\cmm+2}\umach.
        \end{align}

        We can now bound the advertised error. First, if we denote $\matE_{\matA}=\matA-\matQhat\matTtilde\matQhat^*$ then
        \begin{align*}
            \lnorm
                \matTtilde
            \rnorm
            &=
            \lnorm
                \matQhat^{-1}\lpar \matA-\matE_{\matA} \rpar \matQhat^{-*}
            \rnorm
            \\
            &\leq
            \lnorm
            \matQhat^{-1}
            \rnorm^2
            \lpar
                \lnorm \matA \rnorm + \lnorm \matE_{\matA} \rnorm
            \rpar
            \\
            &\leq
            \frac{\lpar
                1 + C_2'\umach n^{\cmm+4}
            \rpar}{1-n^{\cmm+2}\umach}
            \lnorm \matA \rnorm
        \end{align*}
        Denoting by $\matE_{\matQhat}=\matQtilde-\matQhat$, we have that
        \begin{align}
            \lnorm \matA-\matQtilde\matTtilde\matQtilde^* \rnorm
            &\leq
            \lnorm
                \matA-\matQhat\matTtilde\matQhat
            \rnorm
            +
            2\lnorm \matE_{\matQhat}\rnorm \lnorm \matQhat\rnorm
            \lnorm \matTtilde \rnorm
            +
            \lnorm \matE_{\matQhat} \rnorm^2
            \lnorm \matTtilde \rnorm
            \nonumber
            \\
            &\leq
            C_2'\umach n^{\cmm+4}
                \lnorm\matA\rnorm
            +
            \lpar
                2C_Q\umach n^{\cmm+1}
                \sqrt{1+n^{\cmm+2}\umach}
                +
                (C_Q\umach n^{\cmm+1})^2
            \rpar
            \frac{\lpar
                1 + C_2'\umach n^{\cmm+4}
            \rpar}{1-n^{\cmm+2}\umach}
            \lnorm \matA \rnorm
            \nonumber
            \\
            &\leq
            C_{\TRID}\umach n^{\cmm+4}
                \lnorm\matA\rnorm,
            \label{eq:theorem_tridiagonal_reduction_bound_q_tilde}
        \end{align}
        where $C_{\TRID}$ is a constant.
        Note also that
        \begin{align*}
            \lnorm
                \matQtilde\matQtilde^* - \matI
            \rnorm
            &\leq
            \lnorm
                \matQtilde\matQtilde^*-\matQhat\matQhat^*
            \rnorm
            +
            \lnorm
                \matQhat\matQhat^* - \matI
            \rnorm
            \\
            &\leq
            2\lnorm \matE_{\matQhat}\rnorm \lnorm \matQhat\rnorm
            +
            \lnorm \matE_{\matQhat} \rnorm^2
            +
            \umach n^{\cmm+4}
            \\
            &\leq
            2C_Q\umach n^{\cmm+1}\sqrt{1+n^{\cmm+2}\umach}
            +
            (C_Q\umach n^{\cmm+1})^2
            +
            \umach n^{\cmm+4}
            \\
            &\leq
            C_Q'\umach n^{\cmm+4}.
        \end{align*}

        It remains to bound the complexity. 
        Depending whether the matrix $\matQtilde$ is returned or not, we have the following two cases based on Lemma \ref{lemma:bandwidth_halving_floating_point_appendix}.
        \begin{enumerate}[(i)]
            \item If the matrix $\matQtilde$ must be returned, 
            $O\lpar
                n^{2}\lpar 
                    S_{\omega}(\log(n)-1) - S_{\omega}(k)
                \rpar
            \rpar
            $ floating point operations are required during each iteration $k=2,\ldots,\log(n)-2$, which gives a total of 
            \begin{align*}
                O\lpar
                    n^2 
                    \sum_{k=2}^{\log(n)-2}
                    \lpar
                        S_{\omega}(\log(n)-1) - S_{\omega}(k)
                    \rpar
                \rpar
                =
                O\lpar
                    n^2 
                    C_{\omega}(n)
                \rpar
                .
            \end{align*}
            floating point operations, where we used the notation $C_{x}(n) := 
                \sum_{k=2}^{\log(n)-2}
                    \lpar
                        S_{x}(\log(n)-1) - S_{x}(k)
                    \rpar$.
            \item If we ignore $\matQtilde$ and only return $\matTtilde$, only $O(n^{2}(2^{\omega-2})^k)$ floating point operations are executed per iteration $k$. In this case the total complexity is
            \begin{align*}
                O(n^2)\sum_{k=2}^{\log(n)-2} 
                (2^{\omega-2})^k
                =
                O\lpar
                    n^2 S_{\omega}(\log(n))
                \rpar
            \end{align*}
            floating point operations.
        \end{enumerate} 
        If one day it turns out that $\omega=2$, the corresponding complexities in the limit $\eta\rightarrow 0$ become as follows.
        \begin{enumerate}[(i)]
            \item If we return both $\matTtilde$ and $\matQtilde$, the total complexity is
            \begin{align*}
                O\lpar
                    n^2 \sum_{k=2}^{\log(n)-2} S_2(\log(n)-1)-S_2(k)
                \rpar
                =
                O(n^2\log^2(n)).
            \end{align*}
            \item On the other hand, if only $\matTtilde$ is returned, then the complexity becomes \begin{align*}
                O(n^2S_2(\log(n))=O(n^2\log(n)).
            \end{align*}
            Moreover, if $\matA$ is banded with bandwidth $1\leq b \leq n-1$ (number of off-diagonals), then we start the bandwidth-halving recursion from $k=\min\{ j\in\mathbb{N} | 2^j>b\}$. This gives a complexity of $O(n^2S_2(\log(n))=O(n^2\log(d))$. 
        \end{enumerate}
    \end{proof}
\end{theorem}
We have the following direct Corollary.

\begin{corollary}
    \label{corollary:tridiagonal_reduction_eigenvalues}
    The eigenvalues of the matrix $\matTtilde$ returned by the algorithm of Theorem \ref{theorem:stable_tridiagonal_reduction} satisfy
    \begin{align*}
        \labs \lambda_i(\matTtilde)-\lambda_i(\matA) \rabs \leq \epsilon \|\matA\|.
    \end{align*}
    \begin{proof}
        Since $\matQtilde$ is approximately unitary, it can be written as $\matQtilde = \matQ+\matE_{\matQ}$, where $\matQ$ is unitary (exactly) and $\|\matE_{\matQ}\|\leq C_Q'\umach n^{\beta+4}$. Then we can finally bound the error
        \begin{align*}
            \lnorm \matA - \matQ\matTtilde\matQ^* \rnorm
            &\leq \lnorm \matA - \matQtilde\matTtilde\matQtilde^*\rnorm
            + 2 \lnorm \matE_{\matQ} \rnorm \lnorm \matTtilde \rnorm 
            +
            \lnorm \matE_{\matQ} \rnorm^2
            \lnorm
            \matTtilde
            \rnorm
            \\
            &\leq
            \umach n^{\cmm+4}
            \lnorm\matA\rnorm
            \lpar
                C_{\TRID}
                +
                C_Q'
                \lpar 
                    2
                    +
                    C_Q'\umach n^{\cmm+4}
                \rpar
                \frac{\lpar
                    1 + C_2'\umach n^{\cmm+4}
                \rpar}{1-n^{\cmm+2}\umach}
            \rpar
            \\
            &\leq
            C_{\TRID}' \umach n^{\cmm+4}
            \lnorm\matA\rnorm
            \\
            &\leq \epsilon \lnorm \matA \rnorm,
        \end{align*}
        where the last holds by assuming that $\umach=\Theta(\epsilon/n^{\cmm+4})$, like in Theorem \ref{theorem:stable_tridiagonal_reduction}. Since $\matQ$ is unitary, it means that the matrix $\matQ\matTtilde\matQ^*$ is similar to $\matTtilde$, and the eigenvalues of $\matQ\matTtilde\matQ^*$ satisfy $\labs \lambda_i(\matQ\matTtilde\matQ^*) - \lambda_i(\matA)\rabs \leq \epsilon \lnorm \matA\rnorm$ from Weyl's inequality.
    \end{proof}
\end{corollary}

\end{document}